\documentclass[usenatbib,useAMS,fleqn]{mnras}

\usepackage{newtxtext,newtxmath}

\usepackage[T1]{fontenc}
\usepackage{ae,aecompl}

\usepackage{tikz, graphicx}
\usepackage{amsmath}	% Advanced maths commands
\usepackage{amssymb}	% Extra maths symbols
\usepackage{multicol}        % Multi-column entries in tables
\usepackage{threeparttable}
\usepackage{xspace}
\usepackage{times}

\newcommand{\mic}{\,$\mu$m} % micron
\newcommand{\cm}{\,cm$^{-3}$} % per cubic cm
\newcommand{\Ha}{H$\alpha$\xspace}
\newcommand{\HH}{H$_2$}
\newcommand{\HII}{H{\sc ii}\xspace}     %  HII
\newcommand{\Nii}{{\rm [N}$\,${\sc ii}{\rm ]}\xspace}
\newcommand{\aco}{${\alpha}_{\rm CO}$}   %Xco
\newcommand{\tstar}{$t_{\rm star}$}   %tstar
\newcommand{\tstarref}{$t_{\rm star, ref}$}   %tstar,ref
\newcommand{\tHa}{$t_{\rm H\alpha}$}   %tHalpha
\newcommand{\tHaref}{$t_{\rm H\alpha, ref}$}   %tHalpha,ref
\newcommand{\tgas}{$t_{\rm gas}$}   %tgas
\newcommand{\tCO}{$t_{\rm CO}$}   %tCO
\newcommand{\tover}{$t_{\rm fb}$}   %tover
\newcommand{\tref}{$t_{\rm ref}$}   %tref
\newcommand{\tgal}{$\tau_{\rm gal}$}   %tgal
\newcommand{\tcloud}{$t_{\rm cloud}$}   %tcloud
\newcommand{\tff}{$t_{\rm ff}$}   %t_free-fall
\newcommand{\tcr}{$t_{\rm cr}$}   %t_crossing
\newcommand{\Msun}{\mbox{M$_\odot$}}
\newcommand{\msun}{\mbox{M$_\odot$}}

\newcommand{\myr}{\mbox{${\rm Myr}$}}
\newcommand{\gyr}{\mbox{${\rm Gyr}$}}
\newcommand{\pc}{\mbox{${\rm pc}$}}
\newcommand{\mpc}{\mbox{${\rm Mpc}$}}
\newcommand{\kpc}{\mbox{${\rm kpc}$}}
\newcommand{\kms}{\mbox{${\rm km}~{\rm s}^{-1}$}}

\usepackage{etoolbox}
\makeatletter
\patchcmd\@combinedblfloats{\box\@outputbox}{\unvbox\@outputbox}{}{\errmessage{\noexpand\@combinedblfloats could not be patched}}
\makeatother
\title[The molecular cloud lifecycle]{The\hspace{-0.2mm} lifecycle\hspace{-0.2mm} of\hspace{-0.2mm} molecular\hspace{-0.2mm} clouds\hspace{-0.2mm} in\hspace{-0.2mm} nearby\hspace{-0.2mm} star-forming\hspace{-0.2mm} disc\hspace{-0.2mm} galaxies}

\author[Chevance et al.]{M\'elanie~Chevance,$^{1}$\thanks{E-mail: \href{chevance@uni-heidelberg.de}{chevance@uni-heidelberg.de}} 
J.~M.~Diederik Kruijssen,$^{1}$
Alexander~P.~S.~Hygate,$^{2,1}$
\newauthor
Andreas Schruba,$^{3}$
Steven N.~Longmore,$^{4}$
Brent Groves,$^{5}$
Jonathan~D.~Henshaw,$^{2}$
\newauthor
Cinthya N.~Herrera,$^{6}$
Annie Hughes,$^{7,8}$
Sarah M.~R.~Jeffreson,$^{1}$
Philipp Lang,$^{2}$
\newauthor
Adam K.~Leroy,$^{9}$
Sharon E.~Meidt,$^{10}$
J\'{e}r\^{o}me Pety,$^{6,11}$
Alessandro Razza,$^{12}$
\newauthor
Erik Rosolowsky,$^{13}$
Eva Schinnerer,$^{2}$
Frank Bigiel,$^{14}$
Guillermo A.~Blanc,$^{12,15}$
\newauthor
Eric Emsellem,$^{16,17}$
Christopher M.~Faesi,$^{2}$
Simon C.~O.~Glover,$^{18}$
Daniel T.~Haydon,$^{1}$
\newauthor
I-Ting Ho,$^{2}$
Kathryn Kreckel,$^{2,1}$
Janice C.~Lee,$^{19}$
Daizhong Liu,$^{2}$
Miguel Querejeta,$^{16,20}$
\newauthor
Toshiki Saito,$^{2}$
Jiayi Sun,$^{9}$
Antonio Usero$^{20}$
and Dyas Utomo$^{9}$
\\\\
Affiliations are listed at the end of the paper
}

\date{Accepted X{\sevensize xxxx} XX. Received 2019 November 6; in original form 2019 August 31}

\pubyear{2019}

\begin{document}
\label{firstpage}
\pagerange{\pageref{firstpage}--\pageref{lastpage}}
\maketitle

\begin{abstract}
It remains a major challenge to derive a theory of cloud-scale ($\la100$~pc) star formation and feedback, describing how galaxies convert gas into stars as a function of the galactic environment. Progress has been hampered by a lack of robust empirical constraints on the giant molecular cloud (GMC) lifecycle. We address this problem by systematically applying a new statistical method for measuring the evolutionary timeline of the GMC lifecycle, star formation, and feedback to a sample of nine nearby disc galaxies, observed as part of the PHANGS-ALMA survey. We measure the spatially-resolved ($\sim100$~pc) CO-to-\Ha flux ratio and find a universal de-correlation between molecular gas and young stars on GMC scales, allowing us to quantify the underlying evolutionary timeline. GMC lifetimes are short, typically $10{-}30~\myr$, and exhibit environmental variation, between and within galaxies. At kpc-scale molecular gas surface densities $\Sigma_{\rm H_2}\geq8~\msun~\pc^{-2}$, the GMC lifetime correlates with time-scales for galactic dynamical processes, whereas at $\Sigma_{\rm H_2}\leq8~\msun~\pc^{-2}$ GMCs decouple from galactic dynamics and live for an internal dynamical time-scale. After a long inert phase without massive star formation traced by \Ha ($75{-}90$~per~cent of the cloud lifetime), GMCs disperse within just $1{-}5~\myr$ once massive stars emerge. The dispersal is most likely due to early stellar feedback, causing GMCs to achieve integrated star formation efficiencies of $4{-}10$~per~cent. These results show that galactic star formation is governed by cloud-scale, environmentally-dependent, dynamical processes driving rapid evolutionary cycling. GMCs and \HII\ regions are the fundamental units undergoing these lifecycles, with mean separations of $100{-}300~\pc$ in star-forming discs. Future work should characterise the multi-scale physics and mass flows driving these lifecycles.
\end{abstract}


%%%%%%%%%%%%%%%%%%%%%%%%%%%%%%%%%%%%%%%%%%%%%%%%%%

\section{Introduction}
The lifecycle of giant molecular clouds (GMCs) resides at the heart of the physics driving star formation and stellar feedback in galaxies. Star formation takes place in GMCs \citep[e.g.][]{Kennicutt2012} and the stellar feedback from the newly-formed stars deposits mass, metals, energy and momentum into the GMCs, eventually leading to their disruption \citep[e.g.][]{Dobbs2014,Krumholz2014} and regulating the galaxy-wide star formation rate \citep[SFR; e.g.][]{Ostriker2011,Hayward2017,Krumholz2018}. These cloud-scale ($\la100$~pc) processes determine how galaxies evolve and form stars \citep[e.g.][]{Scannapieco2012,Hopkins2013,Semenov2018,Kruijssen2019}, implying that an understanding of galaxy evolution requires describing a rich variety of physics over a wide range of spatial scales.

Recent simulations of galaxy formation and evolution are now able to resolve the scales of GMCs \citep[e.g.][]{Grand2017,Hopkins2018}, but observations have long been unable to match this step outside a small number of very nearby galaxies, mostly confined to the Local Group \citep[e.g.][]{Bolatto2008,Kawamura2009,Miura2012,Hughes2013,Corbelli2017,Faesi2018,Kruijssen2019,Schruba2019}. It is critical to obtain an empirical census of the GMC lifecycle across a wider range of galactic environments, spanning the main sequence of galaxies at $z=0$ \citep[e.g.][]{Brinchmann2004}. Covering a wide range of environments is important, because the cosmic star formation history peaked at redshift $z\sim2{-}3$ \citep{Madau2014} and it is currently unclear if the GMC lifecycle proceeded differently under the high-pressure and high-gas fraction conditions prevalent in high-redshift galaxies \citep[e.g.][]{Genzel2011,Swinbank2011,Swinbank2012,Tacconi2013,Tacconi2018}, with claimed lifetimes of up to several 100\,Myr \citep{Zanella2019}. Analytical and numerical studies predict that the GMC lifecycle likely varies with the galactic environment \citep[e.g.][]{Dobbs2013,Dobbs2015,Fujimoto2014,Jeffreson2018,Meidt2018, Meidt2019}. Due to a crucial lack of observational constraints on GMC scales across a variety of environments, it is therefore not known how most stars in the Universe formed and how they affect galaxy evolution through feedback. Thanks to the construction of large sub-mm interferometers such as the Atacama Large Millimeter/submillimeter Array (ALMA) and the Northern Extended Millimeter Array (NOEMA), it is now possible to overcome this problem.

Observationally, galaxies globally follow a `star formation relation', linking the gas surface density and the SFR surface density \citep[e.g.][]{Silk1997,Kennicutt1998}. This has been observed in a large range of galaxies, from nearby spiral galaxies \citep[e.g.][]{Bigiel2008, Blanc2009, Schruba2011, Kennicutt2012, Leroy2013} to high redshift galaxies \citep[e.g.][]{Daddi2010, Genzel2010, Tacconi2013}. These empirical, large-scale relations are often used in galaxy formation simulations to describe the relation between gas mass and SFR. However, these relations do not apply universally; they are observed to break down at scales $\la1~\kpc$ \citep[e.g.][]{Onodera2010,Schruba2010,Leroy2013,Kreckel2018,Kruijssen2019} as well as in low-density environments, such as in low surface brightness galaxies or in galaxy outskirts \citep[e.g.][]{Kennicutt1989, Martin2001, Boissier2003, Bigiel2010, Goddard2010}. 

As demonstrated by \citet{KL14}, the small-scale breakdown of the star formation relation is driven by evolutionary processes taking place at the scale of molecular clouds. The details of how the star formation relation breaks down differ between different galaxies \citep{Leroy2013}, which suggests that the evolution of individual clouds depends on the galactic environment. Such an environmental dependence has been predicted by theory. Galaxy dynamics, interstellar medium (ISM) pressure, and disc structure modify the balance of cloud formation and destruction \citep[e.g.][]{Dobbs2013,Dobbs2014,Fujimoto2014,Jeffreson2018,Krumholz2018,Meidt2018,Meidt2019} and therefore influence the population and lifecycle of GMCs. One of the major challenges in understanding the parsec-scale physics of star formation and feedback within GMCs and their impact on galaxy evolution is to resolve the scales of individual clouds within galaxies and empirically constrain their lifecycles as a function of the galactic environment \citep[e.g.][]{Lada2010, Hopkins2013}. This requires a large ($>100$) sample of GMCs and star-forming regions for a wide variety ($\gtrsim10$) of galaxies covering different ISM conditions (e.g.\ densities, pressures) and kinematics (e.g.\ dynamical time-scales) to obtain sufficiently representative statistics. In this paper, we address this problem by characterising the GMC lifecycle across nine star-forming disc galaxies spanning a range of properties.

There are two main competing theories describing the cloud lifecycle, which predict strong differences in the time evolution of individual clouds. In one theory, clouds are described as long-lived, stable objects, supported by magnetic fields, such that star formation proceeds over long time-scales \citep[$\sim100~\myr$; e.g.][]{Mckee1989, VazquezSemadeni2011}. In a second theory, clouds are transient objects, undergoing gravitational free-fall or dynamical dispersal, in which star formation proceeds on a dynamical time-scale \citep[$\sim10~\myr$; e.g.][]{Elmegreen2000, Hartmann2001,Dobbs2011}. Measuring the molecular cloud lifetime is a key step to distinguish between these two theories, but so far observations have only been made for small samples and have yielded a variety of different outcomes, largely due to differences in experiment design and the use of differing, subjective ways of defining objects (i.e.\ GMCs and \HII regions).

GMC lifetimes are well in excess of a human lifetime, requiring the use of indirect methods to constrain their lifecycles. Long cloud lifetimes ($\sim 100$\,Myr) have been suggested by the presence of molecular clouds in between spiral arms \citep[i.e.\ `inter-arm' GMCs, see e.g.][]{Scoville1979,Scoville2004, Koda2009}. Short cloud lifetimes ($10{-}50$\,Myr) have been measured by classifying the clouds based on their star formation activity \citep{Engargiola2003, Blitz2007, Kawamura2009, Murray2011, Miura2012, Corbelli2017}, or by quantifying the fraction of CO-bright versus \Ha-bright lines of sight across each galaxy \citep{Schinnerer2019}. Finally, evolution along orbital streamlines has been used to infer cloud lifetimes, leading to values ranging from $\sim 1$~Myr in the Central Molecular Zone of the Milky Way \citep{Kruijssen2015, Henshaw2016, Barnes2017, Jeffreson2018b} to $20{-}50$~Myr in the central $\sim$ 4\,kpc of M51 \citep{Meidt2015}. While the classification of clouds based on their star formation activity is the most promising method due to its general applicability, the subjective definition of cloud categories and the fact that the cloud structure needs to be resolved to classify them limits the application of this method to very nearby galaxies, mostly confined to the Local Group. This can potentially be overcome by describing star formation in galaxies as a multi-scale process, such that the cloud lifecycle is inferred without needing to resolve individual GMCs \citep[see below]{Kruijssen2018}.

In addition to the overall cloud lifetime, the co-existence (or overlap) time-scale of GMCs and \HII regions provides an essential diagnostic for probing the cloud-scale physics of star formation and feedback. By measuring how long GMCs survive after the appearance of ionising photons generating \Ha emission, it is possible to identify the feedback mechanism driving GMC dispersal. In principle, GMC dispersal could be driven by a number of processes, including supernovae, stellar winds, photoionisation, and radiation pressure \citep[e.g.][]{Krumholz2014, Dale2015, Krumholz2019}. Crucially, many of these processes act on different time-scales and all of these have different environmental dependences, so that it is possible to determine their relative importance by measuring the characteristic time-scale for gas dispersal as a function of the galactic environment. Other feedback mechanisms, such as protostellar outflows are local mechanisms which are incapable of disrupting entire GMCs \citep{Bally2016, Krumholz2019}. 

Capitalising on the unprecedented resolution and sensitivity achieved by ALMA, the method introduced by \citet{Kruijssen2018} develops a statistical approach for empirically characterising the evolutionary timeline of cloud evolution, star formation, and feedback by describing the multi-scale nature of the star formation relation in galaxies. This method is based on the fact that the breakdown of the star formation relation between the gas mass and the SFR on sub-kpc scales is highly sensitive to the time-scales governing the GMC lifecycle. In brief, it uses cloud-scale variations of the flux ratio between tracers of molecular gas and star formation to determine the relative occurrence of both phases, thus constraining their relative durations. This approach is agnostic about observational criteria often used to define GMCs or \HII regions, and instead defines these empirically as emission peaks that are positioned on the timeline describing their evolutionary lifecycles in a way that is independent from their neighbours. We refer to these as `independent regions' and find that the identified objects resemble classical GMCs and \HII regions in terms of their spatial dimensions. Rather than needing to resolve individual GMCs, as was the case in previous methods, this new technique only requires resolving the mean separation length of the combined population of GMCs and \HII regions (a few 100\,pc). This enables the systematic application of this method across a significant part of the local galaxy population (out to $\sim$ 50\,Mpc with ALMA's currrent capabilities).

As a result, we can now determine the molecular cloud lifetime, the time-scale for cloud dispersal by feedback, as well as the characteristic distance between individual sites of star formation. In turn, these constrain a variety of additional physical quantities, such as the integrated cloud-scale star formation efficiency, the mass loading factor (i.e.\ the feedback-driven mass outflow rate in units of the SFR), and the feedback outflow velocity. The accuracy of the method has been demonstrated using simulated galaxies \citep{Kruijssen2018} and it has been applied to the individual galaxies NGC300 \citep{Kruijssen2019}, the Large Magellanic Cloud \citep{Ward2019} and M33 \citep{Hygate2019b}. \citet{Kruijssen2019} find a de-correlation between gas and star formation in NGC300, which they attribute to the rapid evolutionary cycling between molecular gas, star formation, and cloud destruction by stellar feedback. \citet{Fujimoto2019} build on this empirical result to propose that this de-correlation is a fundamental test of feedback physics in galaxy simulations, as it probes the dispersive effect of stellar feedback on GMCs.

Here, we greatly expand the sample of galaxies analysed, to cover a relevant range of galaxy types and environments in which star formation takes place and obtain representative constraints on the molecular cloud lifecycle. The systematic application of these novel analysis techniques requires a high-resolution, multi-wavelength census of the nearby galaxy population. To date, the main challenge has been to obtain homogeneous sensitivity mapping of the molecular gas across a large number of galaxies at $\sim100$~pc resolution. With the PHANGS\footnote{Physics at High Angular Resolution in Nearby GalaxieS; \url{http://phangs.org}.} collaboration, we have now made this step by carrying out the PHANGS-ALMA survey (A.~K.\ Leroy et al.\ in prep.), which is mapping the CO emission across the star-forming discs of $\sim80$ nearby galaxies at a point-source sensitivity high enough to detect molecular clouds down to $\sim10^5~\msun$. In combination with matched-resolution, ground-based \Ha maps, these observations probe the multi-phase structure of galaxies at 1\arcsec\ resolution ($35{-}162$~pc for our sample), which allows us to characterise the lifecycle of cloud evolution, star formation, and feedback as a function of galactic environment.

In this paper, we present the first systematic characterisation of the molecular cloud lifecycle in a first sample of nine nearby star-forming galaxies. The structure of the paper is as follows. In Section~\ref{sec:obs}, we first present the observational data, describing the distribution of gas and SFR tracers in nine nearby galaxies. In Section~\ref{sec:method}, we summarise the statistical method used to derive the characteristic quantities of star formation and feedback. In Section~\ref{sec:results}, we then present the derived quantities characterising star formation and feedback processes for the nine galaxies, and carry out a detailed comparison of the measured molecular cloud lifetimes with analytical predictions in Section~\ref{sec:tgas_variations}. Finally, we discuss the physical interpretation and implications of the results in Section~\ref{sec:discussion}, and conclude in Section~\ref{sec:ccl}.

\section{Observations}

We now summarise our galaxy sample, describe the observational data used to trace molecular gas and recent star formation, and discuss the procedure used to obtain total SFRs.

\label{sec:obs}
    \subsection{Sample selection}

\begin{table*}
\centering
\caption{Physical and observational properties of the targets.}
\begin{tabular}{lcccccccc}
\hline
Galaxy                  &  Stellar mass$^a$ &  Metallicity$^b$  & CO$^c$   & CO$^{c,d}$ & \Ha   &  \Ha & Spatial  \\
                        &   &  & resolution     & sensitivity  & observations & resolution    & resolution$^e$ \\
                        & [$\log_{10}$\,M$_\odot$]  & [12+log(O/H)] &  [\arcsec]    & [K km s$^{-1}$]  &  &  [\arcsec]  & [pc] \\
                        
\hline
NGC628 (\textit{M74})    & 10.24    &8.65   & 1.12    & 1.3       & WFI        & 0.87  & 53 \\
NGC3351 (\textit{M95})   & 10.28    &8.80   & 1.46    & 1.2       & KPNO$^f$ & 1.16  & 84  \\
NGC3627 (\textit{M66})   & 10.67    &8.33   & 1.57    & 1.6       & WFI        & 1.44  & 109  \\
NGC4254 (\textit{M99})   & 10.52    &8.62   & 1.71    & 0.7       & WFI        & 1.21  & 154  \\
NGC4303 (\textit{M61})   & 10.67    &8.69   & 1.84    & 1.1       & WFI        & 0.81  & 162  \\
NGC4321 (\textit{M100})  & 10.71    &8.69   & 1.64    & 1.0       & KPNO$^f$ & 1.28  & 137  \\
NGC4535                  & 10.49    &8.68   & 1.56    & 0.8       & WFI        & 1.20  & 139  \\
NGC5068                  &  9.36    &8.39   & 1.00    & 1.8       & WFI        & 1.15  & 35  \\
NGC5194 (\textit{M51})   & 10.73    &8.84    & 1.06    & 4.9      & KPNO$^f$  & 1.83  & 79  \\
\hline
\multicolumn{9}{l}{$^a$ Stellar masses are presented in \citet{Leroy2019b} and references therein, with typical uncertainties of 0.1~dex.}\\
\multicolumn{9}{l}{$^b$Mean molecular gas mass-weighted metallicity based on \citet{Pilyugin2014}, with typical uncertainties of 0.03~dex.}\\
\multicolumn{9}{l}{$^c$CO(1-0) for NGC5194 from \citet{Schinnerer2013}; CO(2-1) for all other galaxies.}\\
\multicolumn{9}{l}{$^d$Characteristic $1\sigma$ sensitivity corresponding to the root-mean-squared noise across the integrated intensity CO map at the resolution}\\
\multicolumn{9}{l}{given in the preceding column.}\\
\multicolumn{9}{l}{$^e$Deprojected spatial resolution accounting for inclination, calculated as the maximum of the CO and \Ha maps. The adopted distances}\\
\multicolumn{9}{l}{and inclinations are listed in Table~\ref{tab:input}.}\\
\multicolumn{9}{l}{$^f$\textit{Spitzer} Infrared Nearby Galaxy Survey (SINGS) \citep{Kennicutt2003}.}
\end{tabular}
\label{tab:observations}
\end{table*}

We use a sample of nine galaxies with currently available, high-resolution, multi-wavelength coverage, targeted by the PHANGS-ALMA survey (P.I.\ E.~Schinnerer; A.~K.\ Leroy et al.\ in prep.). One of the main science goals of the PHANGS collaboration is to link the cloud-scale physics governing ISM structure, star formation, and feedback with galaxy evolution. One of the key steps for achieving this is to map the molecular gas distribution in nearby star-forming galaxies at high physical resolution and high sensitivity. An initial sample of 17 galaxies has been observed during ALMA Cycle~3, targeting the $J=2-1$ transition of carbon monoxide (CO) at a resolution of $\sim 1\arcsec$, which is expanded to a total of 74 galaxies in ALMA Cycle~5. The observations are described in more detail in (A.~K.\ Leroy et al.\ in prep.; also see \citealt{Sun2018} and \citealt{Utomo2018}), but we summarise them below. The galaxies have been selected to be nearby ($\la 17$~Mpc), relatively face-on (inclination $\la 75\degr$) and to lie on or near the main sequence of star formation [$\log_{10}$(SFR/$M_\star$)~[yr$^{-1}$]~$\ga -11$ and $\log_{10}(M_\star)$~[$M_\odot$]~$\ga 9.3$]. At these distances, the spatial resolution achieved across our sample of nine galaxies ranges from $35{-}162~\pc$. This spatial scale is close to the typical sizes of GMCs measured in the Milky Way \citep{Solomon1987, Heyer2009, Miville-Deschenes2017}, implying that the galaxy sample is suitable for constraining the GMC lifecycle using our methodology (see Sections~\ref{sec:method} and~\ref{sec:discussion}).

From this initial sample of the PHANGS-ALMA CO survey, we select the objects which also have newly obtained narrow-band \Ha observations with the MPG/ESO 2.2-m Wide-Field Imager (WFI; A.~Razza et al.\ in prep.) or archival high-quality \Ha observations available at a similar resolution. This restricts our final sample to eight nearby star-forming galaxies: NGC628, NGC3351, NGC3627, NGC4254, NGC4303, NGC4321, NGC4535 and NGC5068. In addition to these targets, we also include the galaxy NGC5194 for which archival observational data of \Ha and CO(1-0) are also available at a similar spatial resolution \citep{Pety2013, Schinnerer2013}. The main characteristics of these galaxies and of the observations are summarised in Table~\ref{tab:observations}. We now summarise the properties of the CO and \Ha data used.

    \subsection{Molecular gas tracer}
As discussed previously, we measure molecular cloud lifetimes in a sample of nine star-forming disc galaxies. To ensure the homogeneity of the results, we select the same tracers of molecular gas and recent star formation across the entire galaxy sample (with the exception of NGC5194; see below). The CO~($J$=1-0) transition [denoted as CO(1-0) in the following] and the CO~($J$=2-1) transition [denoted as CO(2-1) in the following] are commonly used to trace molecular gas \citep[e.g.][]{Schuster2007, Leroy2009, Bolatto2013, Sandstrom2013}. The effective critical density for exciting CO(2-1) is higher than for CO(1-0) ($\sim 10^3$\cm\ and $\sim 10^2$\cm, respectively; \citealt{Leroy2017b}), implying that this tracer is less affected by optical depth. In addition, the mapping of CO(2-1) at a given resolution with ALMA is more efficient than for CO(1-0), which makes it a commonly observed transition for extragalactic studies of molecular gas and the tracer of choice in the PHANGS-ALMA survey (A.~K.\ Leroy et al.\ in prep.). While CO(2-1) does not trace specifically the high density molecular gas (traced for example by HCN, HCO$^+$), it is brighter and easier to observe than these high density gas tracers, allowing entire galaxies to be mapped efficiently at arcsecond resolution. We therefore use the CO(2-1) transition as a tracer of the molecular gas for all galaxies except NGC5194, for which only a CO(1-0) map is available at high resolution, observed by the Plateau de Bure Interferometer \citep[PdBI;][]{Pety2013,Schinnerer2013}.

The typical angular resolution is $1{-}2$\arcsec, allowing us to achieve a median physical spatial resolution of $\sim110~\pc$ at the distances of our target galaxies. This is sufficient to resolve the characteristic spatial separation between independent (i.e.\ temporally uncorrelated) regions (see Section~\ref{sec:results}). The angular resolution for each galaxy is listed in Table~\ref{tab:observations}. For the PHANGS-ALMA galaxies, observations have been taken using the 12-m, 7-m, and total power arrays, covering all spatial scales, including short- and zero-spacing data. For NGC5194, the combination of the PdBI with the IRAM 30-m telescope also enables the recovery of all spatial scales.

We now summarise the main steps of the data reduction of the PHANGS data, which are described in detail in A.~K.\ Leroy et al.\ (in prep.). After calibration of the $u{-}v$ data using the ALMA calibration pipeline, line-specific datasets are extracted, for each $u{-}v$ measurement set and each line of interest, for both the 12-m and 7-m array. These are then regridded to a chosen velocity grid and all measurements for a given spectral line are combined. The cubes are set to have a common channel width of $2.5~\kms$ and a typical bandwidth of typically $500~\kms$. The final cubes of the combined 12-m and 7-m data are reconstructed using several iterations of multiscale clean using the algorithm {\tt tclean} in \textsc{CASA}\footnote{See \url{https://casa.nrao.edu/}} \citep[][v5.4.0]{McMullin2007} and are convolved to a round synthesised beam (where the size of the synthesised beam is approximately equal to the original major axis beam size). For the galaxies NGC3627, NGC4254, NGC4321, and NGC5068, which were observed with two separate 150-pointing mosaics, we combine the two mosaics linearly after convolution to match the beams of the two halves. The total power data are reduced using the \textsc{CASA} v5.3.0 software package \citep[see][for details]{Herrera2019}. For each antenna, the spectra are calibrated, the "OFF" position issubtracted from the spectrum, and a first-order polynomial is fitted and subtracted to correct the baseline. The spectra are then convolved to regularly-gridded data cubes. Finally, the 12m+7m cubes are combined with the total power cubes using \textsc{CASA}'s {\tt feather} task, and corrected for the 12m+7m primary beam response. The reduction, imaging and combination of the PAWS data for NGC5194 are presented in \citet{Pety2013}.
We use the "broad" integrated intensity maps of the PHANGS v1.0 data release (A.~K.\ Leroy et al.\ in prep.). These maps recover most of the CO emission present in the data cube, including low signal-to-noise flux, resulting in high completeness \citep{Sun2018}, but also higher noise compared to maps using more restrictive masking of the faint CO emission.

	\subsection{Star formation tracer} \label{sec:sftracer}
We trace massive star formation using the \Ha line, which mostly originates from ionised gas in the vicinity of newly formed massive stars and is therefore commonly used as a tracer of the SFR \citep[see in particular the review by][]{Kennicutt2012}. We select \Ha as a star formation tracer, because it is the most readily observable tracer of young stars \citep[$\lesssim$ 10\,Myr; e.g.][]{Leroy2012, Haydon2019} with the best coverage across our sample, while minimising contamination from other objects. By contrast, the far-UV or near-UV wavelength ranges probe longer time-scales and have larger associated uncertainties \citep{Haydon2019}. The duration of the phase traced by \Ha also has the advantage of being only weakly dependent on metallicity, in contrast to UV filters. Infrared (IR) emission (e.g.\ at 24\mic) is also a common tracer of young star formation and can be used in particular to correct for extinction, which often heavily affects embedded young stars \citep[e.g][]{Kennicutt2009, Hao2011}. However, IR observations generally do not have sufficient spatial resolution for our science goal (except for the most nearby galaxies), and the duration of the IR emission phase is hard to calibrate due to contamination by evolved stars.

The \Ha maps were obtained using ground-based telescopes and include a variety of archival and new data. For NGC628, NGC3627, NGC4254, NGC4303, NGC4535 and NGC5068, we use newly-obtained \Ha data using the WFI instrument on the MPG/ESO 2.2-m telescope at La  Silla Observatory. We also observe the galaxies in the $R$-band to enable the continuum subtraction of the \Ha data. The details of these observations will be presented in A.~Razza et al.\ (in prep.).

For NGC3351, NGC4321 and NGC5194, we use wide-field high-resolution narrow-band \Ha data from the \textit{Spitzer} Infrared Nearby Galaxies Survey (SINGS; \citealt{Kennicutt2003}). The SINGS galaxies we consider here have been observed using the Kitt Peak National Observatory (KPNO) 2.1-m telescope with the CFIM imager. The data are part of IRSA data release 5.\footnote{More details about these observations can be found at \url{http://irsa.ipac.caltech.edu/data/SPITZER/SINGS/doc/ sings_fifth_delivery_v2.pdf}} SINGS also includes $R$-band observations taken with the same telescope under similar observing conditions, which are used to perform the continuum subtraction of the \Ha maps. The origin of the \Ha data and their spatial resolution are detailed in Table~\ref{tab:observations}.

We now summarise the main steps of the data reduction (for details, see A.~Razza et al.\ in prep.\ and \citealt{Schinnerer2019}). For consistency, the same steps have been applied both to the WFI and SINGS data.

\textit{Background subtraction.}
For all galaxies, the sky background is calculated by masking bad pixels and bright sources, and then masking all emission more than $3\sigma$ above the median flux of the masked image. This masked image is then smoothed by convolution with a Gaussian that has a dispersion of $\sim3$ times the full width half maximum (FWHM) of the angular resolution, in order to mask out all diffuse emission from any bright sources or from the galaxy. We then fit the residual sky background with a plane. In the cases where a good plane fit cannot be obtained (this can happen when the galaxies fill a large fraction of the image), the sky background is taken as the median of the masked image.

\textit{Seeing.}
We fit point sources in both the \Ha and $R$-band background-subtracted images with a Gaussian to determine the seeing of the observations. In cases where the results differ by more than 0.5~pixels, the higher-resolution map is convolved with a Gaussian of the appropriate width to match the lower-resolution data.

\textit{Astrometry.}
The analysis presented in Section~\ref{sec:method} carries out a spatial correlation of the CO and \Ha maps to determine the relative durations of the evolutionary phases governing the cloud lifecycle. This requires that both maps share a common astrometric system at high accuracy. Extensive tests of the method using simulated data show that for meaningful constraints on the coexistence time-scale of CO and \Ha emission (i.e.\ the `feedback time-scale' \tover, see Section~\ref{sec:method}), we require that any astrometric offset is less than $1/3$ of either the FWHM of the size of the emission peaks (GMCs and \HII regions), or of the (synthesised) beam if they are not resolved \citep{Hygate2019}. The angular resolution of our observations is $\sim 1\arcsec$. Therefore, considering the conservative case where emission peaks are not resolved, we adopt a target value of $0.3\arcsec$ for the absolute astrometric precision.

The astrometric precision of \Ha maps has been assessed by matching stellar sources to the Gaia DR2 catalogue \citep{Gaia2016, Gaia2018} and fitting $\sim50$ stars per $R$-band image, for both the SINGS data and the WFI data. The resulting astrometric precision is $0.1\arcsec{-}0.2\arcsec$, which comfortably satisfies our conservative target precision of $0.3\arcsec$.

\textit{Flux calibration.}
The flux scale is determined using the median of the flux ratios for a selection of non-saturated stars that are matched between the \Ha and the $R$-band images. Since the $R$-band continuum has to be subtracted from the \Ha line, but the \Ha line also contributes to the $R$-band data, we proceed iteratively to produce the flux-calibrated \Ha images. First, the ratio of the relative flux calibration is used to determine the scale of the $R$-band continuum in the \Ha narrow-band image. With this flux basis, we perform a first estimate of the \Ha flux, which is then used to determine the contribution of the \Ha line to the $R$-band. We repeat this procedure until the successive continuum estimates differ by less than 1~per~cent. To obtain the continuum-subtracted \Ha image, this estimate of the continuum is then subtracted from the narrow-band image.

\textit{Filter transmission and \Nii contamination.}
We correct the measured \Ha flux for the loss due to the filter transmission, using the spectral shape of the narrow-band filter and the position of the \Ha line within the filter. It is also corrected for the contribution of the \Nii lines at 654.8 and 658.3~nm to the narrow-band filter flux. For all galaxies in our sample, we first assume a uniform contamination of 30~per~cent due to the \Nii lines. This value has been calibrated with high-spectral resolution observations of \HII regions in NGC628 with the VLT/MUSE instrument \citep{Kreckel2016} and comparable results are found from similar observations in NGC3627, NGC4254 and NGC4535, where we measure median ratios \Nii/\Ha of 0.30 to 0.32, with standard deviations of 0.05 to 0.06 (K.~Kreckel et al.\ in prep.). In addition, our galaxies span a relatively narrow range in metallicity and our radial coverage is limited to the inner part of the disk \citep{Kreckel2019} and no trend of the \Nii/\Ha ratio with galactocentric radius (or metallicity) is observed (K.~Kreckel et al.\ in prep.), which supports our assumption of a uniform contamination throughout the sample. We then estimate the contribution of the \Ha and \Nii lines to the narrow-band image based on the redshift of the galaxy and on the filter transmission curves \citep[see Table~2 in][]{Schinnerer2019}, before finally subtracting the effective contribution from the \Nii lines to the \Ha\ flux.

\textit{Extinction.}
We correct for foreground Galactic extinction using the calibration from \citet{Schlafly2011} and assuming a \citet{Fitzpatrick1999} reddening law with $R_V = 3.1$. Note that we do not carry out a spatially-resolved correction for internal extinction of the \Ha line, but instead perform a single, global extinction correction. This is achieved by calculating the global SFR across the field of view using far-UV and 22\mic\ emission and re-scaling the \Ha map accordingly (see Section~\ref{sec:sfr} for details). None the less, our lack of a spatially-resolved extinction correction may cause us to underestimate emission from young, embedded \HII regions, or fail to detect them at all. In practice, this means that we trace the unembedded phase of star formation, when \Ha is visible. Previous studies of nearby galaxies have shown a high spatial correlation between 24$\micron$ emission and \Ha\ emission \citep[e.g.][]{PerezGonzalez2006, Prescott2007, Kruijssen2019}. Most importantly, we aim to derive visibility time-scales rather than absolute flux levels. As long as an \HII region is visible above the noise level, it is included in our analysis. The absolute brightness of a region is only used as a weight when calculating the population-averaged gas-to-SFR flux ratio (see Section~\ref{sec:method}).\footnote{The fluxes of all CO and \Ha\ regions are summed before calculating the gas-to-SFR ratio. As a result, bright regions contribute more to these total fluxes. In the following, we therefore refer to our measurements as `flux-weighted averages'.}

Even if we might expect some impact of extinction on the local \Ha\ flux, we stress that we calibrate the measured timeline for cloud evolution and star formation based on the duration of the unembedded phase of star formation (see Section~\ref{sec:Heisenberg}). As a result, neglecting the embedded star-formation phase during which \Ha\ is not yet visible would only result in underestimating the duration of the overlap between the gas and the young stellar phases. However, it would not affect the cloud lifetime, because the sum of the durations of the inert CO-bright phase and the overlap phase is unaffected by extinction, even if the division between both phases may change. For the same reason, the total duration of the evolutionary cycle would not change either. If an embedded \Ha\ phase is present, \citet{Haydon2019b} demonstrate that this could potentially affect the measured duration of the overlap phase, but only for global gas surface densities larger than 20\,\Msun~pc$^{-2}$ at solar metallicity. This conclusion is based on a numerical simulation that overpredicts the effects of extinction and thus represents a lower limit. Extinction thus affects less than half of our sample (see Figure~\ref{fig:gal_prop} -- galaxies that reside above this (highly conservative) threshold of 20\,\Msun~pc$^{-2}$ across more than 4~kpc in galactocentric radius are NGC3627, NGC4254, NGC4303, and NGC5194). We will quantify the impact of extinction further in future works (M.~Chevance et al.\ in prep.; J.~Kim et al.\ in prep.). 

Finally, sources other than \HII regions generating \Ha emission (such as supernova remnants) might contribute to the \Ha flux and thus contaminate our measurements. However, these generally have considerably smaller sizes and lower luminosities than \HII regions \citep[e.g.][]{Kreckel2018}. As a result, their contribution to the flux-weighted average \Ha flux in each aperture is negligible -- \citet{Peters2017} quantify this using three-dimensional radiation-hydrodynamical simulations and estimate that shocks contribute less than 10~per~cent of the total \Ha flux. A large-scale reservoir of \Ha emission tracing diffuse ionised gas is also commonly observed in galaxies \citep[e.g.][]{Monnet1971,Dettmar1990,Hoopes1996,Oey2007,Kreckel2016,Lacerda2018}. We describe how we separate this diffuse emission reservoir from the compact emission tracing \HII\ regions in Section~\ref{sec:diffuse}.

The final CO and \Ha images of all nine galaxies are shown in Figure~\ref{fig:COmaps}. The figures also indicate the field of view used in the analysis (this is mostly limited by the field of view of the CO observations, but it also excludes some map edges where the noise is high, e.g.\ in NGC628 and NGC5194), the galactic centres and bar regions (which are excluded by eye because of blending effects, see below), and the foreground stars and background galaxies that have been masked. These maps are used throughout this paper.

\begin{figure*}
\begin{center}
\includegraphics[trim=6mm 53mm 61mm 130mm, clip=true, width=7.6cm]{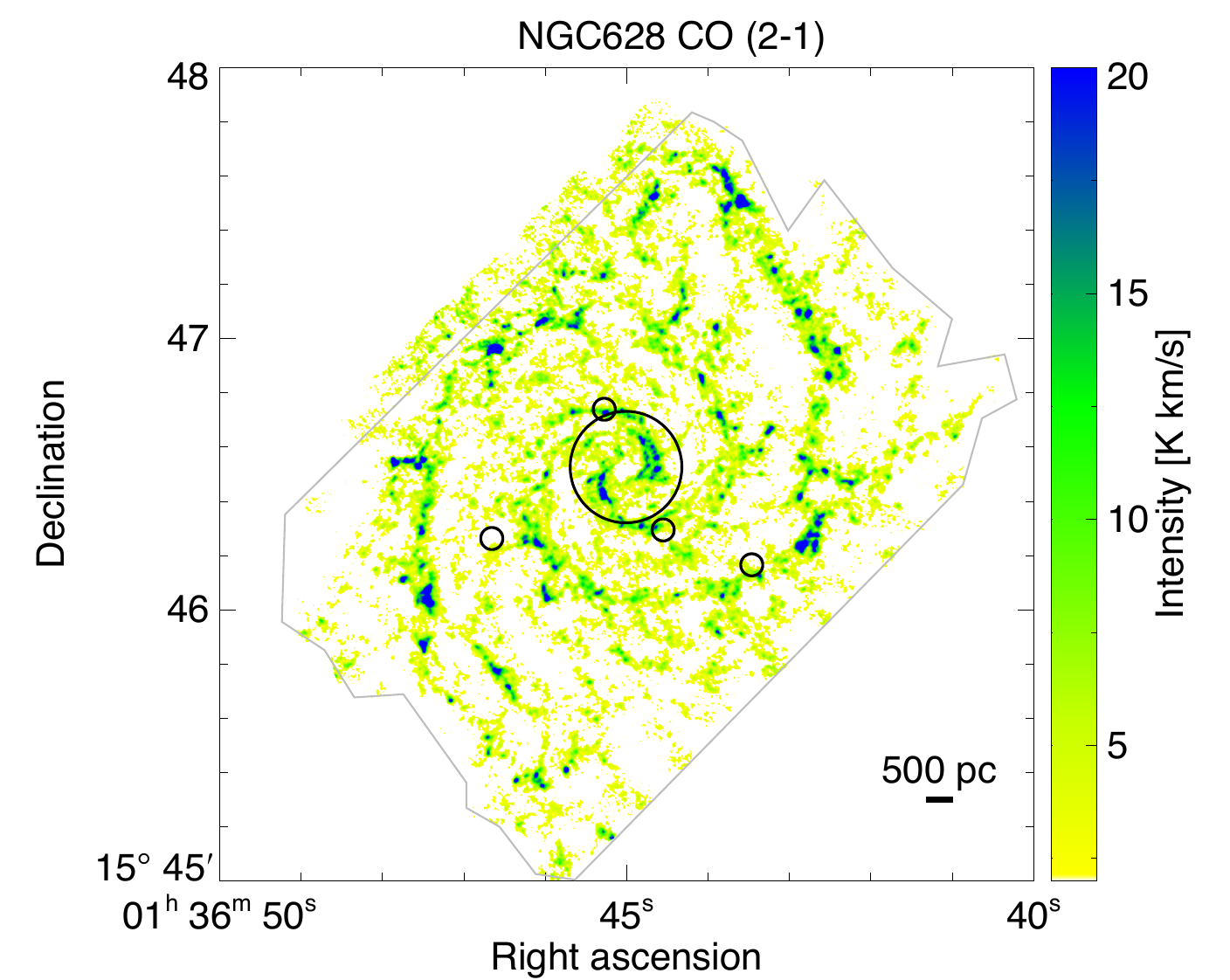}
\includegraphics[trim=6mm 53mm 61mm 130mm, clip=true, width=7.6cm]{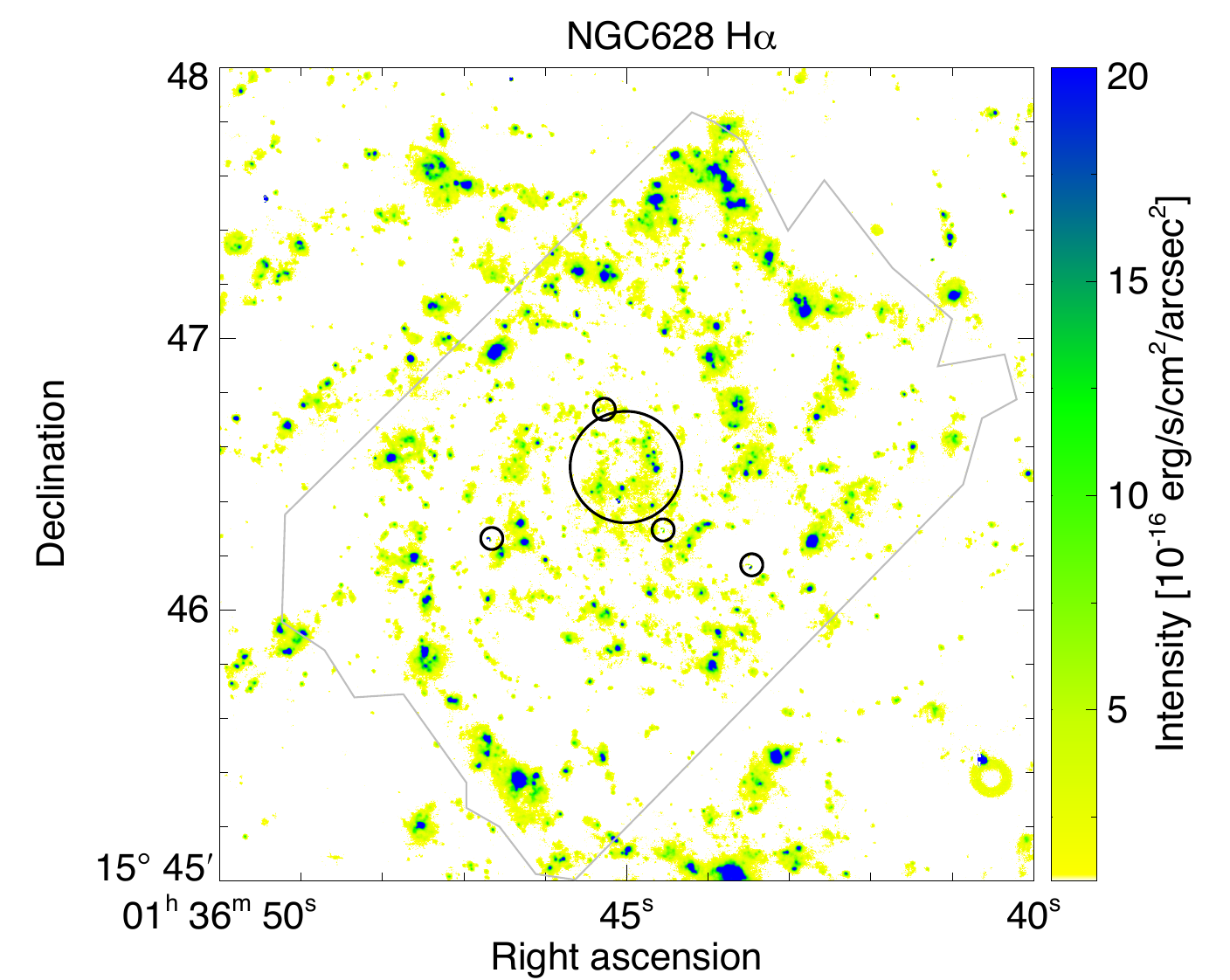}
\includegraphics[trim=2mm 53mm 61mm 130mm, clip=true, width=7.8cm]{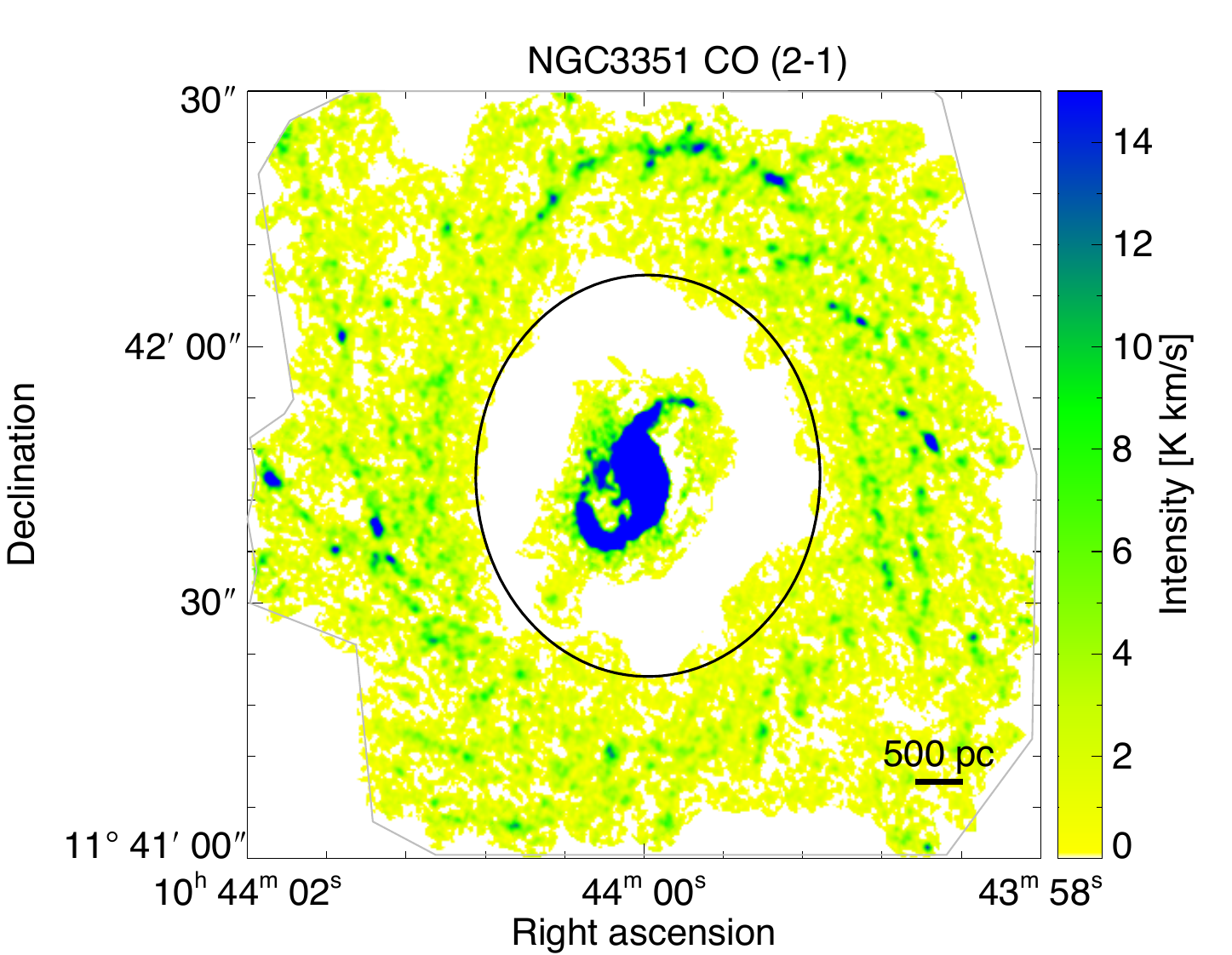}
\includegraphics[trim=2mm 53mm 61mm 130mm, clip=true, width=7.8cm]{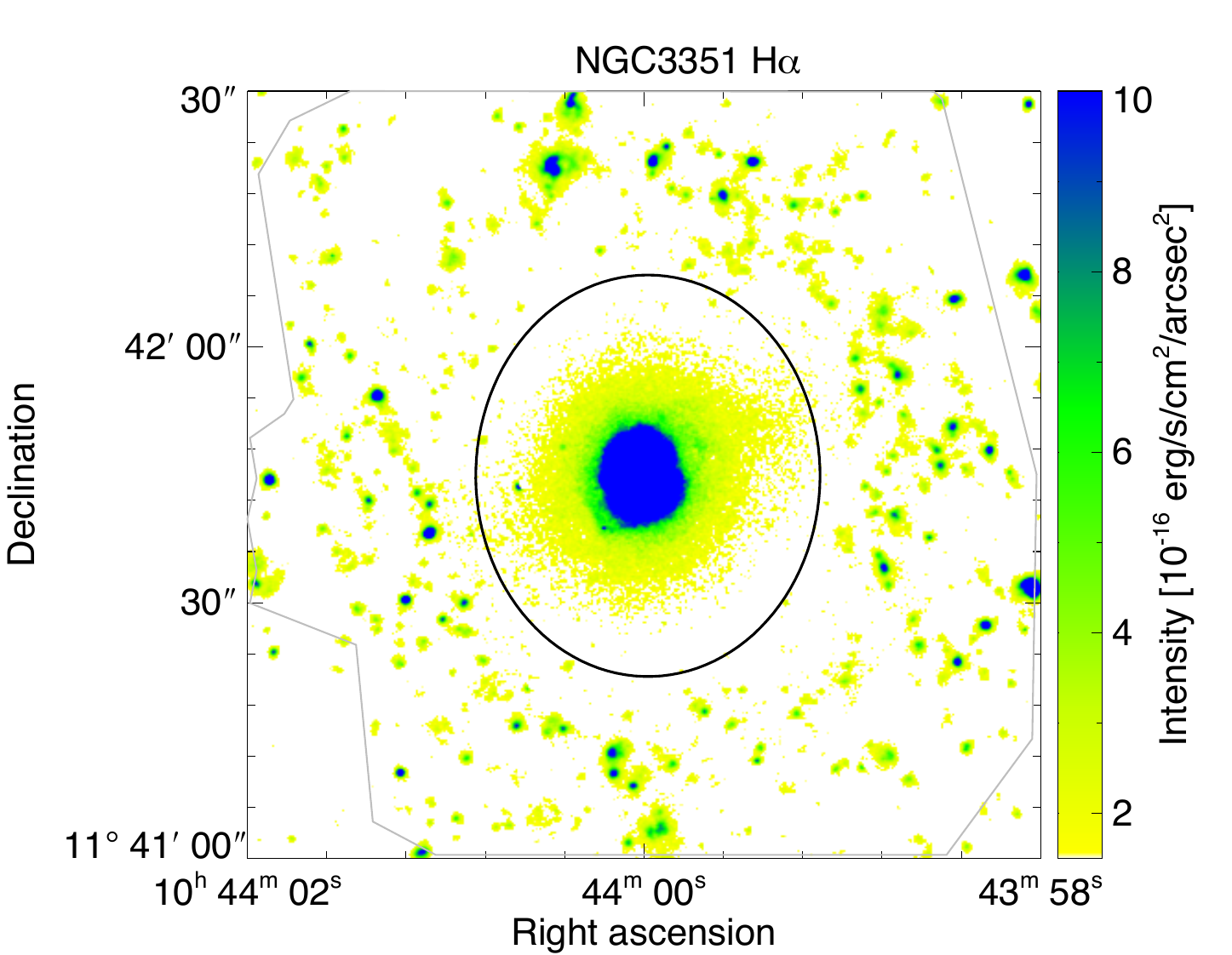}
\includegraphics[trim=3mm 53mm 62mm 58mm, clip=true, width=7.7cm]{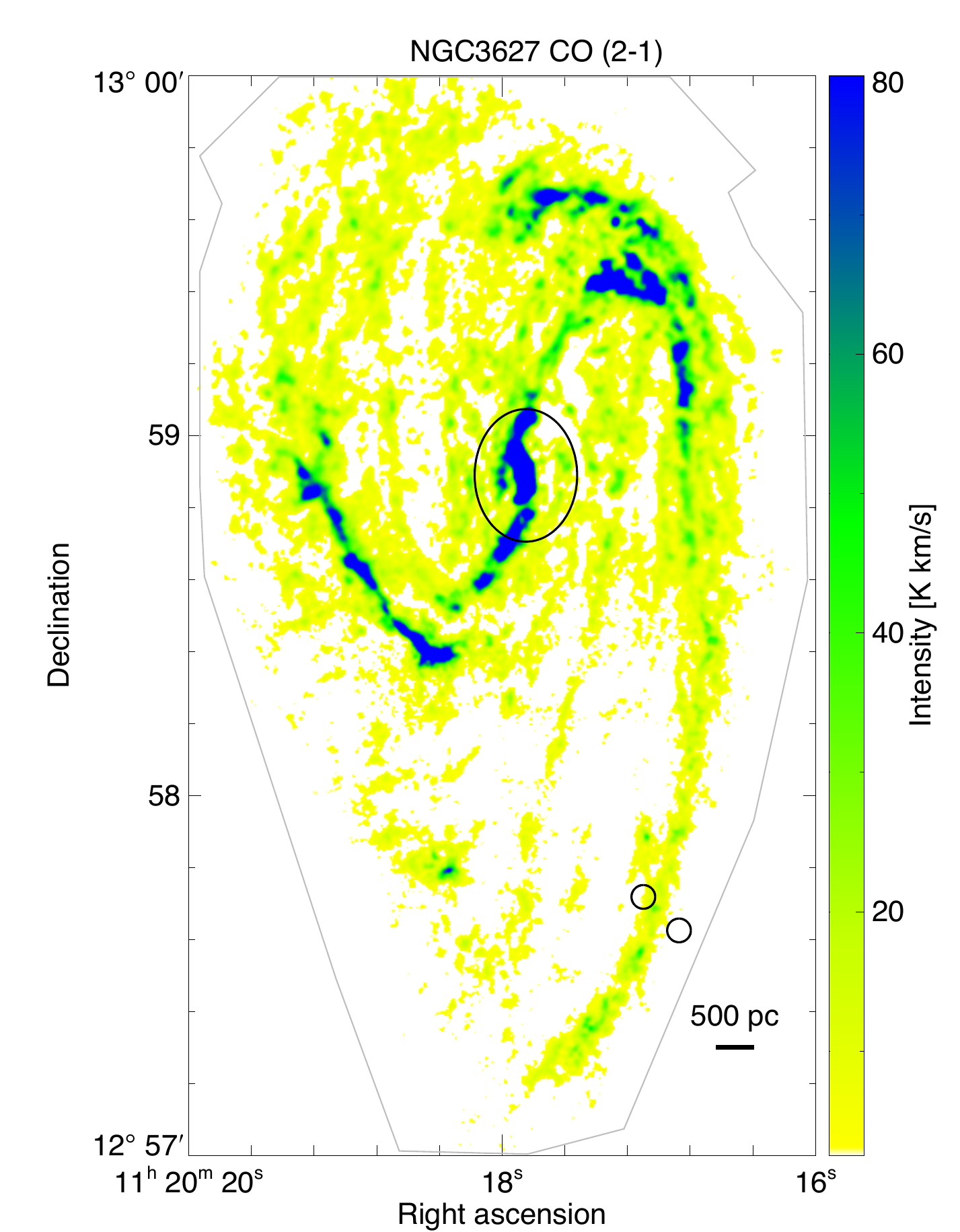}
\includegraphics[trim=3mm 53mm 62mm 58mm, clip=true, width=7.7cm]{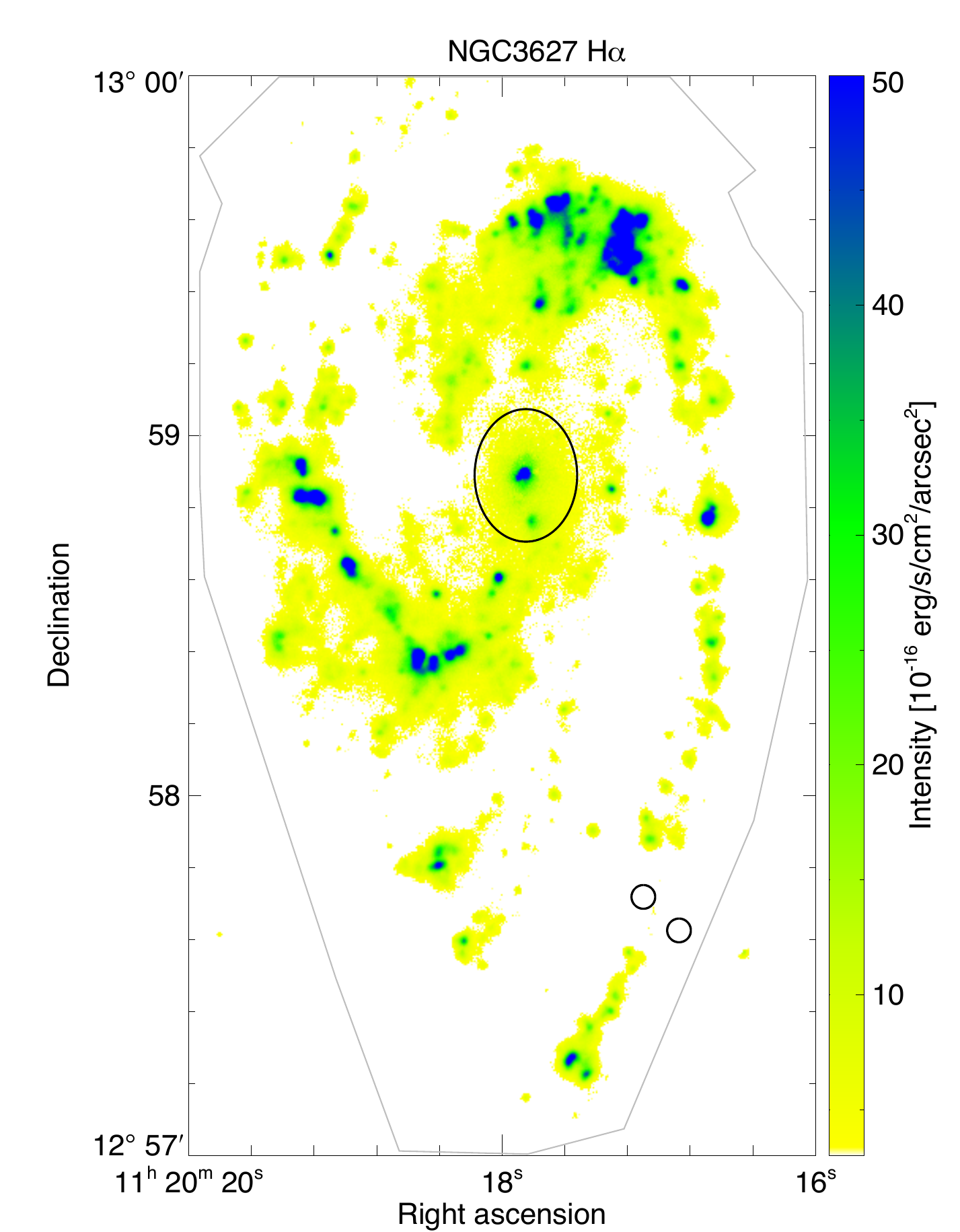}
\caption{Maps of the nine observed galaxies. The left column shows the $^{12}$CO integrated intensity maps ($J=1-0$ transition for NGC5194, $J=2-1$ transition for the other galaxies of the sample; in units of ${\rm K}~\kms$) and the right column shows the \Ha\ intensity maps (in units of $10^{-16}~{\rm erg}~{\rm s}^{-1}~{\rm cm}^{-2}~{\rm arcsec}^{-2}$). To minimise the effects of blending between independent regions, the centre of each galaxy (black central ellipse) is identified by eye and excluded from the analysis. We also mask foreground stars and background galaxies (black circles). The analysis of this work has been performed in the area delineated by the grey line, where both CO and \Ha\ have been observed (the field of view is primarily limited by the size of the CO map, excluding map edges with high noise when necessary). A linear scale of 500\,pc is indicated in each of the CO images.
\label{fig:COmaps}}
 \end{center} 
\end{figure*}
\begin{figure*}
\begin{center}
\includegraphics[trim=2mm 53mm 53mm 110mm, clip=true, width=8.5cm]{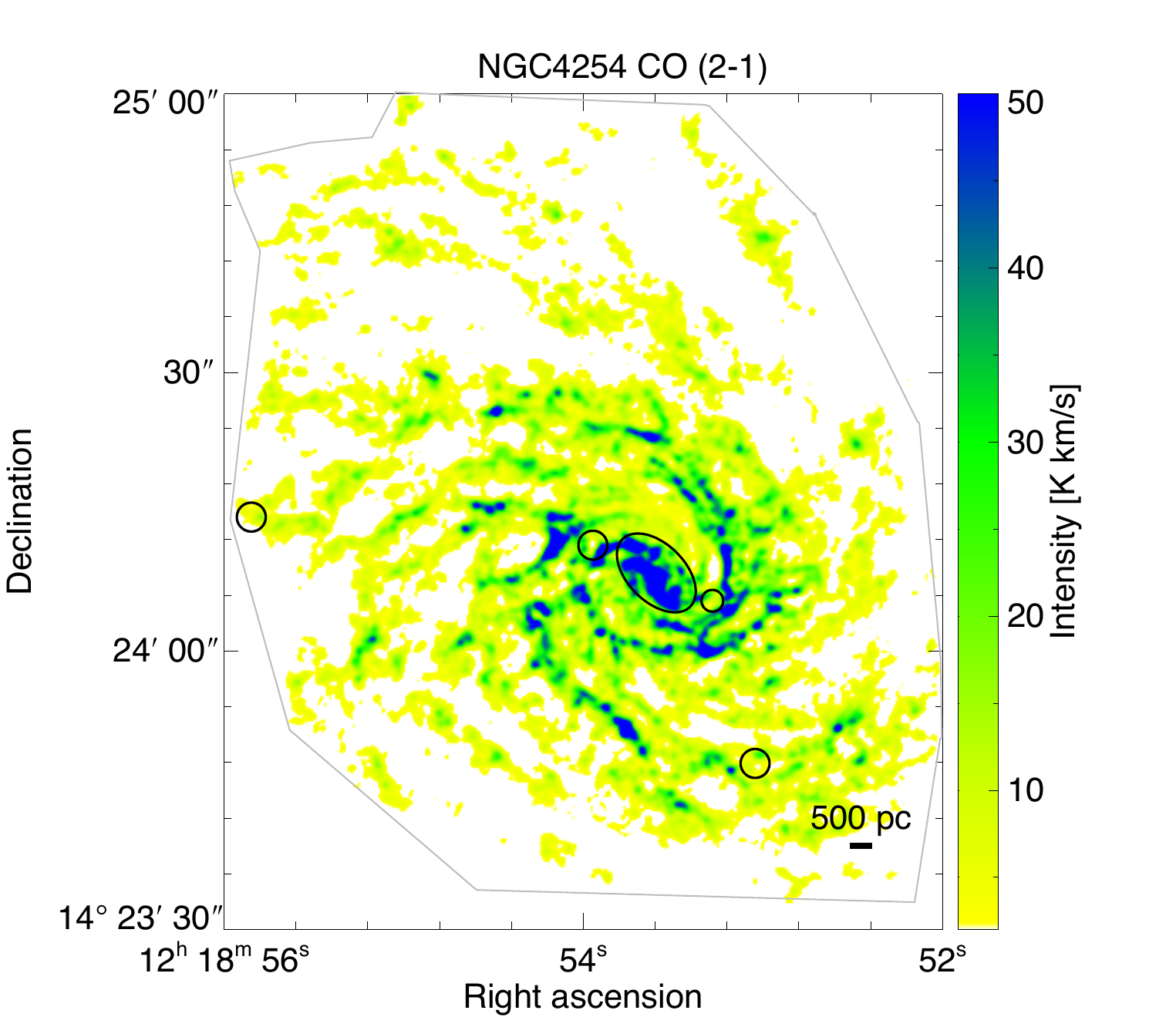}
\includegraphics[trim=2mm 53mm 53mm 110mm, clip=true, width=8.5cm]{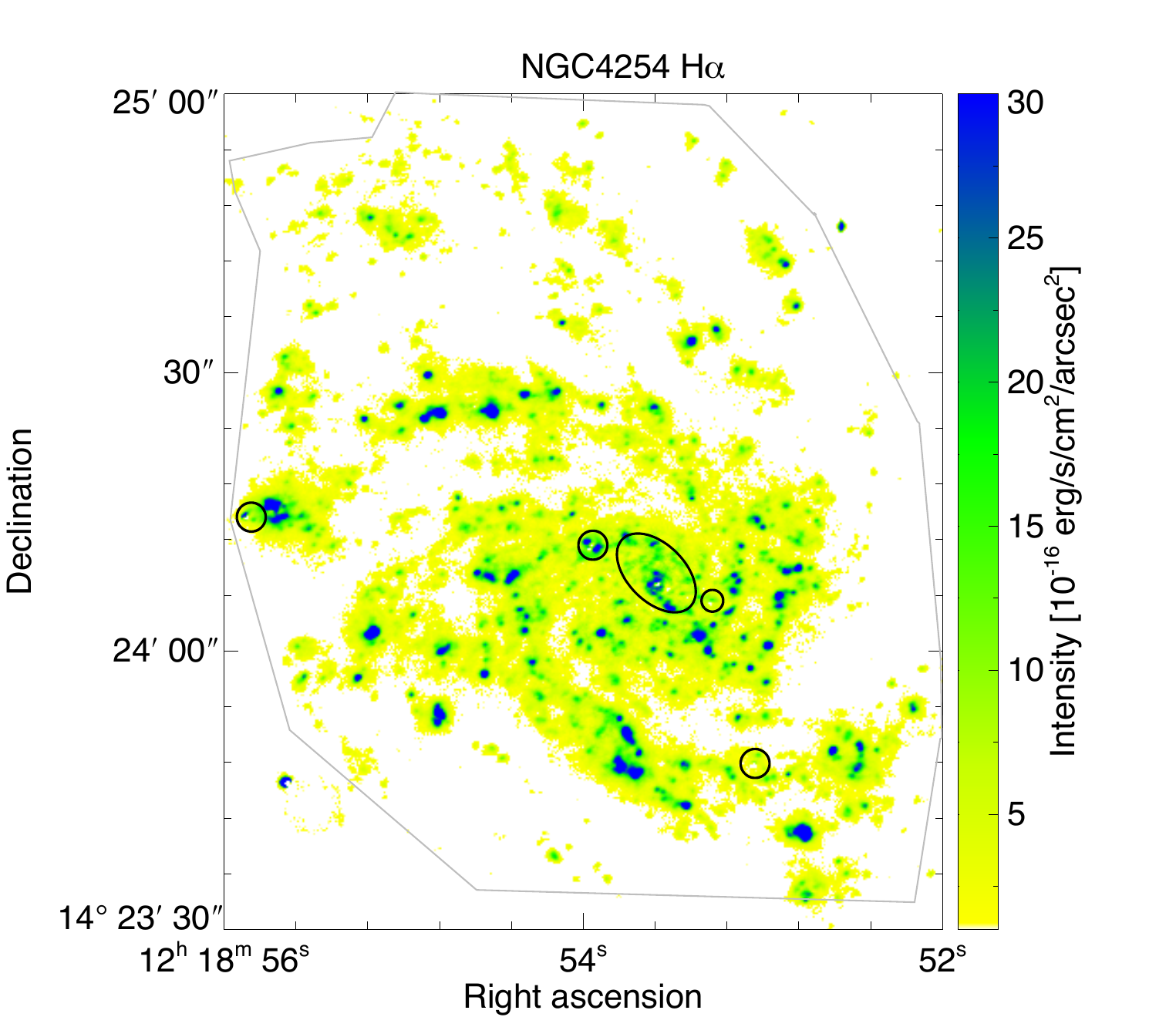}
\includegraphics[trim=2mm 53mm 53mm 120mm, clip=true, width=8.5cm]{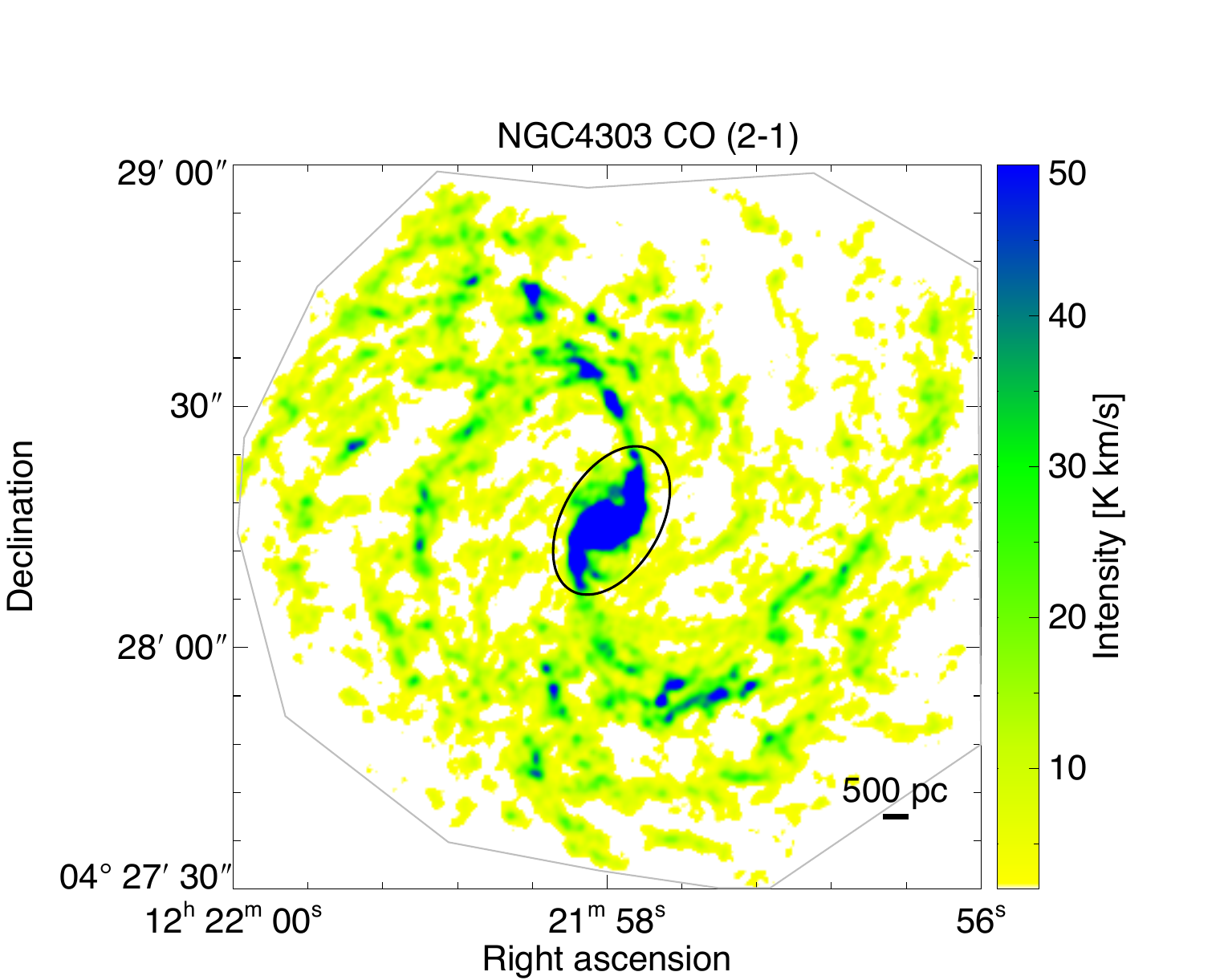}
\includegraphics[trim=2mm 53mm 53mm 120mm, clip=true, width=8.5cm]{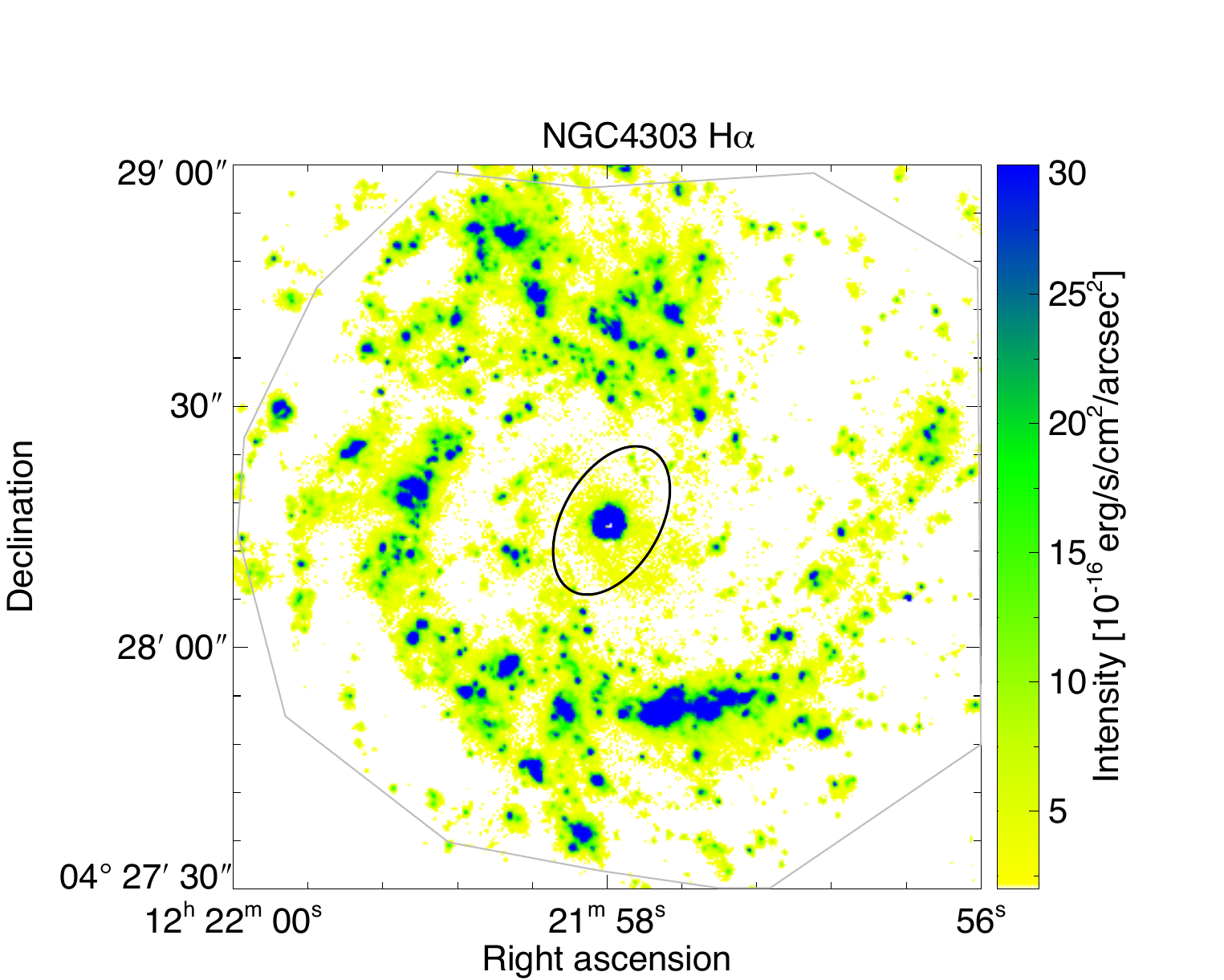}
\includegraphics[trim=2mm 53mm 53mm 130mm, clip=true, width=8.5cm]{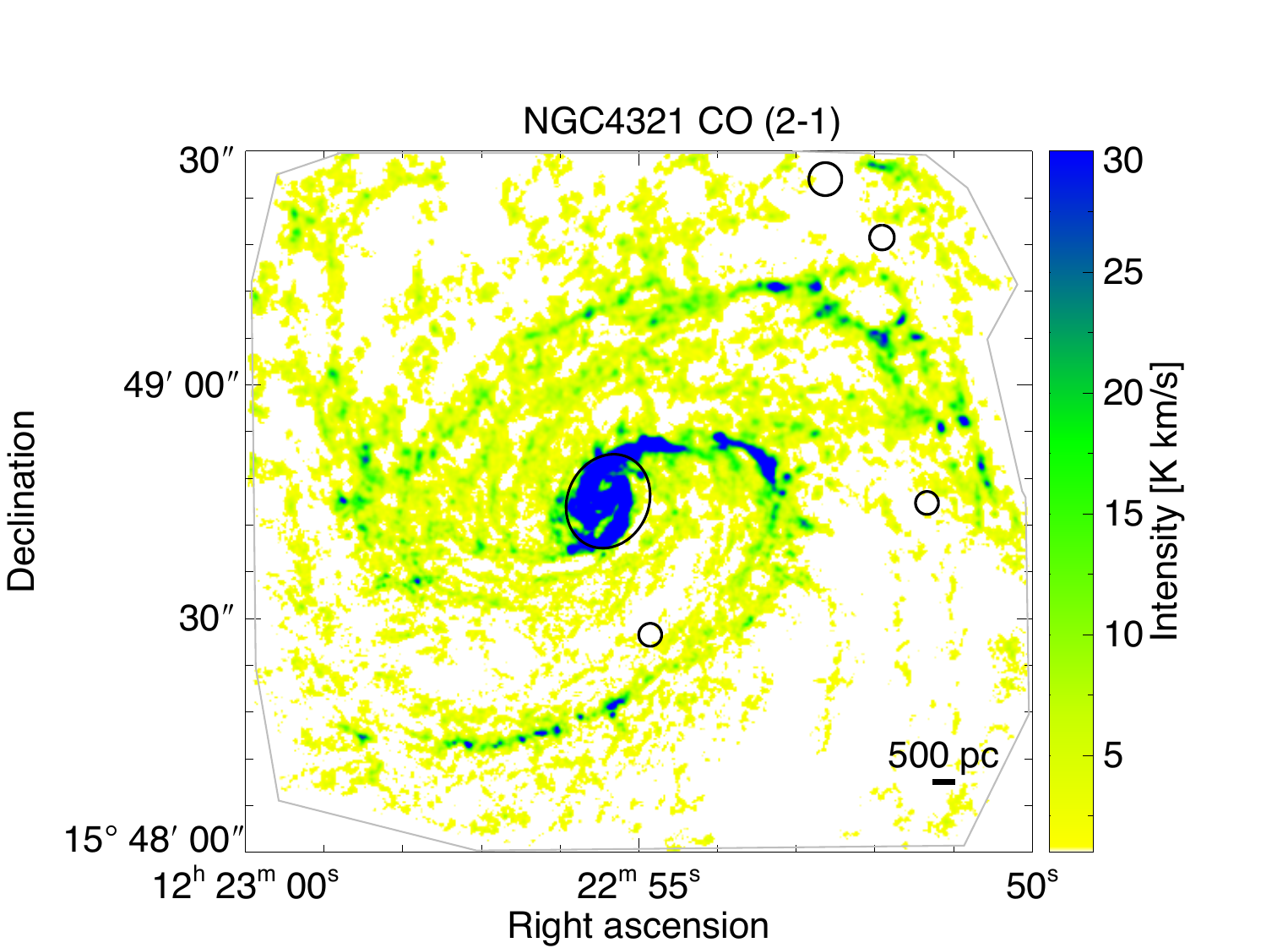}
\hspace{6mm}
\includegraphics[trim=2mm 53mm 53mm 130mm, clip=true, width=8.5cm]{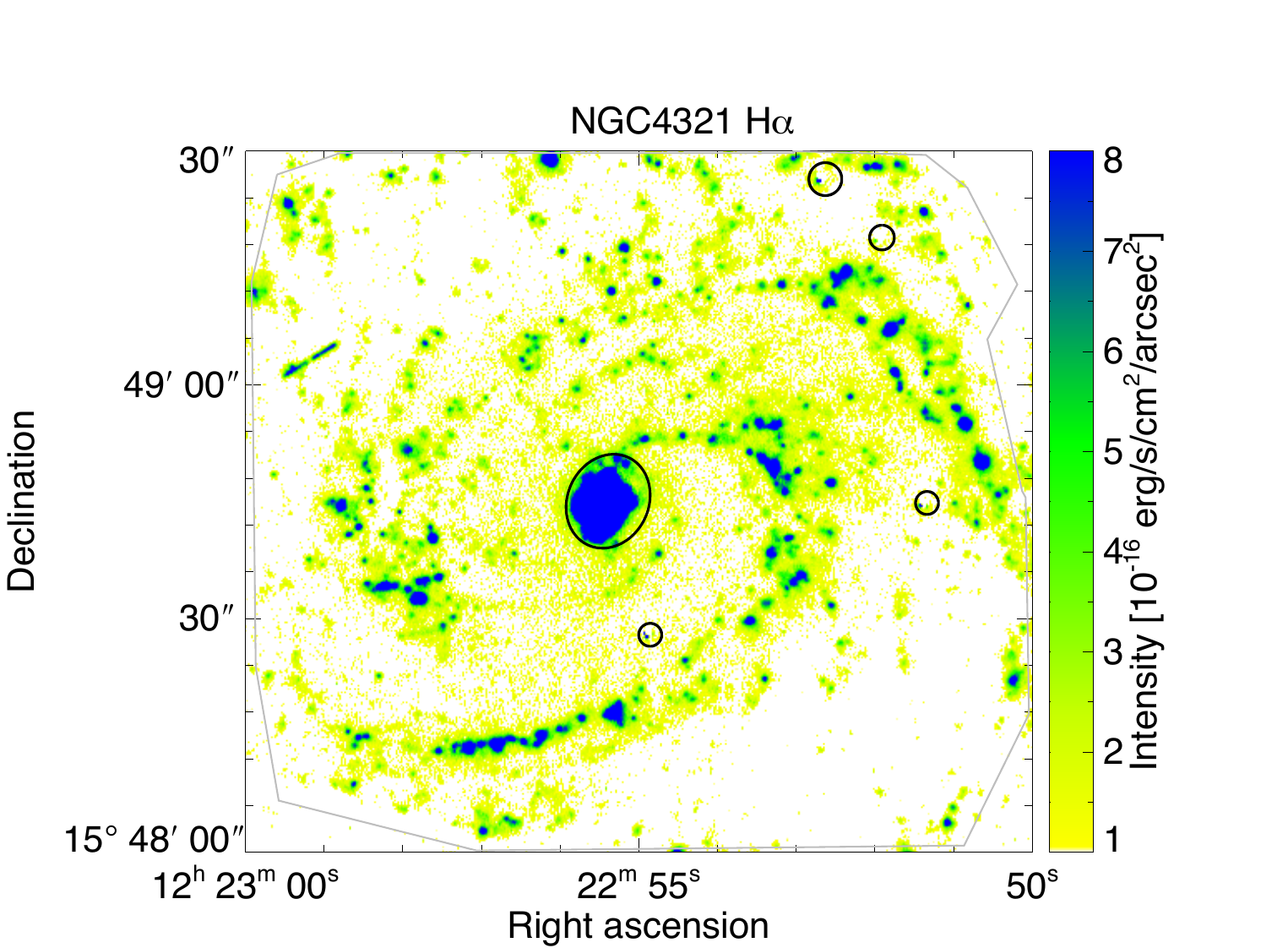}
\contcaption{}
 \end{center} 
\end{figure*}
\begin{figure*}
\begin{center}
\includegraphics[trim=2mm 53mm 62mm 90mm, clip=true, width=8.5cm]{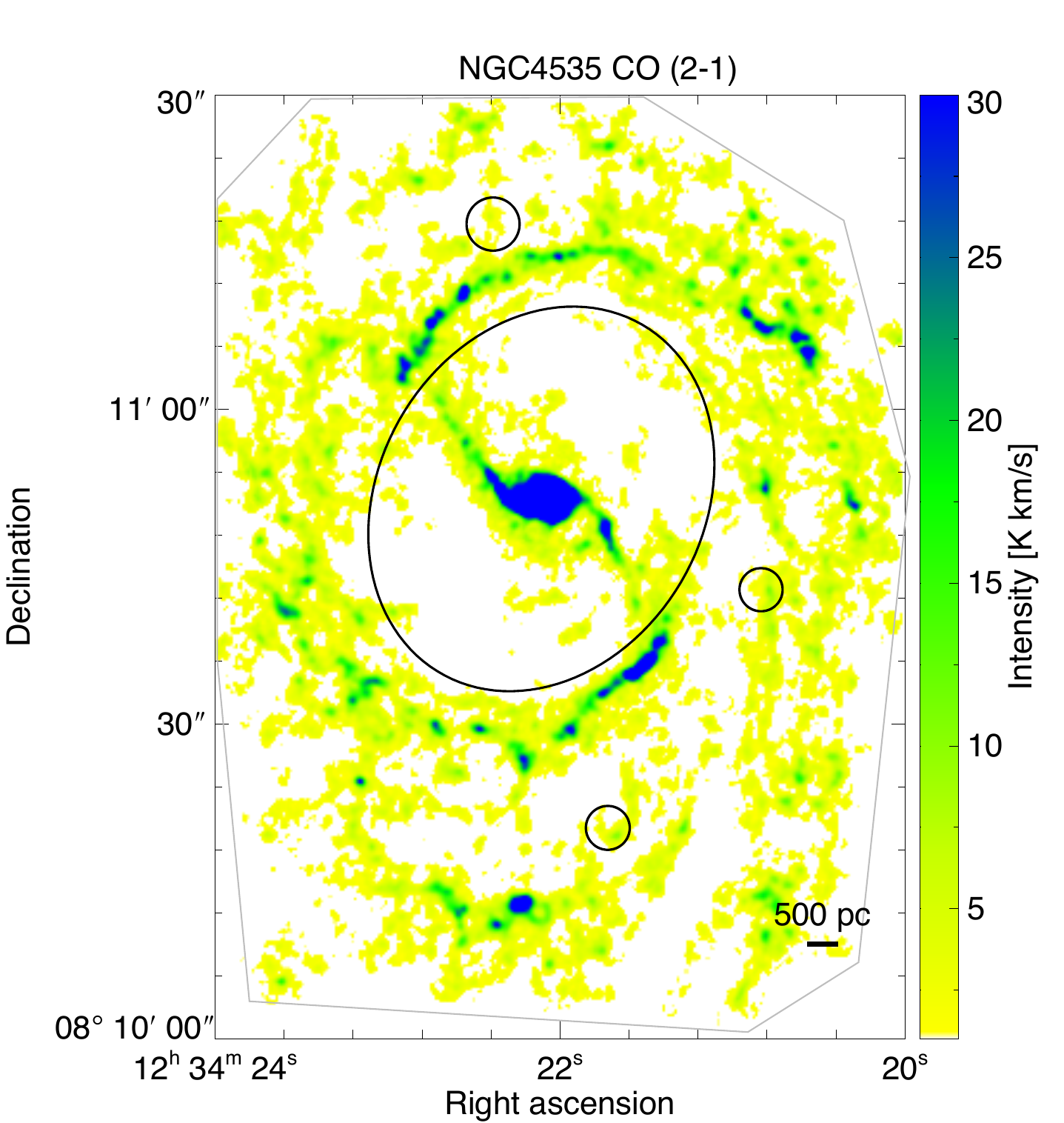}
\includegraphics[trim=2mm 53mm 62mm 90mm, clip=true, width=8.5cm]{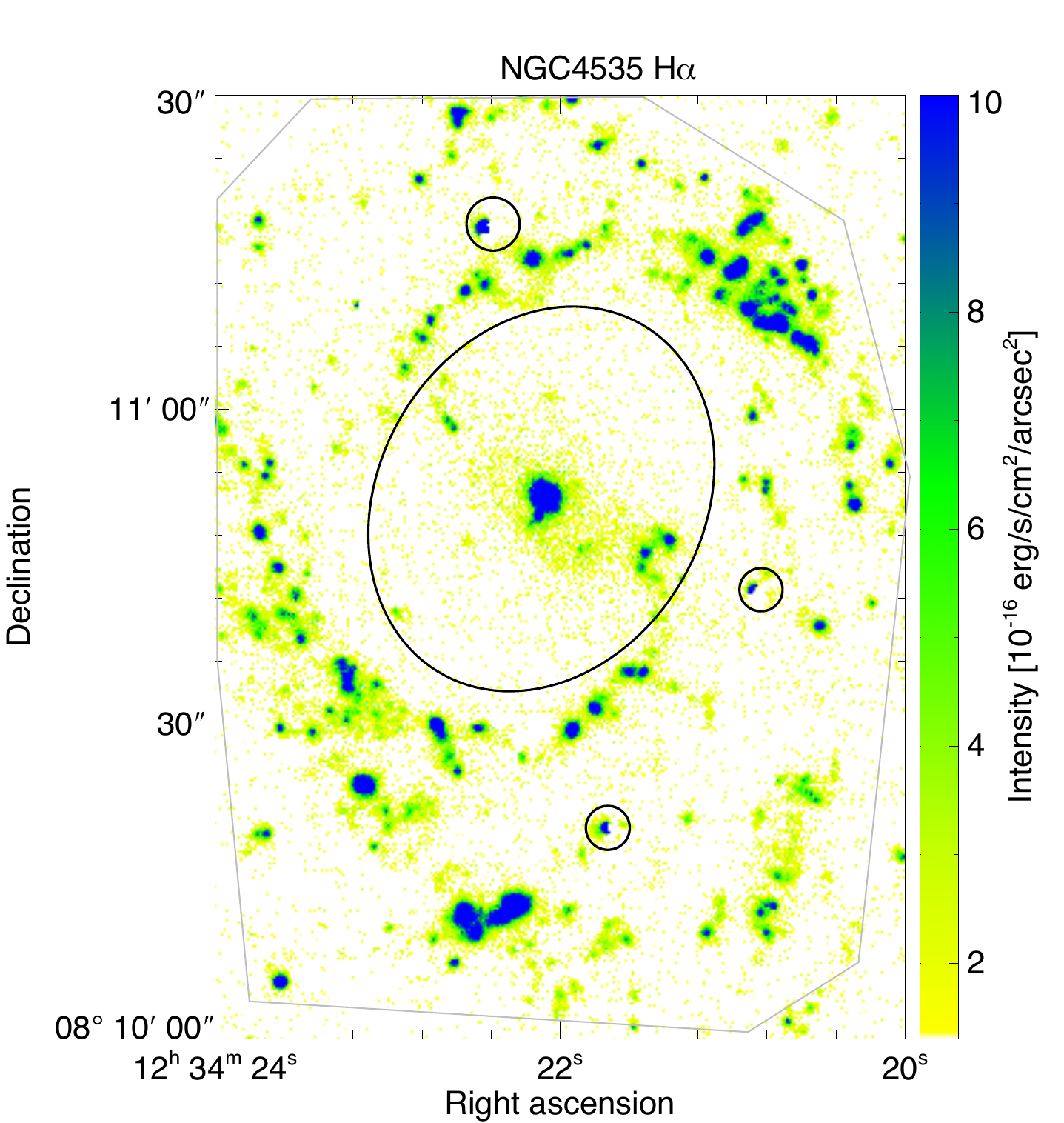}
\includegraphics[trim=2mm 53mm 62mm 120mm, clip=true, width=8.5cm]{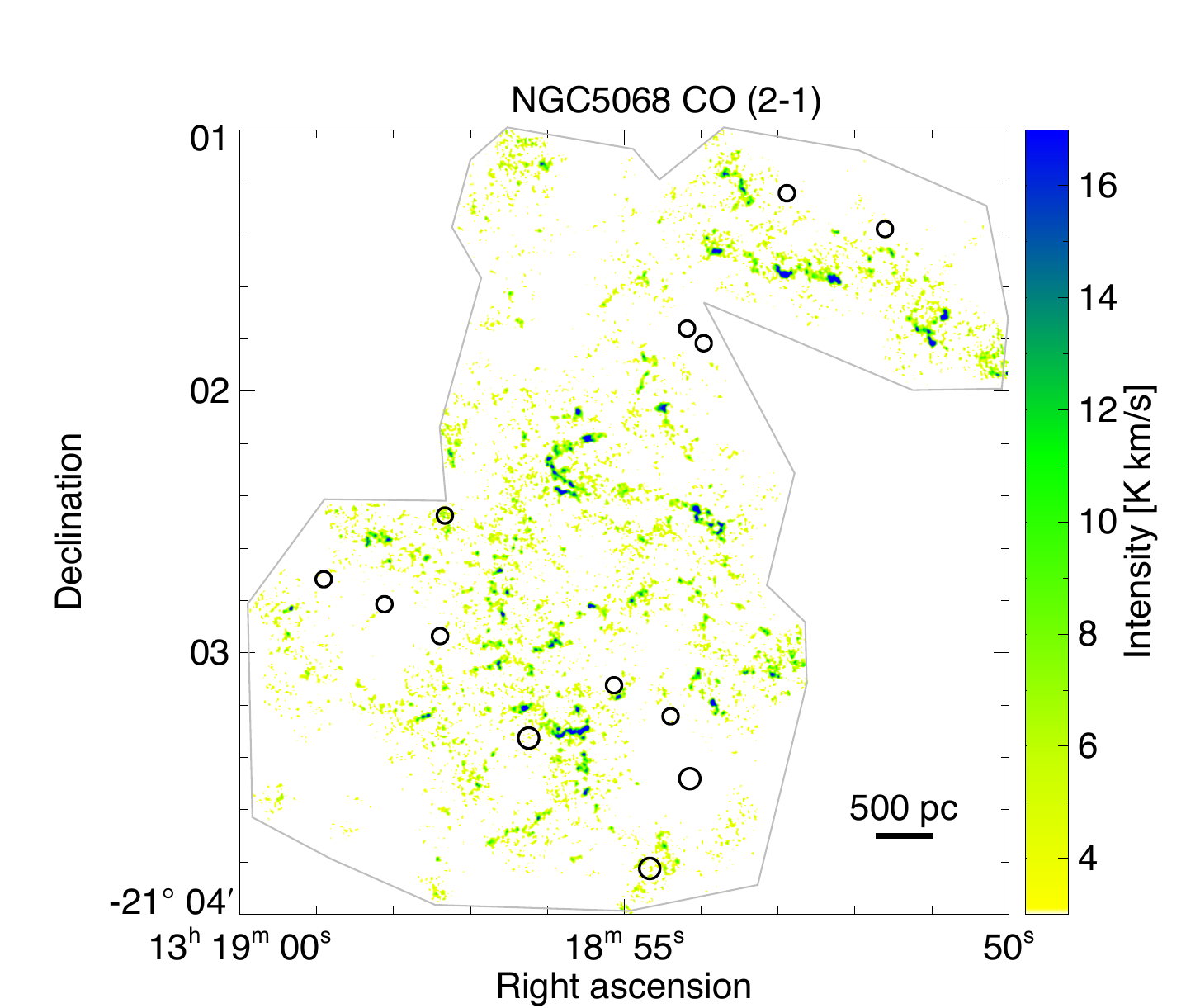}
\includegraphics[trim=2mm 53mm 62mm 120mm, clip=true, width=8.5cm]{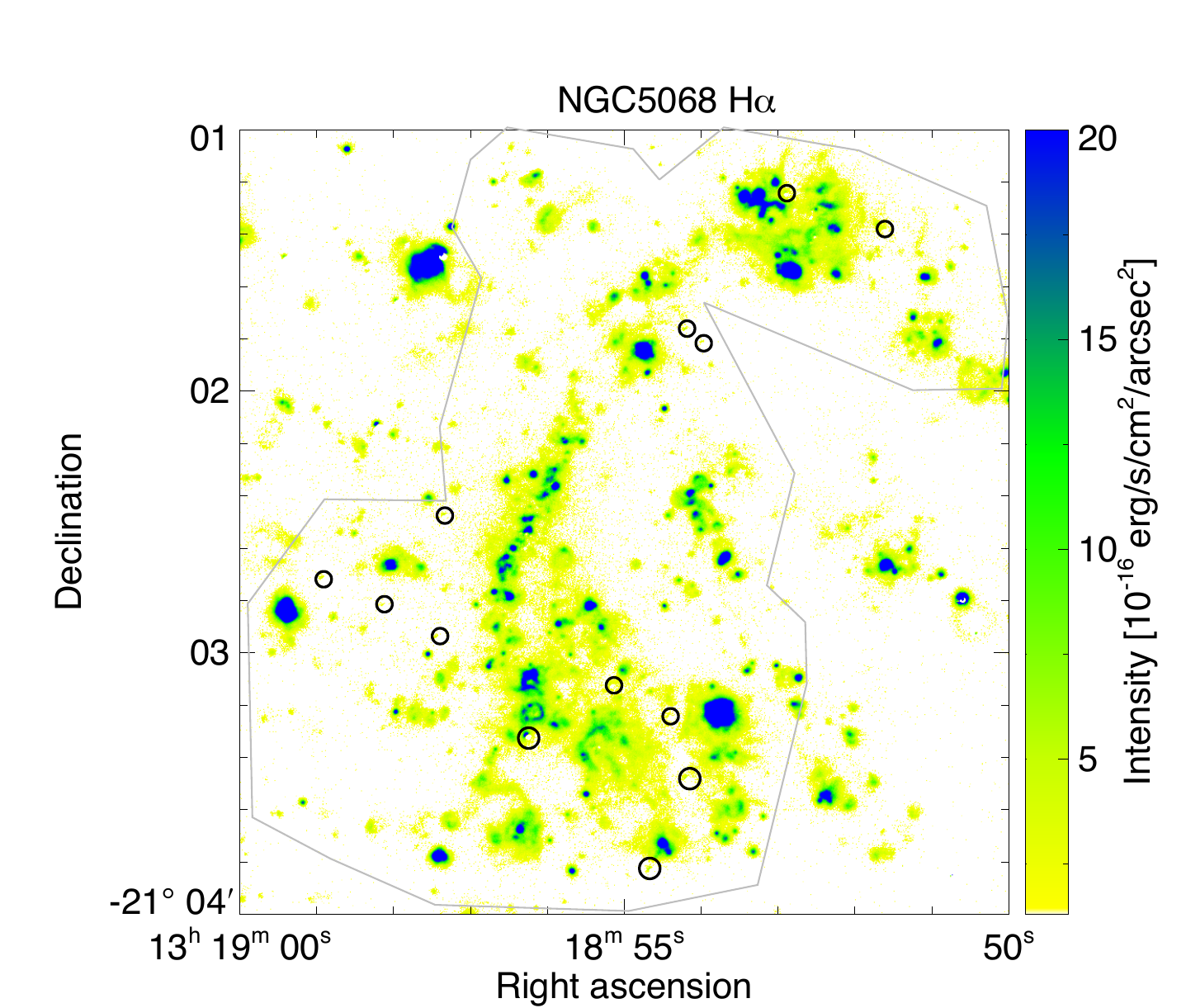}
\hspace{1mm}
\includegraphics[trim=2mm 53mm 60mm 150mm, clip=true, width=8.5cm]{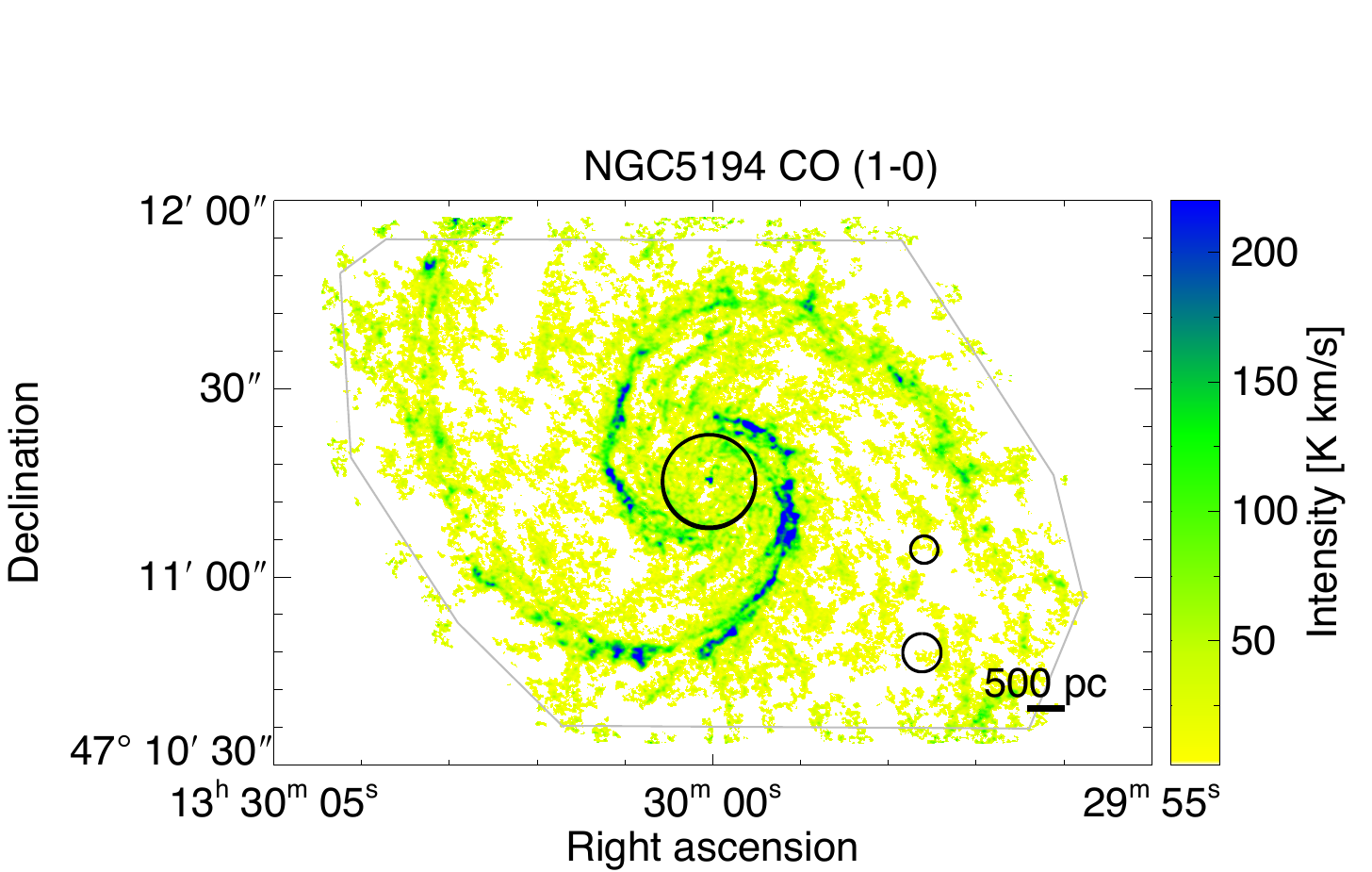}
\hspace{6mm}
\includegraphics[trim=2mm 53mm 60mm 150mm, clip=true, width=8.5cm]{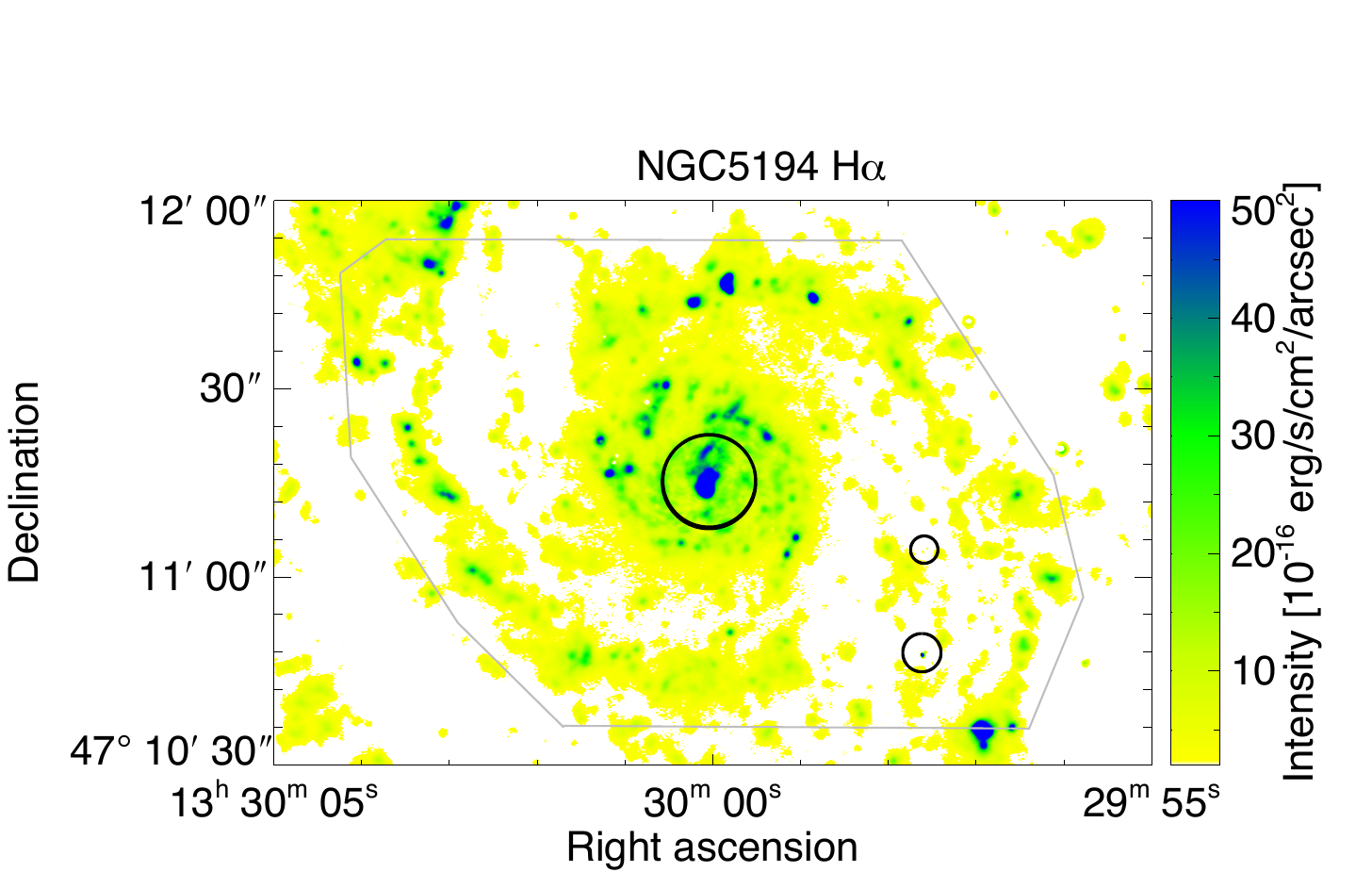}
\contcaption{}
 \end{center} 
\end{figure*}

    \subsection{Global SFR}
    \label{sec:sfr}
As noted above, \Ha line emission can suffer from extinction, implying that the total SFR derived from \Ha alone is underestimated. To correct for extinction, we calculate the SFR from multi-wavelength mapping, combining the Galaxy Evolution Explorer (GALEX) far-ultraviolet band (far-UV; 155\,nm) and the Wide-field Infrared Survey Explorer (WISE) W4 band at 22\mic\ maps \citep{Leroy2019b}, convolved to 15\arcsec\ angular resolution. To convert the observed flux levels to a SFR, we use the SFR prescription provided by \citet{Kennicutt2012} and \citet{Jarrett2013}. The SFR measured this way accross the fields of view used for our analysis are listed in Table~\ref{tab:input}. Finally, we determine the appropriate conversion factor between the flux in the \Ha map and the total extinction-corrected SFR from GALEX and WISE across the same field of view. We note that this conversion factor has no impact on the evolutionary timeline derived in Section~\ref{sec:results} and only plays a role in calculating the integrated star formation efficiency per star formation event (see Section~\ref{sec:Heisenberg}).

\section{Uncertainty principle for star formation}
\label{sec:method}

We now turn to a discussion of our analysis method. We first introduce the general concept and framework, before discussing how it is applied specifically to our sample of nine nearby disc galaxies. This section also includes a summary of the adopted input parameters of the analysis, a discussion of how we filter diffuse emission from the galaxy maps, and a description of how the evolutionary timelines are calibrated.

	\subsection{General concept}
Inspired by the interpretation first proposed by \citet{Schruba2010}, recent work has now demonstrated that the observed small-scale scatter around the global star formation relation \citep[e.g.][]{Bigiel2008, Blanc2009, Onodera2010, Schruba2010, Leroy2013, Kreckel2018, Kruijssen2019} can be understood by assuming that individual regions in a galaxy independently undergo an evolutionary lifecycle during which a molecular cloud assembles, collapses, forms stars, and is disrupted by feedback, with molecular gas and SFR tracers probing different evolutionary phases \citep[e.g.][]{Feldmann2011,KL14}. On small scales, such an independent region is observed at a specific time during this cycle, and therefore does not necessarily satisfy the galactic star formation relation: it is not possible to simultaneously observe a young stellar cluster and the progenitor cloud from which it formed. When focusing on a young, unembedded star-forming region, most of the molecular gas has been consumed or disrupted, leaving an excess of SFR flux compared to the average gas-to-SFR flux ratio. By contrast, when focusing on a non-star-forming GMC, an excess of molecular gas is measured relative to the galactic-scale balance between gas and SFR emission. This means that the gas-to-SFR flux ratio (or gas depletion time) depends strongly on the local evolutionary state of the ISM \citep{Schruba2010,KL14}.

In the context of the above interpretation, the observed scatter around the star formation relation on small scales results from the statistically-insufficient sampling of the different star formation phases. Conversely, the strong correlation between gas mass and SFR observed on galactic scales results from averaging over many regions that collectively sample the full evolutionary lifecycle spanning the successive phases of star formation. In this work, we use the statistical method first presented in \citet{KL14} and developed further in \citet{Kruijssen2018}, which exploits the multi-scale nature of the star formation relation by translating the small-scale variations of the gas-to-SFR flux ratio into the underlying evolutionary timeline of cloud formation, star formation and feedback, as well as deriving the physical quantities describing star formation on the cloud scale.

The evolutionary timeline for star formation is constituted by the lifetime of molecular clouds, \tgas, and the duration of the young stellar phase, \tstar. Here and in the following, we use `\textit{gas}' and `\textit{star}' to refer to molecular gas clouds and young \HII\ regions respectively. These two phases can overlap in time, which defines the duration of the feedback phase, \tover, during which stars and gas coexist within a region. The total duration of this evolutionary timeline, $\tau$, is therefore given by:
\begin{equation}
\tau = t_{\rm star} + t_{\rm gas} - t_{\rm fb} .
\end{equation}
According to this definition, \tstar\ is the complete duration over which the SFR tracer is visible, such that it exceeds the lifetime of massive stars if star formation proceeds over a non-zero time-scale. Likewise, \tgas\ represents the complete duration over which the cloud is visible in the gas tracer. Finally, \tover\ is the time between the moment at which the SFR tracer first becomes visible and the moment at which the gas tracer has completely dispersed. Each of these phases can be probed by a particular observational tracer. Schematically, across a galaxy, the relative abundance (or rarity) of the tracers associated with each of the above phases reflects their relative duration. Therefore, by measuring how common or how rare flux peaks of a given tracer are, we are able to define a relative lifetime between successive phases of the star formation cycle.

In practice, we perform our measurement by centring circular apertures of a certain size on molecular clouds or young \HII regions, and measuring the relative change of the gas-to-SFR flux ratio within these apertures with respect to the galactic average as the aperture size is varied (see e.g.\ Supplementary Video~1 of \citealt{Kruijssen2019}). At large aperture sizes (centred on either emission peak), the galactic average gas-to-SFR flux ratio is recovered. The relative deviation (or `bias') of the gas-to-SFR flux ratio measured at smaller aperture sizes relative to the galactic average directly probes the relative durations of the phases captured by the two tracers. For instance, when placed on the numerous emission peaks of a long-lived tracer, even the smallest apertures will cover most of the galaxy, and will therefore also encompass a large fraction of emission peaks of the other tracer. The resulting flux ratio will be close to the galactic average, resulting in a small bias. By contrast, when placed around the rare emission peaks of a short-lived tracer, small apertures will cover only a small part of the galaxy, and therefore only a small fraction of the emission peaks of the other tracer, leading to a large bias of the flux ratio compared to the galactic average.

To measure the above time-scales for our nine target galaxies, we systematically fit the model from \citet{Kruijssen2018} to the observed gas-to-SFR flux ratios measured as a function of the aperture size, when focusing apertures either on molecular gas emission peaks or on SFR emission peaks. The general steps of the procedure used for this analysis are described in Section~\ref{sec:Heisenberg} and are summarised as follows. We first select two tracers of causally-related phases in a Lagrangian timeline, i.e.\ any individual region visible in one of the tracers will eventually emit in the other tracer.\footnote{\label{ft:cycles}This does not preclude multiple visibility cycles of the first of both tracers before becoming visible in the second, which happens if clouds disperse dynamically without forming massive stars. However, we find in Section~\ref{sec:results} that this is unlikely to occur, because the integrated cloud lifetimes are similar to a (cloud-scale or galactic) dynamical time-scale, leaving insufficient time for multiple cycles. The generality of the method also allows for multiple generations of (or temporally extended) star formation within a single cloud, by allowing $t_{\rm fb}>0$. Because the method identifies `independent regions', which reside on an evolutionary timeline independently of their neighbours, these regions may contain multiple smaller \HII regions or molecular substructure if these have correlated evolutionary ages. For instance, this would apply to a group of \HII regions born from the same molecular cloud.\label{ft:interpretation}} Emission peaks are identified in this pair of maps and the gas-to-SFR flux ratio is measured around these peaks, for a range of different aperture sizes. We then fit a statistical model to these measurements to constrain its three free parameters (these are \tgas, \tover, and the region separation length $\lambda$, see below), propagate the errors on the derived parameters characterising the evolutionary timeline, and derive secondary quantities including their uncertainties. The results of applying this analysis to our galaxy sample are presented in Section~\ref{sec:results}.

	\subsection{Application of analysis method to our galaxy sample}
	\label{sec:Heisenberg}

Our analysis method is formalised in the \textsc{Heisenberg} code, which is presented and described in detail by \citet{Kruijssen2018}. Here we summarise the main steps of the method to measure the duration of the gas phase (\tgas), the duration of the feedback phase (\tover), and the typical separation length between independent regions ($\lambda$). 

\begin{table}
\begin{center}
\begin{tabular}{lcl}
\hline
Flags                             &  Value    & Notes \\
\hline
\texttt{mask{\_}images} &  1           & Mask images on\\
\texttt{mstar{\_}int}        &  1            & Mask the centre of the galaxy\\
\texttt{mgas{\_}ext}        &  1           & Mask outer parts of the galaxy, \\
                                      &               &  where CO is not detected\\
\hline
\end{tabular}
\caption{Flags set to a different value than the default (as listed in table~1 of \citealt{Kruijssen2018}).}
\label{tab:flags}
\end{center}
\end{table}

\begin{table*}
\begin{center}
\caption{Main input parameters of the analysis for each galaxy. The other parameters use the default values as listed in table~2 of \citet{Kruijssen2018}.}
\begin{threeparttable}
\begin{tabular}{lccccccccc}
\hline
Quantity                           &  NGC628    & NGC3351  &   NGC3627  &  NGC4254 &  NGC4303  &  NGC4321  &  NGC4535   &  NGC5068  & NGC5194 \\
\hline
$D$ [Mpc]\tnote{\textit{a}}        &  $9.77$    &  $10.00$ &  $10.57$   &  $16.80$ &  $17.60$  &  $15.20$  &  $15.80$   &  $5.16$   &  $8.60$\\
$i$ [\degr]\tnote{\textit{b}}      &   8.70     &   45.14  &   56.49    &  35.27   &   19.99   &   39.10   &   42.12    &  26.95    & 21.00\\
$\phi$ [\degr]\tnote{\textit{b}}   &   20.82    &   193.24 &  174.04    &  68.51   &   310.60  &   157.65  &   179.35   &  348.96   & 173.0\\  
$l_{\rm ap,min}$ [pc]   & 50  &   80   & 100    &  140    &   150   &   130   &   130   & 30   & 70 \\
$l_{\rm ap,max}$ [pc]  &4800 &   4900 &  5400    & 10700   &   7400  &   6100   &   7200  & 4000 & 3000 \\ 
$N_{\rm ap}$            & 15  &   15   &  15      &  15     &   15    &   15     &   15    & 15   & 12 \\
$N_{\rm pix,min}$       & 10  &   10   &  10      &  10     &  15     &  10      &  10     & 15   & 15 \\
$\Delta \log_{10} \mathcal{F}_{\rm star}$\tnote{\textit{c}} & 1.00 & 1.60 & 2.10 & 2.30 & 2.50 & 1.60 & 2.30 & 1.70 & 2.30\\
$\delta \log_{10} \mathcal{F}_{\rm star}$\tnote{\textit{c}} & 0.06 & 0.25 & 0.20 & 0.20 & 0.15 & 0.25 & 0.25 & 0.50 & 0.20\\
$\Delta \log_{10} \mathcal{F}_{\rm gas}$\tnote{\textit{c}}  & 0.70 & 1.10 & 2.20 & 1.90 & 2.00 & 1.60 & 2.00 & 1.20 & 1.40\\
$\delta \log_{10} \mathcal{F}_{\rm gas}$\tnote{\textit{c}}  & 0.03 & 0.15 & 0.10 & 0.10 & 0.15 & 0.15 & 0.20 & 0.60 & 0.15\\
$t_{\rm star,ref}$ [Myr]                                    & 4.35 & 4.27 & 4.37 & 4.34 & 4.37 & 4.29 & 4.38 & 4.53 & 4.19\\
$\sigma(t_{\rm star,ref})$ [Myr]\tnote{\textit{d}}          & 0.16 & 0.17 & 0.16 & 0.16 & 0.16 & 0.16 & 0.16 & 0.14 & 0.17\\
SFR [\Msun~yr$^{-1}$]\tnote{\textit{e}}                     & 0.87 & 0.22 & 2.81 & 4.50 & 4.37 & 2.50 & 0.92 & 0.22 & 1.91\\
$\sigma({\rm SFR})$ [\Msun~yr$^{-1}$]\tnote{\textit{d}}     & 0.17 & 0.04 & 0.56 & 0.90 & 0.87 & 0.50 & 0.18 & 0.04 & 0.38\\
log$_{10}X_{\rm gas}$\tnote{\textit{f}}                     & 0.81 & 0.67 & 0.33 & 0.78 & 0.82 & 0.45 & 0.82 & 0.92 & 0.59\\
$\sigma_{\rm rel}(X_{\rm gas})$\tnote{\textit{d}}           & 0.40 & 0.63 & 0.50 & 0.31 & 0.50 & 0.50 & 0.50 & 0.50 & 0.50\\
$n_{\lambda \rm,iter}$                                      & 10   &   10 &  12  &  12  &  12  &  12  &  12  & 10   & 12  \\
\hline
\end{tabular}
\begin{tablenotes}
\item[\textit{a}]Distances adopted from A.~K.\ Leroy et al.\ (in prep.) and references therein.
\item[\textit{b}]Inclinations and position angles are preliminary and will be presented by \cite{Lang2019}.
\item[\textit{c}]The parameters for the peak identification listed here are valid for the diffuse-emission filtered maps (see Section~\ref{sec:diffuse}). Different values are used for the first iteration during which emission peaks are identified in unfiltered maps, but we have verified that the choice of these initial parameter values does not significantly affect our results.
\item[\textit{d}]Standard error. The subscript `rel' indicates a relative error.
\item[\textit{e}]This is the SFR measured from GALEX and WISE (see Section~\ref{sec:sfr}) across the field of view considered in this paper, rather than of the entire galaxy.
\item[\textit{f}]The gas conversion factor corresponds to $\alpha_{\rm CO(1-0)}$ for NGC5194 and to $\alpha_{\rm CO(2-1)}$ for all of the other galaxies, in $\Msun~({\rm K}~\kms~\pc^2)^{-1}$.
\end{tablenotes}
\label{tab:input}
\end{threeparttable}
\end{center}
\end{table*}

We provide two galaxy maps of the tracers characterising the evolutionary timeline of interest (CO and \Ha, see Section~\ref{sec:input}). Both maps are convolved to the same resolution and matched to the same pixel grid before running the analysis. We specify as needed if the maps should be partially masked or a galactocentric radius cut should be applied. We define a central region by eye to exclude the galactic centre (where independent regions are the most prone to blending). For NGC3351 and NGC4535, this mask is extended to cover the bar region, because their strong bars have cleared most of the corresponding area of molecular gas and star formation. We also exclude the galaxy outskirts beyond the galactocentric radius of the outermost emission peak identified across both maps (see below). The masking also takes into account the edges of the field of view. If any, masks or radial cuts are applied to both maps. The masked regions (galaxy outskirts, central region, foreground stars and background galaxies) are visible as ellipses in Figure~\ref{fig:COmaps}. To enable a straightforward measurement of the gas-to-SFR flux ratio (here CO-to-\Ha\ flux ratio) at various aperture sizes, we next use a top-hat kernel to convolve both maps to $N_{\rm ap}$ different spatial scales, spaced logarithmically between a minimum ($l_{\rm ap,min}$) and maximum ($l_{\rm ap,max}$) aperture size (see Table~\ref{tab:input}).

The emission peaks on which the apertures are placed are identified in both maps at the best common resolution, using the algorithm \textsc{Clumpfind} \citep{Williams1994}. In brief, the \textsc{Clumpfind} algorithm identifies closed contours for a given set of flux level intervals, defined by a flux range below the maximum flux level, $\Delta \log_{10} \mathcal{F}$, and an interval between flux levels, $\delta \log_{10} \mathcal{F}$. In Table~\ref{tab:input}, these carry subscripts `star' and `gas', referring to the \Ha\ and CO maps, respectively. We set the minimum number of pixels within a closed contour necessary for a peak to be identified to $N_{\rm pix,min}$, to avoid selecting point sources, and the position of the peak is then defined as the pixel with the maximum flux value within this closed contour. For each of the $N_{\rm ap}$ spatial scales, we place apertures on each peak and measure the gas and SFR fluxes within these apertures, as well as the effective average aperture area, which may be smaller than the intended aperture area due to the potential presence of masked pixels.\footnote{For instance, apertures that partially fall outside of the field of view have their area reduced accordingly.} This results in four fluxes per aperture size: the total summed CO flux and total summed \Ha\ flux across the entire sample of CO peaks, and the total summed CO flux and total summed \Ha\ flux across the entire sample of \Ha\ peaks. From these summed fluxes, we then calculate the CO-to-\Ha\ flux ratio around CO peaks or around \Ha\ peaks, at each given aperture size. We then calculate the bias relative to the galactic averaged CO-to-\Ha\ flux ratio for each set of peaks. As a function of the aperture size, this bias for CO and \Ha\ emission peaks takes the characteristic shape of a `tuning fork' diagram (see Section~\ref{sec:results}).

In practice, placing an aperture on each peak would result in counting at least some of the pixels multiple times, because some apertures overlap. This occurs for large aperture sizes and in regions with a high number density of peaks, and leads to inaccurate measurements of the flux ratio bias due to over-representing regions at high number densities. To avoid this effect, the flux ratio bias is calculated 1000 times on different Monte-Carlo realisations of sub-samples of independent, non-overlapping apertures, for each peak type and aperture size. These Monte-Carlo realisations contain the maximum number of non-overlapping apertures obtained by going through the full list of apertures in a different order each time and rejecting those that overlap with any apertures that have already been drawn. The final CO-to-\Ha\ flux ratio is an average over all of the Monte-Carlo realisations. The uncertainties on the flux ratio bias measurements account for both the finite sensitivity and resolution of the maps, as well as for the intrinsic stochasticity of the gas mass and SFR of the different regions. These are then translated into effective uncertainties, which take into account the covariance between the flux measurements at different aperture sizes, and are used when fitting the statistical model to the tuning fork diagram. In the tuning fork diagrams presented in Section~\ref{sec:results}, we show both the individual and effective uncertainties.

The next step is to fit these measurements with a statistical model linking the flux ratio biases to the duration of the different phases of the evolutionary timeline. The mathematical expressions for the flux biases have been derived in \citet[Appendix~C]{KL14} by considering a random spatial distribution of point-like regions at random positions on the evolutionary timeline, and taking into account the possible flux evolution between regions in isolation and regions within which both phases coexist. The model was since updated to account for a spatially-extended profile of the regions. As stated above, the model depends on three independent quantities: \tgas , \tover\ and $\lambda$. As a function of these quantities, it predicts how the CO-to-\Ha\ flux ratio changes as a function of the aperture size when focusing apertures on regions bright in CO or in \Ha. We refer to \citet[sect.~3.2.11]{Kruijssen2018} for the complete details of the model and note that a concise summary is provided in the Methods section of \citet{Kruijssen2019}. The model is fitted to the data points by minimising the reduced-$\chi^2$ over the above three free parameters. These three quantities are non-degenerate, as they affect the predictions of the model in different ways (see Section~\ref{sec:fit}). The resulting three-dimensional probability distribution function (PDF) is marginalised to obtain the one-dimensional PDF for each free parameter. The uncertainty on each free parameter is defined as the 32nd percentile of the part of the PDF below the best-fitting value, and the 68th percentile of the part of the PDF above the best-fitting value. For a Gaussian PDF, this reduces to the $1 \sigma$ uncertainties. We provide the full PDFs of our measured cloud lifetimes in Section~\ref{sec:results}, finding that they are often close to log-normal.

Fundamentally, the above analysis only measures relative time-scales, such that the duration of one of the two phases needs to be provided as a reference time-scale in order to convert the relative time-scales into absolute ones. We use the calibration of the \Ha-emitting phase by \citet{Haydon2019} to convert the relative duration of each phase to an absolute timeline, using a reference timescale (\tstarref). This calibration has been carried out in a self-consistent way, by applying the \textsc{Heisenberg} code to pairs of simulated galaxy maps. The input maps used there are a mass surface density map of star particles within a specified age range (of which the duration is then known)\footnote{By using a mass surface density map with a pre-defined stellar age range as the `reference map' in these calibration experiments, we ensure that the calibration is largely insensitive to the baryonic physics of the simulation. See \citet{Haydon2019} for details.} and a synthetic emission map of a star formation tracer (\Ha\ or UV emission for various filters), the duration of which is then an output of the method. \citet{Haydon2019} generate these maps by post-processing their hydrodynamical disc galaxy simulations with the stellar population synthesis (SPS) code \textsc{SLUG2} \citep{DaSilva2012,DaSilva2014,Krumholz2015}. They sample stars stochastically from a \citet{Chabrier2005} initial mass function (IMF) and use Geneva stellar evolutionary tracks \citep{Schaller1992} with \textsc{Starburst99} spectral synthesis \citep{Leitherer14}.\footnote{Binaries are not included in the adopted SPS model, but they may prolong the emission of ionising photons and increase \tstarref, because stars in binaries may be tidally stripped, thus exposing their hot interiors \citep{Eldridge2017,Goetberg2019}. However, we do not expect this to substantially change our results, because binaries only increase the ionising flux at times when it has already dropped considerably, i.e.\ well after the nominal value of \tstarref\ derived by \citet{Haydon2019}, and are unable to boost it to values similar to the ionising flux predicted at $t<t_{\rm star,ref}$ \citep[see fig.~4 of][]{Goetberg2019}.}

\citet{Haydon2019} calibrate \tstarref\ using an SPS model describing an instantaneous burst of star formation, to avoid any dependence on the duration of star formation, which likely varies in nature. This implies that the reference time-scale (\tstarref) differs from the total duration of the \Ha-bright phase (\tstar) by excluding the feedback phase. This choice of defining $t_{\rm star,ref}=t_{\rm star}-t_{\rm fb}$ thus allows for a continuous star formation history, in which new massive stars can form as long as the region contains molecular gas, and the `clock' defining \tstarref\ only starts when the last massive star forms. The exact value of \tstarref\ varies somewhat with metallicity (see Section~\ref{sec:metallicity}) and the sampling of the IMF. In this work, we account for the dependence of this time-scale on metallicity. Its dependence on IMF sampling is weak in general, and is negligible for the range of region masses probed by our observations (see section~6 of \citealt{Haydon2019}). For reference, the total \Ha visibility time-scales (i.e.\ $t_{\rm H\alpha}\equiv t_{\rm star}=t_{\rm star,ref}+t_{\rm fb}$) obtained in this work range from $5{-}9~\myr$ (see Section~\ref{sec:results}), broadly consistent with previous studies \citep[e.g.][]{Kennicutt2012,Leroy2012}.

Finally, we calculate a wide variety of derived quantities from the three free parameters, including their PDFs. Among others, these include the total star formation tracer lifetime ($t_{\rm star}\equiv t_{\rm star,ref}+t_{\rm fb}$), the total duration of the evolutionary timeline ($\tau$), the region radii ($r_{\rm H\alpha}$ and $r_{\rm CO}$), the region size-to-separation ratios or filling factors ($\zeta_{\rm H\alpha}$ and $\zeta_{\rm CO}$), the feedback outflow or phase transition front velocity ($v_{\rm fb}$), the global gas depletion time ($t_{\rm depl}$), the integrated star formation efficiency per star formation event ($\epsilon_{\rm sf}$), and the region-scale mass loading factor ($\eta_{\rm fb}$). How these quantities are derived from the three free parameters is detailed in section~3.2.14 of \citet{Kruijssen2018}.

In this paper, we apply the \textsc{Heisenberg} code to a sample of nine galaxies, and derive all of the quantities mentioned above. However, we will mainly focus our discussion on the molecular cloud lifetime (Sections~\ref{sec:results} and \ref{sec:tgas_variations}), while future papers will present a detailed investigation of the other derived quantities and their dependence on galactic environment. Before discussing the results of our analysis, we now first describe the input parameters of the \textsc{Heisenberg} code used in this paper, as well as how we determine a number of observational quantities that are required as input for the measurement of the molecular cloud lifetime.

	\subsection{Input maps and parameter choices}
	\label{sec:input}
The requirement that the tracers of the different phases represent causally-related phases along a Lagrangian timeline means that the tracers must be chosen with care. Each independent region needs to be detectable in both tracers at some point in its lifetime, but not necessarily simultaneously. Based on the strong correlation between molecular gas and star formation on galactic scales, we therefore consider the timeline from molecular gas (traced by CO) to young stars (traced by \Ha), under the assumption that young stars form from molecular gas.\footnote{Note that we do not assume the opposite, i.e.\ not every CO emission peak is assumed to host massive star formation at any point of its life. However, the short cloud lifetimes reported in Section~\ref{sec:results} imply that clouds only live for approximately one (cloud-scale or galactic) dynamical time before being associated with \Ha emission, which strongly suggests that most CO peaks in our maps do eventually host massive star formation. See the Methods section of \citet{Kruijssen2019} for further discussion.} This means that GMCs hosting unembedded massive star formation will be visible in both tracers simultaneously.

For the analysis presented here, we trace the first phase (the `gas' phase) with the emission of CO(2-1), except for NGC5194, for which we use the high resolution CO(1-0) PAWS map. In the following, we will use the notation \tCO\ to represent the duration of the gas phase (instead of the more general notation \tgas), which in this context refers to the molecular cloud lifetime. This choice of tracer defines the structures of which the lifetimes are measured: we assume that CO and molecular gas coexist in time and space, so that CO emission can be used to trace molecular gas. As such, the molecular cloud lifetimes presented here represent the `CO visibility' lifetimes of molecular clouds, i.e.\ the flux-weighted, population-averaged time for which an individual molecular cloud emits in CO, for the molecular cloud population above our point source sensitivity limit of $\sim10^5~\msun$. Beyond this definition, an important advantage of the method used here is that the measured time-scales do not explicitly depend on the \aco\ conversion factor, which is uncertain in extragalactic environments \citep[see e.g.][]{Kennicutt2012, Bolatto2013}. Once a molecular gas tracer has been chosen, the assumption of a particular \aco\ conversion factor or of a ratio CO(2-1)/CO(1-0) has no impact on the derived molecular cloud lifetime, nor on the other primary derived quantities, \tover\ and $\lambda$ (see also Section~\ref{sec:metallicity}). This insensitivity to conversion factors arises, because the flux observed near emission peaks is divided by the kpc-scale flux of the same tracer, which means that the conversion factor cancels out on average. However, if there is a considerable \aco\ spread within the galaxy, the flux-averaging nature of our method implies that the measurements may be biased towards regions of low \aco\ (high flux). For the shallow metallicity gradients shown in Figure~\ref{fig:gal_prop}, we expect this effect to be minor.

We select \Ha as a star formation tracer for the second phase and use it to calibrate the obtained timelines. In the following, we will use the notation \tHa\ to represent the duration of the young stellar phase (instead of the more general notation \tstar). The duration of the young stellar phase probed by (continuum subtracted) \Ha has been calibrated by \citet{Haydon2019} to be \tHaref\ = 4.3\,Myr at solar metallicity, for the calibration setup described in Section~\ref{sec:Heisenberg}. In Section~\ref{sec:metallicity}, we quantify the slight dependence of this time-scale on metallicity; the reference time-scales listed in Table~\ref{tab:input} account for the gas mass-weighted mean metallicity of each galaxy.

It is necessary that the observed tracer maps have a  spatial resolution sufficient to resolve the separation length $\lambda$ between independent regions, and that the inclination of the galactic disc is moderate ($i \lesssim $ 75\degr) to avoid confusion between emission peaks. We also assume that the regions are randomly distributed in each other's vicinity, such that the distribution of neighbouring regions is accurately described in two-dimensions, without dominant one-dimensional structures.\footnote{Because the de-correlation between CO and \Ha takes place below a size scale $\sim\lambda$ of typically a few hundreds of pc (see Section~\ref{sec:results}), our methodology is largely insensitive to galactic structure. This means that strong morphological features on the galactic scale do not typically break the assumption of local spatial randomness and two-dimensionality. Even local evolutionary stream lines \citep[e.g.\ across spiral arms,][]{Meidt2013,Querejeta2019,Schinnerer2017} are accommodated by the method, as long as the increase of the number of neighbouring emission peaks with size scale proceeds roughly as expected for a two-dimensional distribution. We have tested the method on simulated galaxies with a flocculent spiral structure to demonstrate this \citep{Kruijssen2018}.} These requirements, as well as the other guidelines listed in \citet[section~4.4]{Kruijssen2018}, have been determined based on experiments on simulated galaxies. We demonstrate in Section~\ref{sec:requirements} that our analysis satisfies these guidelines.

The tracer maps of the two consecutive phases are the primary inputs of the \textsc{Heisenberg} code. Tables~\ref{tab:flags} and~\ref{tab:input} present a selection of flags and input parameters used for our galaxy sample. The other flags and input parameters of the \textsc{Heisenberg} code not listed here have been set to their default values as listed in tables~1 and~2 of \citet{Kruijssen2018}. We note that, while we have optimised the input parameters of the model (such as $N_{\rm ap}$ and $l_{\rm ap,min}$) and of the peak identification ($\Delta \log_{10} \mathcal{F}$ and $\delta \log_{10} \mathcal{F}$) to each of the galaxies in our sample, small variations of these numbers do not strongly affect the constrained quantities, as long as the physically relevant peaks are identified and the criteria listed in Section~\ref{sec:requirements} are satisfied.

	\subsection{Diffuse emission}
	\label{sec:diffuse}

The presence of diffuse emission on large scales in the maps of the observed tracers affects the measured cloud lifetime, by adding a reservoir of emission on scales larger than $\lambda$ that does not belong to the emission peaks identified. This diffuse emission can have different physical origins for different tracers, as described below (e.g.\ diffuse molecular gas not forming massive stars, ionising photons leaking from \HII\ regions and therefore not spatially associated with a star-forming region), and does not participate in the evolutionary cycle of emission peaks described in Section~\ref{sec:method} \citep{Kruijssen2018, Hygate2019}. This large-scale emission therefore needs to be filtered out of the observed maps to ensure an unbiased measurement of the different phases of the molecular cloud lifecycle.

In the case of \Ha, the leaking of ionising photons outside of the \HII regions where they are produced leads to the presence of a diffuse \Ha component in the observed maps \citep[e.g.][]{Mathis1986, Sembach2000, Wood2010}. Other contributions to this diffuse ionised gas include ionisation by post-asymptotic giant branch stars \citep[e.g.][]{Binette1994, Sarzi2010, FloresFajardo2011}, dust scattering \citep{Seon2012}, shocks \citep{Pety2000, Collins2001}, and the presence of small, unresolved \HII\ regions \citep{Lee2016b}. Different methods can be applied to remove the contribution from the diffuse ionised gas, depending on its assumed origin. The simplest methods consist of subtracting an estimate of the diffuse emission based on a smoothed version of the star-formation tracer \citep[e.g.][]{Hoopes1996, Greenawalt1998} or applying a fixed intensity threshold to remove all emission lower than a given value \citep[e.g.][]{Blanc2009, Kaplan2016}. Including information about the spatial extent of \HII\ regions can also help decomposing the emission into a diffuse background and compact sources \citep[e.g.][]{Thilker2002, Oey2007}. However, while most of these approaches are physically motivated, they ultimately rely on subjective choices regarding the intensity threshold, the smoothing scale, the size of \HII\ region and/or the scaling factor applied to the smoothed map. We note that if the main source of diffuse \Ha\ emission results from the leaking of ionising photons, this flux should not be omitted from the global SFR when calculating the star formation efficiency (see Section~\ref{sec:sfe}). 
 
In the case of CO, a diffuse component on large scales can be emitted by truly diffuse, unbound molecular gas, or by an ensemble of small mass, unresolved clouds. Our observations have the point source sensitivity to detect cloud masses down to $10^5$\,\Msun\, which for the star formation efficiencies reported in Section~\ref{sec:results} corresponds to a few $10^3~\Msun$ in stellar mass over the duration of an evolutionary cycle $\tau$. Few massive stars are expected in lower-mass regions \citep[e.g.][]{Weidner2006,DaSilva2012} and our measurements represent flux-weighted population averages \citep{Kruijssen2018}. Because the cloud mass function follows an exponentially truncated power law with a slope below the truncation mass ($M_{\rm GMC,\star}$) that is shallower than $-2$ \citep[e.g.][E.~Rosolowsky et al.\ in prep.]{Freeman2017}, this means that the lifecycles inferred here mostly describe the cloud population near the truncation mass. For the galaxies considered here, this is $M_{\rm GMC,\star}=10^{6}{-}10^{7}~\msun$ (E.~Rosolowsky et al.\ in prep.). We can therefore filter out the lower mass clouds, which do not strongly contribute to the flux-weighted average evolutionary cycle constrained here.\footnote{Lower-mass clouds could potentially represent an accretion flow onto more massive clouds and therefore also participate at some level in the high mass formation process. However, these must represent a small gas reservoir, as we only filter out $\sim 15$~per~cent of the CO emission on average (see Table~\ref{tab:diffuse}), and therefore do not constitute the main units for massive star formation.}

For both the CO and \Ha\ maps, we filter out diffuse emission from the input images using the method of \citet{Hygate2019}, which uses the mean separation length between independent regions obtained with \textsc{Heisenberg} ($\lambda$) to iteratively filter out emission in Fourier space on scales larger than a fixed multiple of the separation scale. This approach avoids making assumptions about the physical scale of \HII\ regions or a flux threshold to separate \HII\ regions from the diffuse background. Instead, it uses the characteristic separation length between independent regions as a physically-motivated scale for separating the diffuse emission from compact emission. This is achieved by filtering out the emission in Fourier space on spatial scales larger than those of the independent regions undergoing the evolutionary lifecycle of interest, and doing this consistently for the SFR map and the gas map. While this approach does not presume a fixed scale for diffuse emission, it does a posteriori introduce a spatial scale over which diffuse emission is thought to exist. However, given that the separation length is larger than the typical \HII\ region size by definition, this should not introduce a large bias (even though it may not remove all of the diffuse emission). A key advantage of this method is that it also deals well with a diffuse background that varies across the map, as long as the variations manifest themselves over a size scale larger than the region separation length.

The influence of the size and type of filter used are fully described in \citet{Hygate2019}. For our analysis, we use a Gaussian high-pass filter, which is the best compromise between the selectivity of the filter and the undesired appearance of artefacts around compact regions. We then set the characteristic cut-off wavelength of this Gaussian filter to be between $10{-}12\times\lambda$ (see Appendix~\ref{sec:app_diffuse} for details), with $\lambda$ the characteristic separation length between independent clouds or star-forming regions, as measured with our analysis method (see Sections~\ref{sec:method} and~\ref{sec:results}). The multiples of $\lambda$ ($n_{\lambda \rm,iter}$) used are listed in Table~\ref{tab:input}. This choice of $n_{\lambda \rm,iter}$ ensures that the large-scale diffuse emission is filtered, while minimising the impact of the filter on the compact regions \citep{Hygate2019}. After filtering, we again measure $\lambda$ for the filtered maps and iterate this process until convergence is reached (when $\lambda$ varies by less than 5~per~cent from the previous iteration, for at least four successive iterations). The resulting compact emission fractions ($f_{\rm H\alpha}$ and $f_{\rm CO}$; and by complement the diffuse fractions $1-f_{\rm H\alpha}$ and $1-f_{\rm CO}$) are presented for our nine target galaxies in Appendix~\ref{sec:app_diffuse}.

        \subsection{Metallicity and reference time-scale}
	\label{sec:metallicity}

In this section, we quantify how metallicities of the target galaxies affect the input quantities and derived quantities of our analysis. While our method itself is not directly affected by changes in metallicity, accounting for metallicity variations allows us to calibrate the measured timeline more accurately and to calculate additional derived quantities as described below.

Firstly, the absolute calibration of the reference lifetimes of the young stellar phase (see Section~\ref{sec:Heisenberg}) depends weakly on the metallicity as \citep{Haydon2019}: 
\begin{equation}
\label{eq:tref}
t_{\rm star,ref} = (4.32\pm0.16~\myr)\times\left(\frac{Z}{{\rm Z}_{\sun}}\right)^{-0.086\pm0.017} ,
\end{equation}
where we define
\begin{equation}
\frac{Z}{{\rm Z}_{\sun}} \equiv \frac{({\rm O}/{\rm H})}{({\rm O}/{\rm H})_{\sun}} ,
\end{equation}
with $12+\log{({\rm O}/{\rm H})}_{\sun} = 8.69$ \citep{Asplund2009}. We therefore scale the reference time-scale for each galaxy by the mean gas mass-weighted metallicity, based on the metallicity gradients  measured in \citet{Pilyugin2014}. This measurement is available for all galaxies in our sample, except NGC3627. For this galaxy, we therefore use the slope of the metallicity gradient as measured from MUSE observations (see \citealt{Kreckel2019}, which use the S-calibration method from \citealt{Pilyugin2016}). Because the calibration method is different than the one used in \citet{Pilyugin2014}, we compare the average metallicities of the galaxies present in both samples and scale the absolute values in \citet{Kreckel2019} to match the average values in \citet{Pilyugin2014}. For the three galaxies in common between the samples, this correction is smaller than 0.1~dex over the radial intervals considered in this work. For each galaxy, the resulting metallicities are shown as a function of galactocentric radius in Figure~\ref{fig:gal_prop} and the adopted average metallicities are presented in Table~\ref{tab:observations}. The corresponding reference time-scales calculated using equation~\ref{eq:tref} are listed in Table~\ref{tab:input}, and shown in Figure~\ref{fig:Bin_inputs} as a function of galactocentric radius. Over the entire sample, the average metallicity ranges between $12+\log{({\rm O}/{\rm H})} = 8.39$ (for NGC5068) and $12+\log{({\rm O}/{\rm H})} = 8.84$ (for NGC5194), which translates into a narrow range of associated reference time-scales of $t_{\rm star,ref}=4.19{-}4.53$\,Myr. Within individual galaxies, the reference time-scale also varies by less than 10~per~cent across the range of radii considered.

Secondly, the total molecular gas mass surface density scales directly with the value of the CO-to-\HH\ conversion factor, \aco. Therefore, the choice of the conversion factor affects a small subset of the quantities derived through our analysis, such as the molecular gas depletion time, the integrated star formation efficiency per star formation event, and the region-scale mass loading factor \citep[see][]{Kruijssen2018}. For the conversion factor from CO(2-1) to total molecular gas mass, we adopt a fixed ratio CO(2-1)/CO(1-0) = 0.7 (e.g.\ \citealt{Gratier2010, Leroy2011}; T.~Saito et al.\ in prep.) and use the \aco\ factors provided by \citet{Sandstrom2013} when available (i.e.\ for NGC628, NGC3351, NGC3627, NGC4254 and NGC4321). For all other galaxies, we simply scale the conversion factor with metallicity as suggested by \citet{Bolatto2013}:
\begin{equation}
    \alpha_{\rm CO} = \left[2.9~\Msun~({\rm K}~\kms~\pc^2)^{-1}\right] \times \exp\left(\frac{0.4{\rm Z}_{\sun}}{Z}\right), 
\end{equation}
where \aco\ is the conversion factor from CO(1-0) flux to total molecular gas mass, including the contribution of heavy elements. The adopted CO-to-\HH\ conversion factors are listed in Table~\ref{tab:input}. When dividing the galaxies into several bins of galactocentric radius (see Section~\ref{sec:tgas_variations}), we use the appropriate values of \aco\ and \tref\ corresponding to the mean metallicity in each bin (see Figure~\ref{fig:Bin_inputs} for the profiles of \aco\ and \tstarref\ as a function of galactocentric radius). We note that these global values may deviate considerably on the scales of individual clouds \citep[e.g.][]{Schruba2017}.

We note again that the absolute metallicity value has no direct influence on the primary parameters of the model (\tgas, \tover\ and $\lambda$), which are based on the relative change of the gas-to-SFR flux ratio compared to the galactic average, and not on the absolute values of the gas mass or SFR. The only way in which it affects the first two of these quantities is through the (slight) metallicity dependence of the reference time-scale in equation~(\ref{eq:tref}), which causes the reference time-scale to vary by less than 10~per~cent across all galactic environments considered here (see Figure~\ref{fig:Bin_inputs}).

\section{The molecular cloud lifecycle averaged across nearby galaxies}
\label{sec:results}

We now apply the methodology described in Section~\ref{sec:method} to the data presented in Section~\ref{sec:obs}. We first show that our galaxy sample exhibits a universal de-correlation between molecular gas and star formation on the cloud scale, before translating this de-correlation into the evolutionary timeline of cloud evolution, star formation, and feedback. We conclude the section by giving brief summaries of other inferred quantities, each of which will be the subject of a more detailed analysis in follow-up work.

    \subsection{A universal de-correlation between gas and star formation on the cloud scale}
	\label{sec:fit}
We apply the analysis described in Sections~\ref{sec:Heisenberg} and \ref{sec:input}, using \Ha as a star formation tracer and CO as a gas tracer as discussed in Section~\ref{sec:obs}. For each galaxy, we measure the gas-to-SFR flux ratio compared to the galactic average, focusing on gas peaks and SFR peaks, as a function of varying aperture sizes and then fit these measurements with a model describing how this observable changes as a function of the underlying evolutionary time-scales and region separation length.

\begin{figure*}
\includegraphics[width=\linewidth]{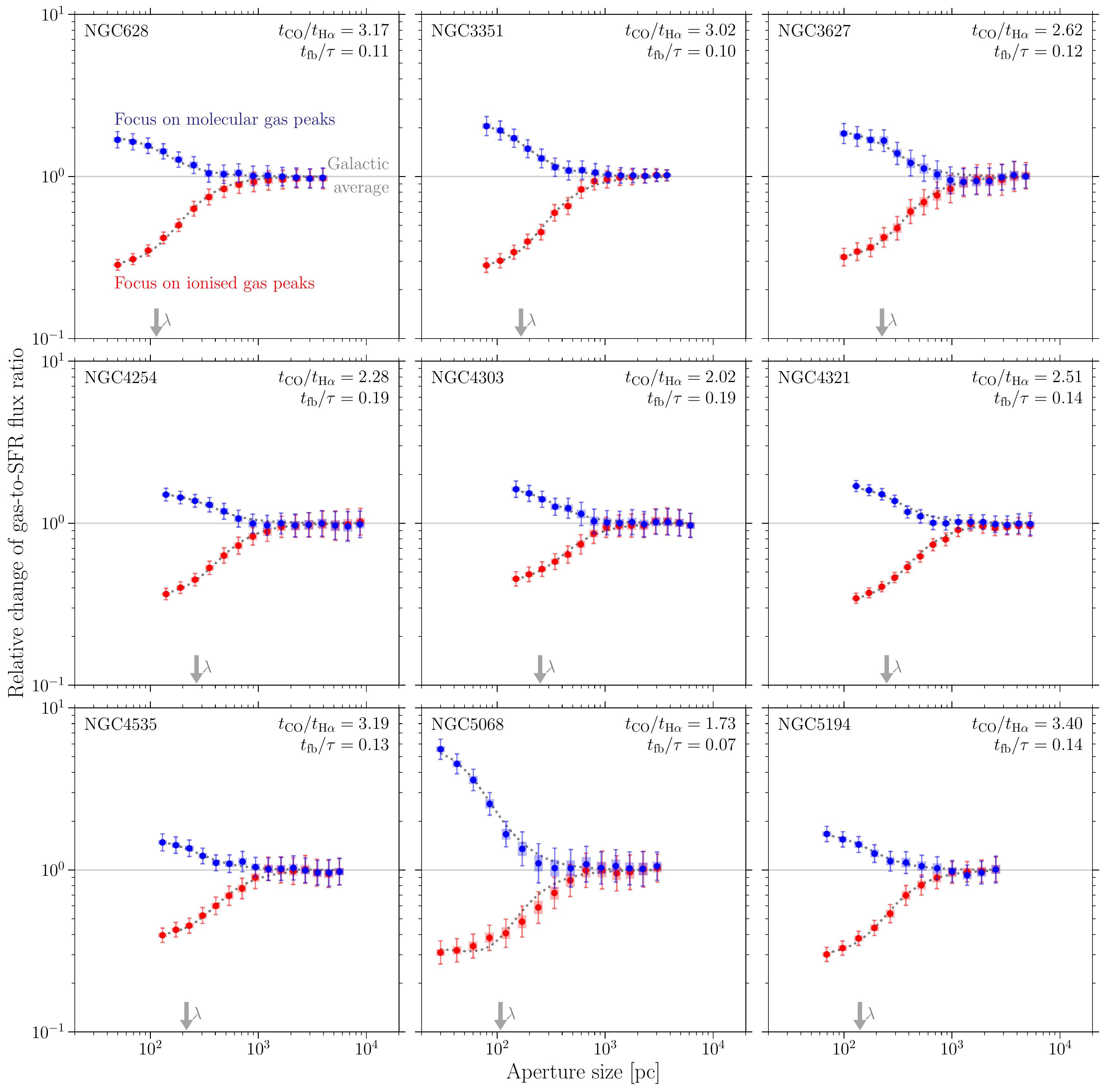}
\caption{Relative change of the gas-to-SFR (CO-to-\Ha) flux ratio compared to the galactic average as a function of aperture size, for apertures placed on CO emission peaks (blue) and \Ha emission peaks (red). The error bars indicate the $1\sigma$ uncertainty on each individual data point, whereas the shaded areas indicate the effective $1\sigma$ uncertainty range that accounts for the covariance between the data points and should be used when visually assessing the quality of the fit. The horizontal solid line indicates the galactic average and the dotted line is the best-fitting model \citep{Kruijssen2018}, which allows us to constrain the GMC lifecycle. The arrows indicate the best-fitting values of the region separation length $\lambda$, which is always resolved given the minimum aperture sizes. The ratios \tCO/\tHa\ (controlling the asymmetry between the two branches) and \tover/$\tau$ (controlling the flattening of the branches) are indicated in the bottom-right corner of each panel.}
\label{fig:tuningfork}
\end{figure*}
Figure~\ref{fig:tuningfork} shows the measured gas-to-SFR flux ratios as a function of the aperture size for each galaxy, together with the best-fitting model. All galaxies in our sample exhibit a pronounced de-correlation between gas emission and SFR emission, which becomes stronger as the aperture size decreases. This leads to two distinct branches, diverging from the galactic average. The cloud-scale de-correlation between gas and star formation was first observed in M33 by \citet{Schruba2010} and we find that it is a universal feature of the galaxies studied here. As discussed in \citet{Kruijssen2019}, this de-correlation implies the rapid evolutionary cycling between molecular gas, star formation, and cloud destruction by stellar feedback. \citet{Fujimoto2019} builds on our empirical results to show that the de-correlation represents a fundamental test of feedback physics in galaxy simulations, as it probes the dispersive effect of stellar feedback on GMCs.

While we find a universal de-correlation between gas and star formation tracers on $\sim100~\pc$ scales, Figure~\ref{fig:tuningfork} also reveals quantitative variation between galaxies. This variation is caused by differences between the underlying evolutionary timelines. The mathematical expression of the model depends on three independent quantities: \tCO, \tover\ and $\lambda$, from which we can also derive secondary quantities as described in Section~\ref{sec:Heisenberg}. These three quantities are non-degenerate and affect the shape of the model in very different ways. The characteristic scale at which the branches diverge from the galactic average is set by $\lambda$. The ratio $t_{\rm CO}/t_{\rm H\alpha}$ governs the asymmetry between the branches, and the ratio $t_{\rm fb}/\tau$, as well as the finite size of the CO and \Ha\ peaks, regulate the flattening of the branches at small aperture sizes (see \citealt{Kruijssen2018} for more details). The best-fitting values of the above time-scale ratios are indicated in the bottom-right corner of each panel of Figure~\ref{fig:tuningfork}, with $\lambda$ marked with an arrow along the $x$-axis in each panel. The figure clearly shows the impact of the above quantities on the shape of the tuning fork diagram describing the de-correlation between gas and star formation. First, galaxies with a small value of $\lambda$ show a de-correlation at smaller aperture sizes (compare e.g.\ NGC628 and NGC4321). Secondly, galaxies with a small value of \tCO/\tHa\ have tuning fork diagrams with steeper top branches, reflecting a shorter cloud lifetime (compare e.g.\ NGC628 and NGC5068). Finally, galaxies with a small value of $t_{\rm fb}/\tau$ have less flattened branches at small aperture sizes (compare e.g.\ NGC4254 and NGC5068).

Table~\ref{tab:results} summarises the best-fitting values for \tCO , \tover\ and $\lambda$, as well as the implied feedback outflow velocity ($v_{\rm fb}$) and the integrated cloud-scale star formation efficiency ($\epsilon_{\rm sf}$). Together, these describe the molecular cloud lifecycle in the nine star-forming disc galaxies considered here. We now turn to a more detailed discussion of these results.
\begin{table}
\begin{center}
{\def\arraystretch{1.5}
\begin{tabular}{lccccc}
\hline
                     Galaxy &                       \tCO\  &                        $t_{\rm fb}$  &                           $\lambda$  & $v_{\rm fb}$  &                 $\epsilon_{\rm sf}$  \\
                                    &                                 [Myr]  &                                 [Myr]  &                                  [pc]  &                         [km s$^{-1}$]  &                                  [per~cent]  \\
\hline
                            NGC0628   & $     24.0_{-    2.5}^{+    3.6} $   & $      3.2_{-    0.4}^{+    0.6} $   & $      113_{-     14}^{+     22} $   & $      8.5_{-    1.1}^{+    1.0} $   & $      6.1_{-    2.2}^{+    3.7} $  \\
                            NGC3351   & $     20.6_{-    3.0}^{+    3.4} $   & $      2.5_{-    0.6}^{+    0.8} $   & $      166_{-     16}^{+     25} $   & $     14.8_{-    3.2}^{+    4.3} $   & $      5.2_{-    2.6}^{+    5.0} $  \\
                            NGC3627   & $     18.9_{-    3.2}^{+    3.4} $   & $      2.8_{-    0.7}^{+    0.8} $   & $      225_{-     34}^{+     55} $   & $     20.9_{-    3.8}^{+    5.9} $   & $     10.2_{-    4.5}^{+   7.7} $  \\
                            NGC4254   & $     20.9_{-    2.3}^{+    3.9} $   & $      4.8_{-    1.0}^{+    1.1} $   & $      267_{-     44}^{+     53} $   & $     14.7_{-    2.4}^{+    2.6} $   & $      4.2_{-    1.3}^{+    2.2} $  \\
                            NGC4303   & $     16.9_{-    2.2}^{+    4.6} $   & $      4.0_{-    1.0}^{+    1.8} $   & $      250_{-     44}^{+     87} $   & $     17.4_{-    4.2}^{+    4.0} $   & $      4.3_{-    1.7}^{+    3.7} $  \\
                            NGC4321   & $     19.1_{-    2.2}^{+    2.3} $   & $      3.3_{-    0.6}^{+    0.7} $   & $      248_{-     26}^{+     33} $   & $     19.6_{-    2.9}^{+    3.8} $   & $      7.1_{-    4.1}^{+    5.2} $  \\
                            NGC4535   & $     26.4_{-    3.6}^{+    4.7} $   & $      3.9_{-    0.9}^{+    1.2} $   & $      216_{-     37}^{+     65} $   & $     15.4_{-    2.7}^{+    2.6} $   & $      3.8_{-    1.6}^{+    2.9} $  \\
                            NGC5068   & $      9.6_{-    1.8}^{+    2.9} $   & $      1.0_{-    0.3}^{+    0.4} $   & $      107_{-     11}^{+     19} $   & $     15.6_{-    4.3}^{+    5.8} $   & $      4.3_{-    1.8}^{+    3.7} $  \\
                            NGC5194   & $     30.5_{-    4.8}^{+    9.2} $   & $      4.8_{-    1.1}^{+    2.1} $   & $      140_{-     17}^{+     25} $   & $      7.9_{-    2.2}^{+    1.9} $   & $      4.0_{-    1.6}^{+    3.5} $  \\
\hline
\end{tabular}
}
\caption{Physical quantities describing the lifecycle of molecular cloud evolution, star formation, and feedback, obtained with the analysis described in Section~\ref{sec:method}. Each of these values represents the flux-weighted average for the corresponding galaxy. The uncertainties account for the finite sensitivity and resolution of the maps, as well as for the intrinsic stochasticity of the gas mass and SFR of the different regions.}
\label{tab:results}
\end{center}
\end{table}

	\subsection{Measured molecular cloud lifetime}
	\label{sec:tgas}
	
\begin{figure*}
\includegraphics[width=\linewidth]{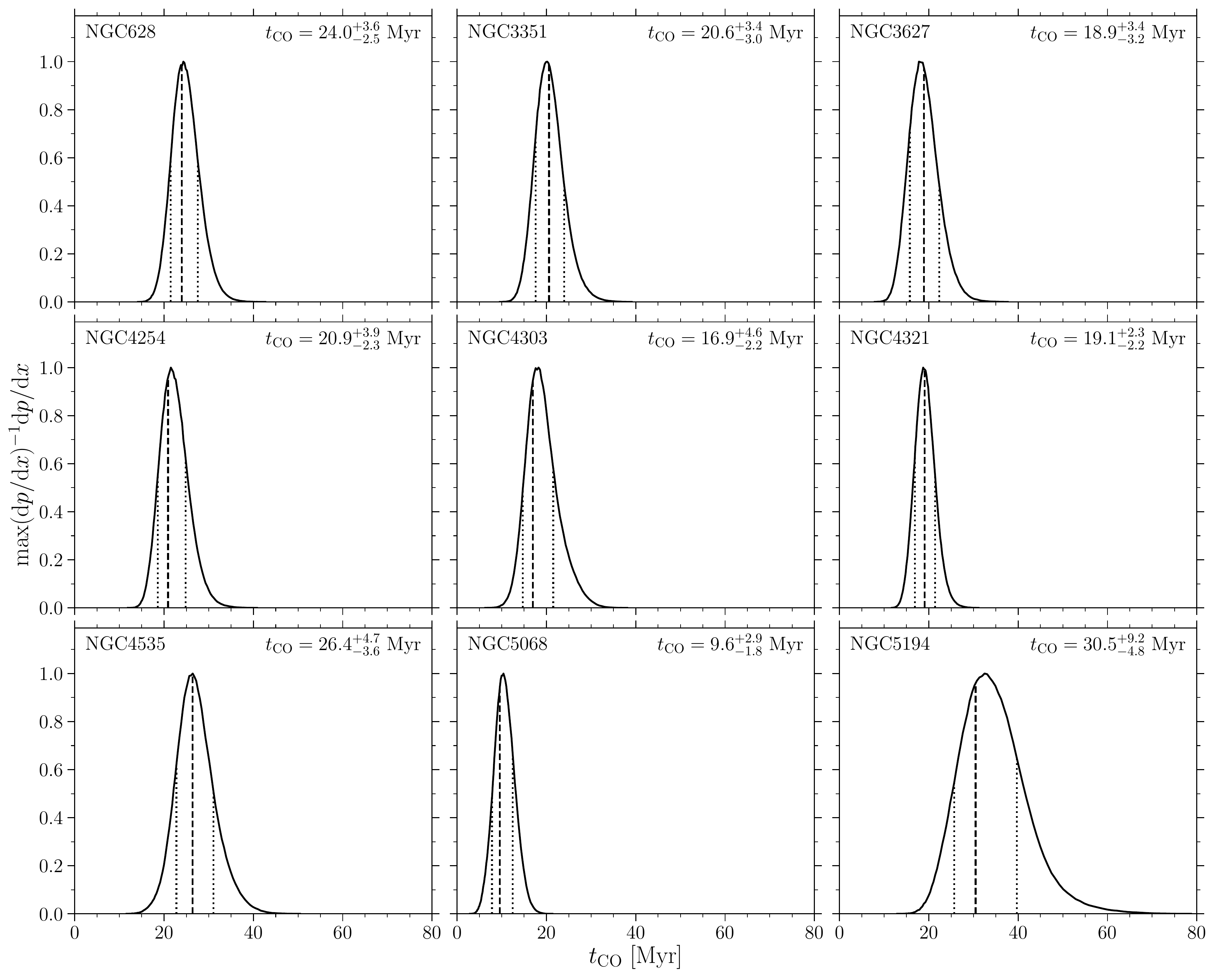}
\caption{One-dimensional PDFs of \tCO\ for each galaxy. The vertical dashed lines indicate the best-fitting values and the dotted lines indicate the 1$\sigma$ uncertainties, defined as the 32nd percentile of the part of the PDF below the best-fitting value, and the 68th percentile of the part of the PDF above the fitted value. The uncertainties account for the finite sensitivity and resolution of the maps, as well as for the intrinsic stochasticity of the gas mass and SFR of the different regions. The best-fitting values and their uncertainties are also indicated in the top-right corner of each panel.}
\label{fig:tgmc}
\end{figure*}

When applying the model to CO and \Ha as tracers of the gas and young stellar phases, respectively, \tCO\ represents the duration of the molecular cloud lifetime during which CO emission is visible as local enhancement. The one-dimensional PDFs of the constrained \tCO\ are presented in Figure~\ref{fig:tgmc}. The figure shows that \tCO\ is well constrained for all galaxies in our sample, with relative uncertainties ($\sigma_{\text{\tCO}}$/\tCO) in the range of $10{-}40$~per~cent.\footnote{We note that the uncertainties on \tCO\ in Figure~\ref{fig:tgmc} appear to be asymmetric and tend to increase with increasing \tCO . This is caused by two effects. First, the uncertainties are largely log-normal because we fundamentally measure relative time-scales, so that the ratio $\sigma_{\text{\tCO}}/\text{\tCO}$ is roughly constant. This manifests itself as an extended positive wing of the PDF when shown in linear space and generally broader PDFs for galaxies with longer GMC lifetimes. Secondly, the results of our analysis are more accurate when the time-scales of both phases (CO and \Ha) are similar \citep{Kruijssen2018}. Because \tHa\ is always the shortest, this means that the smallest relative uncertainties are typically found in galaxies with the shortest GMC lifetimes.} We find that the derived molecular cloud lifetimes are relatively short and vary with galactic environment: they range between 10 and 30 Myr across our galaxy sample. This range of values for the molecular cloud lifetime is consistent with those found in previous studies combining region classification with statistical incidence arguments \citep[e.g.][]{Engargiola2003, Kawamura2009, Meidt2015, Corbelli2017} and those based on the same statistical method used here \citep{Kruijssen2019, Hygate2019b}. This is discussed in more detail in Section~\ref{sec:literature}.

The above results have two important implications. First, they favour theories suggesting that molecular clouds are short-lived, transient objects that form, evolve, and disperse on a (cloud-scale or galactic) dynamical time \citep[e.g.][]{Elmegreen2000, Dobbs2011, Grudic2018, Jeffreson2018, Semenov2018}. Secondly, the strong variation of the cloud lifetime between different galaxies suggests that cloud formation and collapse does not proceed on a universal time-scale but is plausibly governed by environment, such as galactic dynamics, either by directly setting the time-scale or indirectly, by changing the properties of the clouds \citep[e.g.][]{Leroy2017}. We will explore this hypothesis in Section~\ref{sec:tgas_variations}. For the galaxy-wide quantities discussed in this section, the potential importance of environmental variations implies that the presented numbers are a flux-weighted average representation of the cloud lifecycle across the field of view covering each galaxy.

\begin{figure*}
\includegraphics[trim=50mm 10mm 15mm 0mm, clip=true, width=\columnwidth, angle =90]{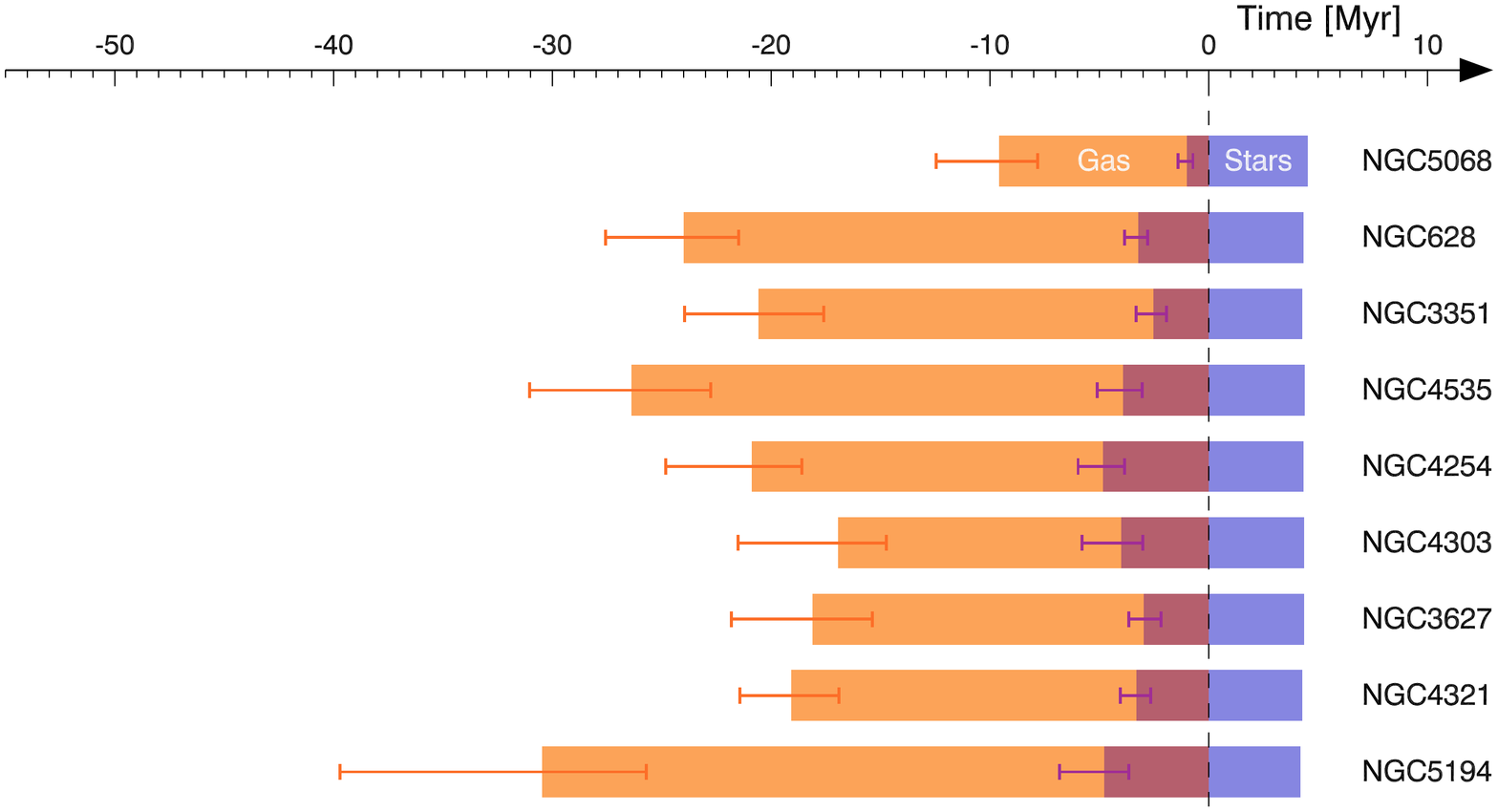}
\caption{Evolutionary timeline of molecular clouds, star formation, and feedback for each of the nine galaxies. From top to bottom, the galaxies are ordered by increasing galaxy stellar mass. Orange indicates when only CO emission is visible (with duration $t_{\rm CO}-t_{\rm fb}$), purple indicates when only \Ha emission is visible (with duration $t_{\rm star,ref}=t_{\rm H\alpha}-t_{\rm fb}$), and maroon indicates the `overlap' phase, when the region emits both in CO and \Ha (with duration $t_{\rm fb}$). The error bars on the left indicate the uncertainty on \tCO, whereas the error bars in the middle indicate the uncertainty on \tover.}
\label{fig:timeline}
\end{figure*}

The full timelines from molecular clouds emitting in CO to young stellar populations emitting in \Ha are shown in Figure~\ref{fig:timeline}, including the time for which CO and \Ha coexist. This `overlap' time shows when massive stars have started to emerge, but have not completely dispersed the parent molecular cloud yet. As such, it reflects the time over which stellar feedback acts on the cloud. We see that the evolutionary timelines all show evidence for a long `inert' (or `isolated' CO-bright) phase, during which the molecular clouds are brightly emitting in CO, but no signs of massive star formation have appeared yet in \Ha . This is consistent with the results obtained using the same method for NGC300 \citep{Kruijssen2019} and M33 \citep{Hygate2019b}, as well as with the results of GMC--\HII region catalogue matching \citep{Kreckel2018} and pixel statistics \citep{Schinnerer2019}, which all find large numbers of CO-bright clouds or pixels unassociated with \Ha emission. While this does not exclude the (potentially prevalent) formation of low-mass stars during this inert phase, it seems inescapable to conclude that unembedded high-mass star formation arrives late during the cloud lifecycle, after $75{-}90$ per~cent of the cloud lifetime (corresponding to $9{-}26$~Myr for the lifetimes measured here).\footnote{In principle, some clouds may condense and disperse more than once before massive star formation occurs. However, this does not seem very likely, because the measured cloud lifetimes are similar to a (cloud-scale or galactic) dynamical time, which leaves little time for multiple cycles prior to massive star formation (also see footnote~\ref{ft:cycles}.}

The complete timelines shown in Figure~\ref{fig:timeline} also demonstrate that unembedded massive star formation correlates strongly with cloud dispersal, as indicated by the short overlap times in Figure~\ref{fig:timeline} (see Section~\ref{sec:tfb} below), which constitute $9{-}18$~per~cent of the entire timeline (also see Figure~\ref{fig:tuningfork}). In principle, some massive stars may form earlier and remain embedded, so that they are not visible in \Ha. We reiterate here that this would not affect our measurements of the cloud lifetimes, but only increase the duration of the overlap phase, because the reference time-scale \tstarref\ to which our results are calibrated refers to the duration of the unembedded \Ha-bright phase, without associated CO emission. We note that this holds under the assumption that new massive stars form as long as the region contains CO-bright molecular gas. In this context, the duration \tstarref\ is not affected by extinction and the extent to which embedded massive star formation would extend the duration of the overlap phase can be determined by applying the same methodology to galaxies for which high-resolution $24\micron$ maps are available. For NGC300, including embedded star formation would increase the duration of the overlap phase ($t_{\rm fb}/\tau=0.1$) by only a few per~cent \citep{Kruijssen2019}. Because NGC300 is a low-mass, half-solar metallicity galaxy, we might expect a stronger effect in more massive galaxies with higher cloud column densities. In particular, if massive stars are forming in a CO-dark environment, or if star formation stops before the CO gas has been cleared, \tstarref\ might be affected by extinction. In a future paper, we plan to systematically address the impact of embedded massive star formation on the relative durations of the inert, isolated CO phase and the overlap phase (J.~Kim et al.\ in prep.). Without further evidence, the strong correlation between massive star formation and the end of the CO-bright phase that we find here suggests a causal relation (see Section~\ref{sec:tfb}). This extends the result previously obtained for NGC300, i.e.\ that stellar feedback is a likely, if not dominant driver of molecular cloud dispersal \citep{Kruijssen2019}, to a wide variety of nearby star-forming galaxies.

	\subsection{Other derived quantities}
	\label{sec:others}
	
In addition to the molecular cloud lifetime, our analysis allows us to constrain a wide variety of other physical quantities. These will be described in more detail in follow-up papers, but here we already summarise some of the key results. 

\subsubsection{Feedback time-scale} \label{sec:tfb}
The duration of the feedback phase (\tover), during which molecular clouds and \HII regions coexist, is relatively short, with $t_{\rm fb}=1{-}5~\myr$, and also exhibits environmental variation between galaxies. For four of the galaxies in our sample (NGC628, NGC3351, NGC3627, and NGC5068), this feedback time is significantly shorter than the typical lower limit of 4\,Myr at which the first supernovae explode \citep[e.g.][]{Leitherer14}, whereas for another two (NGC4321 and NGC4535) it is marginally shorter or consistent with 4\,Myr. Under the assumption that the embedded phase of massive star formation is short (i.e.\ $\la1~\myr$, see the discussion in Sections~\ref{sec:sftracer} and~\ref{sec:tgas}, as well as e.g.~\citealt{Prescott2007,Hollyhead2015,Kruijssen2019}), this implies that, in these environments, early feedback mechanisms such as winds, photoionisation or radiation pressure must be the dominant processes driving the destruction of molecular clouds.

The short feedback time-scales are not achieved by dynamical cloud dispersal without associated massive star formation. As explained in Section~\ref{sec:method}, our methodology fundamentally constrains the time spent in a CO-bright phase until associated \Ha emission. If clouds would disperse dynamically without massive star formation, this `starless' cycle would be added onto a future one during which massive stars do form. The fact that this integrated cloud lifetime is found to be similar to a (cloud-scale or galactic) dynamical time-scale (see Section~\ref{sec:tgas_variations}) means that there is very little time to go through multiple cycles of dynamical dispersal and (re-)formation. While we cannot formally reject such a scenario, the above time-scale argument makes it unlikely that clouds go through multiple lifecycles prior to experiencing massive star formation. We therefore propose that the close correlation between the appearance of massive stars and rapid cloud dispersal is physical in nature.

In addition, the short feedback time-scales provide evidence against multiple generations of massive star formation within GMCs taking place (and ceasing) prior to the (potentially extended) star formation episode that drives cloud dispersal (see footnote~\ref{ft:interpretation}). The reason is that the feedback time-scale represents the total time spent by a region in a combined CO-bright and \Ha-bright state. Because these overlap time-scales are of a similar duration as the time-scale over wich \Ha\ is emitted by a massive star-forming region, $t_{\rm star,ref}\approx4.3~\myr$, allowing even a single earlier, unembedded massive star formation episode would leave little or no time for the final massive star formation episode to coexist with a CO-bright cloud. The only alternative is that \HII\ regions born during any earlier episodes of massive star formation would be ejected from the cloud on a short ($\sim1~\myr$) time-scale. This would require velocities of $\sim30~\kms$, well in excess of the typical cloud-scale velocity dispersion observed in these galaxies \citep{Sun2018}. Therefore, the most plausible interpretation is that massive star formation is temporally clustered towards the end of the cloud lifecycle.

When comparing the measured feedback time-scales to the physical resolutions listed in Table~\ref{tab:observations}, we see a suggestion of a weak trend of increasing \tover\ towards coarser resolutions (also see the minimum aperture sizes $l_{\rm ap,min}$ in Table~\ref{tab:input}). We have combined our results with other studies performing the same analysis for NGC300 and M33 \citep{Hygate2019b,Kruijssen2019}, which all have resolutions of 50~pc or better, to determine whether this constitutes a systematic trend. We find that the feedback time-scale is uncorrelated with resolution for $l_{\rm ap,min}<120$~pc, but a very weak trend starts to appear for $l_{\rm ap,min}>120$~pc, in that no feedback time-scales $t_{\rm fb}<3.3~\myr$ are found for galaxies with observations at these resolutions. We therefore advise some caution in the interpretation of the feedback time-scales measured for NGC4254, NGC4303, NGC4321, and NGC4535. It is possible (though not necessarily likely) that these represent upper limits.

Using the same statistical method applied to \Ha\ and CO(1-0) observations, a similarly short feedback time of 1.5\,Myr has been measured in NGC300, for which \citet{Kruijssen2019} infer that molecular clouds are predominantly destroyed by photoionisation and stellar winds. Other studies, most of which rely on different methodological approaches, have also found evidence that GMCs are dispersed within a few Myr after the onset of massive star formation \citep[e.g.][]{Kawamura2009, Whitmore2014, Hollyhead2015, Corbelli2017, Grasha2019, Hannon2019, Hygate2019b}. A detailed comparison between the measured feedback time-scales and theoretical expectations for different feedback mechanisms is investigated in more detail in a companion paper (M.~Chevance et al.\ in prep.). The results of that work confirm the importance of `early', pre-supernova feedback highlighted here.

\subsubsection{Region separation length}
In addition to the evolutionary timeline discussed so far, we also measure the separation length between independent regions ($\lambda$). This length scale is not an area-weighted mean separation length (which would be inflated by large empty space in galaxies, such as inter-arm regions), but instead describes the length scale in the immediate vicinity of a region over which a sufficiently large number of neighbouring regions is found to wash out the decorrelation seen in Figure~\ref{fig:tuningfork}. As such, it reflects the local number density of regions around emission peaks and does so in a way that combines both maps (CO and \Ha in this case). Physically, $\lambda$ defines the separation length between independent building blocks that each undergo the evolutionary lifecycles visualised in Figure~\ref{fig:timeline} and together determine how galaxies form stars.

We find that the region separation length ranges between $\lambda=100{-}300~{\rm pc}$. Similar values have been found in NGC300 \citep[$\lambda=104^{+22}_{-18}~{\rm pc}$,][]{Kruijssen2019} and M33 \citep[$\lambda=164^{+37}_{-24}~{\rm pc}$,][]{Hygate2019b}, but our measurements extend this range. \citet{Elmegreen2018} find a separation length of $\lambda_{\rm IR}=410$~pc for infrared-bright ($3.6{-}8~\micron$) clumps situated along 27 filaments in NGC4321. While this appears to be larger than the separation length measured here for the same galaxy ($\lambda=248^{+33}_{-26}~{\rm pc}$), we note that infrared emission tracing embedded stars only spans part of the timelines in Figure~\ref{fig:timeline}, such that the resulting separation length is increased by a factor of $\sqrt{\tau/t_{\rm IR}}$, with $t_{\rm IR}$ the visibility lifetime of the infrared emission. Therefore, these two values match each other to within the uncertainties on $\lambda$ if $t_{\rm IR} = \tau(\lambda/\lambda_{\rm IR})^2$ falls in the range $6.2{-}12.1$~Myr. This is not an unreasonable range, because it requires that the IR emission traces the overlap phase, part of the `isolated young stellar' phase,\footnote{Based on observations of NGC300, the Large Magellanic Cloud, and M33, 24\mic\ emission and \Ha\ emission seem to largely trace the same part of the timeline (J.~Kim et al.\ in prep.), so we expect IR emission to also partly trace the isolated \Ha\ phase.} as well as a short embedded phase of (at most) a few Myr.
Finally, we note that the other difference between \citet{Elmegreen2018} and this work is that we consider the entire galaxy, whereas \citet{Elmegreen2018} focus on the separation along dominant filamentary structures. Excluding peaks that do not closely follow these structures likely results in a longer measured separation scale. A similarly larger separation scale between individual CO peaks ($\sim 400$\pc) is indeed observed by \citet{Henshaw2019} along the southern spiral arm of NGC4321. 

When comparing the measured separation lengths to the physical resolutions listed in Table~\ref{tab:observations}, we typically find larger separation lengths at coarser resolution (also see the discussion in Section~\ref{sec:tfb}). We have combined our results with other studies performing the same analysis for NGC300, M33, and the LMC \citep[with the latter measuring the separation length for H\textsc{i} clouds and \HII\ regions]{Hygate2019b,Kruijssen2019,Ward2019}, which all have resolutions of 50~pc or better, to determine whether this constitutes a systematic trend. We find that the region separation length mirrors the behaviour of the feedback time-scale. It is uncorrelated with resolution for $l_{\rm ap,min}<120$~pc, but a trend starts to appear for $l_{\rm ap,min}>120$~pc. We therefore advise some caution in the interpretation of the separation lengths measured for NGC4254, NGC4303, NGC4321, and NGC4535. It is possible (though not necessarily likely) that these represent upper limits.

It remains to be determined which physical mechanisms set the region separation length across our galaxy sample. For NGC300, \citet{Kruijssen2019} compare the region separation length to the gas disc scale height and the \citet{Toomre1964} instability length and find that $\lambda$ matches the gas disc scale height across the full extent of the star-forming disc. A future paper will present this comparison for the nine galaxies considered here, and will investigate whether this correlation applies across the nearby galaxy population (M.~Chevance et al.\ in prep.).

\subsubsection{Feedback velocity}
Having measured the time-scale over which stellar feedback disperses molecular clouds, combining this with the typical spatial extent of the clouds results in a characteristic velocity scale. This `feedback outflow velocity' is defined as
\begin{equation}
    v_{\rm fb} = \frac{r_{\rm CO}}{t_{\rm fb}}, 
\end{equation}
where $r_{\rm CO}$ is the mean radius of the CO emission peaks determined with \textsc{Heisenberg} as the standard deviation of a two-dimensional Gaussian (see eq.~95 in \citealt{Kruijssen2018} and the discussion in Appendix~A1 of \citealt{Hygate2019}; we show $r_{\rm GMC}=1.91r_{\rm CO}$ in Figure~\ref{fig:GMC_prop}).\footnote{While some of the CO emission peaks may represent unresolved groups of molecular clouds, the measured feedback velocity is quite robust against such blending effects, because the CO peak radius and the feedback time-scale exhibit similar dependences on blending (see fig.~3 of \citealt{Kruijssen2019}). As a result, the uncertainties increase towards coarser resolution, but the feedback velocity itself remains largely consistent with its true value.} Depending on the nature of molecular cloud dispersal, e.g.\ whether it is kinetic or takes place by a phase transition, this velocity may represent the speed of the kinetic removal of molecular gas or the speed of the phase transition front. We obtain values in the range $v_{\rm fb}=8{-}21~\kms$, with a mean of $v_{\rm fb}\approx15~\kms$. These velocities fall within the range of typical expansion velocities found in nearby \HII regions in the Milky Way, LMC, NGC300, and M33 \citep[$6{-}30~\kms$, see e.g.][]{Bertoldi1990,Murray2010,Hygate2019b,Kruijssen2019,Mcleod2019,Mcleod2019b} and in numerical simulations of expanding \HII regions \citep[e.g.][]{Dale2014,Kim2018}. These predictions can be tested independently by measuring the ionised gas kinematics through (integral-field) spectroscopy (for instance with MUSE) for these galaxies.

\subsubsection{Star formation efficiency}
\label{sec:sfe}
On galactic scales, the star formation relation between the gas mass ($M_{\rm gas}$) and the SFR implies a gas depletion time $t_{\rm dep}\equiv M_{\rm gas}/{\rm SFR}$, which is observed to be $t_{\rm dep}\approx2~\gyr$ in nearby star-forming galaxies \citep{Bigiel2008,Bigiel2011,Leroy2008,Leroy2013,Blanc2009,Schruba2011} and represents the time necessary to convert the entire reservoir of molecular gas into stars at the current SFR. Because the SFR can be expressed as ${\rm SFR}=\epsilon_{\rm sf}M_{\rm gas}/t_{\rm CO}$, where $t_{\rm CO}$ is the cloud lifetime and $\epsilon_{\rm sf}$ is the mean star formation efficiency per unit cloud lifetime, the gas depletion time is also given by $t_{\rm dep}=t_{\rm CO}/\epsilon_{\rm sf}$. This expression highlights that, at fixed depletion time, there exists a degeneracy between the star formation efficiency and the cloud lifetime. The long depletion time measured on galactic scales (i.e.\ $t_{\rm dep}\approx2~\gyr$ being much larger than a dynamical time, see e.g.\ \citealt{Zuckerman1974}) can either be a result of a small cloud-scale star formation efficiency or of a long cloud lifetime. By directly measuring the characteristic time-scale on which individual clouds within galaxies live and form stars, $t_{\rm CO}$, we break this degeneracy. As noted above, our results qualitatively indicate that only a small fraction of the gas mass is converted into stars, with clouds being short-lived and disrupted by stellar feedback before they reach a high star formation efficiency. This is consistent with theoretical and numerical predictions \citep[e.g.][]{Semenov2017,Grudic2018,Kim2018}.

Quantitatively, we calculate the integrated star formation efficiency per star formation event as \citep[eq.~143]{Kruijssen2018}:
\begin{equation}
\label{eq:esf}
\epsilon_{\rm sf} = \frac{t_{\rm CO} \Sigma_{\rm SFR}}{\Sigma_{\rm gas}},
\end{equation}
where $\Sigma_{\rm SFR}$ is the SFR surface density and $\Sigma_{\rm gas}$ the molecular gas surface density across the field of view of each galaxy where we carry out our analysis. We calculate $\Sigma_{\rm SFR}$ as described in Section~\ref{sec:sfr} and we obtain $\Sigma_{\rm gas}$ from the filtered CO map, using the conversion factor $X_{\rm gas}$ from Table~\ref{tab:input}. For $\Sigma_{\rm gas}$, we thus take only the compact CO emission into account, because this is the emission for which \tCO\ describes the lifetime. This choice assumes that most of the diffuse CO emission\footnote{This is $\sim25$~per~cent on average, see Table~\ref{tab:diffuse} for all measurements of the diffuse CO and \Ha emission fractions across our galaxy sample.} originates from truly diffuse molecular gas or from small molecular clouds that do not participate in the formation of massive stars generating \Ha emission. With these assumptions in mind, we measure small star formation efficiencies per star formation event, ranging between $\epsilon_{\rm sf}=4{-}10$~per~cent. The combination of a short \tCO\ and low $\epsilon_{\rm sf}$ indicates that star formation is fast and inefficient, for all galaxies in our sample.

In closing, we note the difference in definition between the integrated star formation efficiency per star formation event from equation~(\ref{eq:esf}) to the star formation efficiency per free-fall time, which is given by $\epsilon_{\rm ff}=t_{\rm ff}\Sigma_{\rm SFR}/\Sigma_{\rm gas}$ (where $t_{\rm ff}$ is the free-fall time). \citet{Utomo2018} measure $\epsilon_{\rm ff}$ for all nine galaxies in our sample. Because we find cloud lifetimes of 1--3 free-fall times, $\epsilon_{\rm sf}\equiv\epsilon_{\rm ff}t_{\rm CO}/t_{\rm ff}$ is higher than $\epsilon_{\rm ff}$ by a factor of a few. Another difference is that we measure the star formation efficiency of compact clouds, i.e.\ after removing diffuse CO emission from the maps, whereas \citet{Utomo2018} measure $\epsilon_{\rm ff}$ from the unfiltered CO maps, resulting in a lower efficiency, appropriate for the entire molecular gas reservoir rather than for the clouds considered here. This also contributes to $\epsilon_{\rm sf}>\epsilon_{\rm ff}$.

\section{Variation of the molecular cloud lifetime as a function of galactic environment}
\label{sec:tgas_variations}

We now discuss how the cloud lifetime depends on the galactic environment, by applying our analysis to bins in galactocentric radius. We then compare to analytical models for cloud evolution to determine whether the cloud lifetime is set by internal or external processes. We also discuss the influence of galactic morphological features on the measured cloud lifetimes. In conclusion, we find that both internal and external processes can set the cloud lifetime, and propose a rough separation between both regimes in terms of a critical value of the large-scale gas surface density (i.e.\ the area-average across radial rings or annuli within the galaxies). 

\subsection{Radial profiles of the molecular cloud lifetime}
\label{sec:binning}

After having identified a variation of the integrated cloud lifetime between galaxies in Section~\ref{sec:tgas}, we investigate potential variations of \tCO\ within galaxies to test the hypothesis that the cloud evolution process depends on galactic environment, and determine which mechanisms are playing a role in this process. For each galaxy in our sample, we apply our analysis to successive radial bins around the galactic centre (see Figure~\ref{fig:Bin_images} for images of the galaxies showing how the radial bins are defined). To do this, we divide each galaxy into non-overlapping radial bins of a minimum width of 1\,kpc. This condition is set to satisfy the requirement of having a random two-dimensional distribution on a scale $\sim \lambda$ (typically a few 100\,pc, see Table~\ref{tab:results}) for the application of the method. In addition, we require that each bin contains a minimum of 50 peaks identified in our full-galaxy runs for each tracer, to ensure sufficient statistics to constrain the derived quantities to sufficiently high precision \citep{Kruijssen2018}. If this condition is not satisfied for bins that are 1\,kpc in width, we increase their width (and therefore decrease the total number of bins) in order to satisfy this condition.

\begin{figure*}
\includegraphics[trim=00mm 50mm 0mm 45mm, clip=true, width=\linewidth]{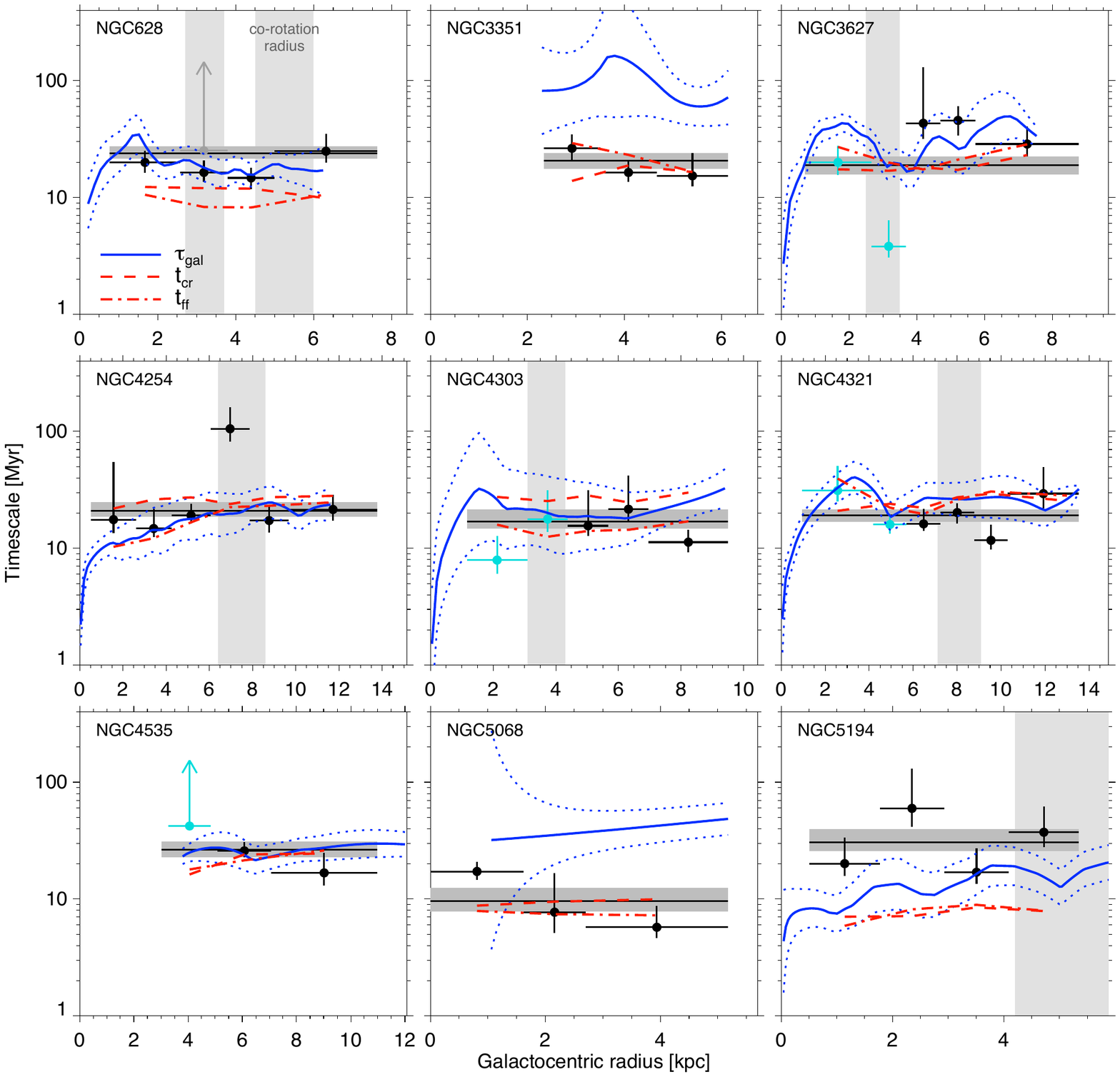}
\caption{
Measured molecular cloud lifetime (\tCO) as a function of the galactocentric radius for each galaxy (data points with error bars) and for the full galaxy (horizontal black lines with shaded area representing the 1$\sigma$ range of uncertainties). For each data point, the horizontal bar spans the range of radii within which \tCO\ is measured and the vertical bar represents the 1$\sigma$ uncertainties. We note that the uncertainties on \tCO\ for the individual bins are generally larger than the uncertainties for the full galaxies. This results from a larger degree of stochasticity due to a lower number of regions per bin, while the uncertainties on the full galaxies mostly reflect observational uncertainties on the (well-constrained) mean value. The shaded data point in NGC628 represents the value of \tCO\ measured in the second radial bin if the `headlight' cloud (which vastly dominates the CO emission from the cloud population in that radial bin, see the text and \citealt{Herrera2019}) is not masked. The light blue data points indicate the bins coinciding with the bar or residing at the tip of the bar for the barred galaxies NGC3627, NGC4303, NGC4321, and NGC4535. The lines indicate the predictions from simple theoretical prescriptions. Red dash-dotted and dashed lines indicate the profiles of the cloud free-fall time and the cloud crossing time, respectively. Blue solid lines represent the cloud lifetime due to various galactic dynamical processes as predicted by the analytical model of \citet{Jeffreson2018}, with dotted lines indicating the $1\sigma$ uncertainties on the prediction. The positions of the co-rotation radii (see text and Table~\ref{tab:co-rotation}) are indicated by vertical shaded areas. All measurements of the cloud lifetime shown in this figure are listed in Table~\ref{tab:all_tgas}.}
\label{fig:radprof}
\end{figure*}

As the peak identification is normally done by stepping down in flux density relative to the brightest peak in the image (which is different in each radial bin, possibly causing the identified peaks to be different than in the full maps; see Section~\ref{sec:method}), we supply the peaks identified across the full maps as input for the analysis in each radial bin. As input maps, we use the filtered maps, from which diffuse emission has been removed through the iterative filtering process applied to the full field of view (as described in  Section~\ref{sec:diffuse}). This approach is validated a posteriori by the fact that $\lambda$ (which sets the filtering scale) is approximately constant between each bin for a given galaxy. The resulting \tCO\ profiles as a function of the galactocentric radius for each galaxy are presented in Figure~\ref{fig:radprof} and in Table~\ref{tab:all_tgas}. 

We find that, given the uncertainties, the cloud lifetime is often consistent with being constant within galaxies, although some radial variations can be identified. For example, \tCO\ is relatively constant in some galaxies (e.g.\ NGC628, NGC4303 and NGC4321), but in others it peaks at a certain radius (e.g.\ NGC3627, NGC4254 and NGC5194), or decreases outwards (e.g.\ NGC3351, NGC4535 and NGC5068). To understand the origin of these variations, we will now compare the measured cloud lifetimes with analytical predictions for cloud-scale dynamical time-scales and galactic dynamical time-scales.

	\subsection{Comparison with analytical models}
	\label{sec:analytical}

To understand the origin of the environmental variation of \tCO , we compare our observations with analytical predictions. If molecular clouds are gravitationally bound and globally collapsing, their evolution is governed by the gravitational free-fall time \tff. We define the cloud free-fall time as:
\begin{equation}
\label{eq:tff}
  t_{\rm{ff}} = \sqrt{\frac{\pi^2 r_{\rm GMC}^3}{10 G M_{\rm GMC}}},
\end{equation}
for spherical clouds of radius $r_{\rm GMC}$ and molecular gas mass $M_{\rm GMC}$, where $G$ is the gravitational constant. The GMC radius and mass are derived using the output from \textsc{Heisenberg} by defining $r_{\rm GMC}\equiv1.91r_{\rm CO}$ \citep{Kruijssen2019} and $M_{\rm GMC}\equiv{\cal E}_{\rm gas}\Sigma_{\rm H_2}\pi(\lambda/2)^2$, where ${\cal E}_{\rm gas}$ is the surface density contrast on a size scale $\lambda$ relative to the surface density measured across the field of view, $\Sigma_{\rm H_2}$ \citep{Kruijssen2018}. The choice to take these from the output of \textsc{Heisenberg} is mainly self-consistency -- this way, the masses and radii are obtained for the units that are inferred to undergo the evolutionary lifecycles characterised in this work. By contrast, using a cloud catalogue would rely on subjective classification. It would also be more strongly affected by the finite resolution of the observations. For example, applying the cloud characterisation algorithm {\sc CPROPS} \citep{Rosolowsky2006} to the CO maps leads in some cases to the identification of GMC complexes of several 100~pc in size (larger than $\lambda$), rather than individual GMCs. This is remedied in \textsc{Heisenberg} by using a sub-resolution model to infer GMC sizes from the surface brightness contrast of a subsample of emission peaks against the large-scale background. By definition, these emission peaks are then separated by the separation length of independent regions (see eqs.~94 and~95 of \citealt{Kruijssen2018}), such that their radii cannot exceed $\lambda$. In this context, $r_{\rm GMC}$ and $M_{\rm GMC}$ represent the CO flux-weighted average for each radial bin.

We show the median profile of \tff\ as a function of the galactocentric radius in Figure~\ref{fig:radprof}. For all the galaxies in our sample, \tff\ is relatively constant within galaxies, exhibiting variations of less than a factor of two. In general, \tff\ is close to or shorter than the measured molecular cloud lifetime, both for the global measurements and in individual bins, which is expected given that it represents the extreme case of free-fall collapse. Quantitatively, we find that clouds live for $1{-}3$ free-fall times. However, in some cases \tCO\ appears shorter than \tff\ by more than the uncertainty. This happens in 6 out of 39 radial bins and could potentially be caused by a biased measurement of \tCO\ due to the effect of galaxy morphology (see the discussion below in Section~\ref{sec:morphology}), or by the fact that the clouds are not resolved, resulting in an underestimated value of \tff\ due to beam dilution. However, at least some of these six bins with short-lived clouds should simply result from the uncertainties. We find 12 bins for which $t_{\rm CO}\approx t_{\rm ff}$ to within the uncertainties, implying that for a normal distribution we expect three bins where \tCO\ falls significantly below \tff.

We also compare the measured cloud lifetime to the GMC crossing time, which is defined as:
\begin{equation}
\label{eq:tcr}
t_{\rm cr} = \frac{r_{\rm GMC}}{\sigma_{\rm vel}},
\end{equation}
where $\sigma_{\rm vel}$ is the one-dimensional cloud velocity dispersion. Because \textsc{Heisenberg} does not provide kinematic information, we use the individual cloud velocity dispersions determined by E.~Rosolowsky et al.\ (in prep.) using {\sc CPROPS} \citep{Rosolowsky2006}. Before inserting $r_{\rm GMC}$ and $\sigma_{\rm vel}$ into equation~(\ref{eq:tcr}), we calculate the CO flux-weighted average for each radial bin. For a large fraction of the galaxies, $t_{\rm{cr}}$ is similar to the free-fall time and the measured molecular cloud lifetimes, both globally and in individual bins.

Despite the rough similarity between \tCO\ and \tff, in many cases (e.g.\ NGC628, NGC3627, NGC5194) the measured molecular cloud lifetimes cannot be simply explained by the local cloud dynamical time, as visible in Figure~\ref{fig:radprof}. We pursue the alternative hypothesis that molecular cloud lifetimes are environmentally dependent and can be affected by galactic dynamics, which has been shown to hold for other cloud properties, such as surface density, velocity dispersion, and boundedness \citep[e.g.][]{Leroy2017,Sun2018,Schruba2019}. We therefore compare our measurements with the predictions of the analytical theory for GMC lifetimes from \citet{Jeffreson2018}. Within this theory, the cloud lifetime is set by the large-scale dynamics of the ISM and calculated as the harmonic average of characteristic time-scales associated with the gravitational collapse of the ISM ($\tau_{\rm{ff,g}}$) counter-acted by galactic shear ($\tau_{\beta}$), cloud-cloud collisions ($\tau_{\rm{cc}}$), density wave perturbations ($\tau_{\Omega_{\rm P}}$), and epicyclic perturbations ($\tau_{\kappa}$). As galactic shear is a dynamically dispersive process, while the other mechanisms are dynamically compressive, it competes with gravitational collapse and the resulting lifetime can be written as
\begin{equation}
    \label{eq:galdyn}
    \tau_{\rm gal} = |\tau_{\rm{ff,g}}^{-1}-\tau_{\beta}^{-1}+\tau_{\rm{cc}}^{-1}+\tau_{\Omega_{\rm P}}^{-1}+\tau_{\kappa}^{-1}|^{-1} .
\end{equation}
We show the predictions of this model with blue lines in Figure~\ref{fig:radprof}, including the uncertainties on the predictions obtained by propagating the uncertainties on the input quantities (see below). We note that in the inner part of NGC3351, the $\tau_{\beta}$ term becomes as large as all other mechanisms combined, resulting in an extremely large \tgal\ with a large downward uncertainty. This most likely reflects the morphology of NGC3351, with a strong bar in the center, and a prominent gas ring between $\sim 2$ and 5\,kpc.

All of the time-scales taken into account in \tgal\ depend on observable parameters. Specifically, these are the angular velocity $\Omega$ (all time-scales depend inversely on $\Omega$), the Toomre $Q$ parameter, the surface densities and velocity dispersions of gas ($\Sigma_{\rm g}$ and $\sigma_{\rm g}$, respectively) and stars ($\Sigma_{\rm s}$ and $\sigma_{\rm s}$, respectively), which are combined into the single quantity $\phi_{\rm P} = 1+(\Sigma_{\rm s} \sigma_{\rm g} / \Sigma_{\rm g} \sigma_{\rm s})$, the shear $\beta\equiv {\rm d}\ln{\Omega}/{\rm d}\ln{R}+1$, the number of spiral arms $m$ and their pattern speed $\Omega_{\rm P}$. Different (regions of the) galaxies are therefore likely to cover different areas of the ($\beta$, $Q$, $\Omega$, $\phi_{P}$, $m$, $\Omega_{\rm P}$) parameter space, where cloud evolution is predicted to be governed by different processes, resulting in different values of the cloud lifetime \citep{Jeffreson2018}. We describe how these quantities are derived in Appendix~\ref{sec:appradial}. Figure~\ref{fig:radprof} shows that in most galaxies (except for NGC3351 and NGC5068) there exists a broad agreement between the measured molecular cloud lifetimes and the analytical predictions. However, we also note discrepancies in some individual radial bins, which we explore below.

\subsection{Influence of galaxy morphology}
\label{sec:morphology}

  \subsubsection{Co-rotation radius}
At the co-rotation radius, the velocity of the material in the disc equals the pattern speed of the spiral structure. Molecular clouds located at the co-rotation radius are therefore likely to permanently reside in a deep potential well provided by the spiral arm, which potentially facilitates sustained gas inflow and extends cloud lifetimes (i.e.\ the duration of the CO-visible phase). Table~\ref{tab:co-rotation} summarises the available measurements of co-rotation radii for the galaxies in our sample, and these are indicated as vertical grey-shaded bands in Figure~\ref{fig:radprof}. For NGC3351, NGC4535 and NGC5068, either the co-rotation radius falls outside of the range of radii considered here, or no clear pattern speed was available from previous measurements.

\begin{table}
\begin{center}
\begin{tabular}{lcc}
\hline
                                Galaxy  & $R_{\rm CR}$ [kpc] & Reference  \\
\hline
                            NGC628    & $2.7{-}3.7$ & 1 \\
                                      & $4.5{-}6.0$ & 2 \\
                            NGC3627   &  $2.5{-}3.5$ & 3\\
                            NGC4254   &  $6.4{-}8.6$ & 4,5,6\\
                            NGC4303   &  $3.1{-}4.3$ & 7 \\
                            NGC4321   &  $7.1{-}9.1$ & 8,9\\
                            NGC5194   &  $4.2{-}6.5$ & 8,10,11\\
\hline
\end{tabular}
\caption{Position of co-rotation radii for the galaxies of our sample.
References: (1) \citet{Herrera2019}, (2) \citet{Cepa1990}, (3) \citet{Rand2004}, (4) \citet{Elmegreen1992}, (5) \citet{Gonzalez1996}, (6) \citet{Kranz2001}, (7) \citet{Schinnerer2002}, (8) \citet{Elmegreen1989}, (9) \citet{Garcia-Burillo1998}, (10) \citet{Scheepmaker2009}, (11) \citet{Querejeta2016}. The range of values given takes into account 15~per~cent uncertainties for single references, or the range of values found in the literature when several references exist. Where necessary, we scale $R_{\rm CR}$ to be consistent with the distances tabulated in Table~\ref{tab:input}.}
\label{tab:co-rotation}
\end{center}
\end{table}

Some influence of co-rotation on the molecular cloud lifetime is suggested in several galaxies in Figure~\ref{fig:radprof}, with longer cloud lifetimes compared to the galactic average in the bins located at the co-rotation radius. This is most prominent in NGC4254, NGC3627 and NGC628 (lower limit on the cloud lifetime shown by the grey data point), but also somewhat in NGC5194 in the sense that the last bin falls above the median of all bins. One of the most striking examples is found in NGC628, with the presence of a very bright cloud (referred to as the `headlight' cloud in \citealt{Herrera2019}), located at the intersection of a spiral arm and the co-rotation radius at 3.2\,kpc. We have masked this particular cloud in our analysis in Section~\ref{sec:results}. The reason is that this cloud is three times brighter in CO(2-1) than any other cloud in the galaxy and it would thus dominate our flux-weighted cloud-lifetime, therefore strongly biasing the results towards this particular gas-dominated environment and increasing the apparent cloud lifetime, especially in the second radial bin. If left unmasked, the average cloud lifetime for the full galaxy increases slightly from $t_{\rm CO}=24.0^{+3.6}_{-2.5}~\myr$ to $t_{\rm CO}=25.1_{-2.8}^{+5.0}~\myr$ and becomes unconstrained in the bin including the `headlight' cloud, with a lower limit of 25.1\,Myr (in the top-left panel of Figure~\ref{fig:radprof}, compare the grey symbol to the black symbol at the same galactocentric radius). We have verified that other galaxies are not dominated by a single cloud, meaning that the headlight cloud in NGC628 represents an exception due to its extreme mass. In a less extreme way, the low value of \tCO\ measured at the co-rotation radius in NGC4321 likely results from a similar effect, but at a later stage of the star formation cycle. Some of the brightest \Ha\ peaks (including the brightest peak of our map) are located at this co-rotation radius, indicating an accumulation of recent massive star formation events. This violates our requirement of an approximately constant SFR and biases the measured \tCO\ towards a low value.

  \subsubsection{Influence of the bar}
Bars are known to drive large local variations of the molecular gas depletion time due to gas transport and bursty star formation. Specifically, they generate accumulations of material at the bar ends, where massive clouds and bursty star formation are commonly observed \citep[e.g.][]{Beuther2017}, they induce strong radial transport and suppress star formation \citep[e.g.][]{Khoperskov2018,Sormani2019}, and they drive nuclear starbursts \citep[e.g.][]{Peeples2006}. A small number of bins in Figure~\ref{fig:radprof} exhibit very high or very low values of the molecular cloud lifetime (e.g.\ the second bin of NGC3627 or the inner bin of NGC4535), which is plausibly caused by the presence of a bar in these galaxies. In Figure~\ref{fig:radprof}, we highlight in light blue the data points corresponding to the bins including the bar or the end of the bar for the four barred galaxies in our sample (these are NGC3627, NGC4303, NGC4321, and NGC4535, but excludes NGC3351, for which we do not cover the bar; see also Figure~\ref{fig:Bin_images}). For NGC3627, NGC4303 and NGC4321, the inner bin covers the bar of the galaxy, whereas the second bin covers the tip of the bar. For NGC4321 we note an elevation (by $\sim60$~per~cent) of the molecular cloud lifetime in the inner bin compared to the galactic average. This is caused by a lack of \Ha emission in the bar noted by \citet{Schinnerer2019}. We do not measure a significant elevation of \tCO\ in NGC3627 and NGC4303, where the star formation in the bar is not as strongly suppressed.

For the bins at the end of the bar (i.e.\ the second bins of NGC3627, NGC4303, and NGC4321, as well as the inner bin of NGC4535), we expect highly bursty and localised star formation. This can lead to strongly enhanced or deficient star formation, depending on the moment of observation, and would violate one of the fundamental assumptions of our methodology, which is that the star formation history over the recent time interval $\tau$ must have been relatively constant \citep{Kruijssen2018}. As a consequence, it is not surprising that the data point at the tip of the bar in NGC3627 seems to be an outlier in Figure~\ref{fig:radprof}. In NGC3627, \citet{Beuther2017} suggest that the interaction between the end of the bar and the spiral arms might induce strong star formation events. The particularly short cloud lifetime ($t_{\rm CO}=3.8_{-0.8}^{+2.6}$\,Myr) measured in the second bin of NGC3627 (as well as a short $t_{\rm fb}=0.8_{-0.4}^{+0.7}$\,Myr) is due to the fact that two very bright regions both in \Ha and in CO dominate this bin. This likely traces a recent burst of star formation and biases the average duration of the different phases towards low values. By contrast, the long \tCO\ (lower limit of 42.1\,Myr) measured in the innermost bin of NGC4535, also covering the end of the bar, indicates a low SFR over the recent time interval $\tau$. Interpreted in the context of bursty star formation, this reflects the same physical mechanism as in NGC3627, but observed at a different moment in time. While a starburst has recently taken place at the tip of the bar in NGC3627, gas is currently accumulating at the tip of the bar in NGC4535. Both of these extremes bias the measured cloud lifetimes.

\subsection{Galactic dynamics versus internal dynamics}
\label{sec:galactic dynamics}

We now investigate the variation of the measured molecular cloud lifetimes in regions that are not affected by galaxy morphology as described in Section~\ref{sec:morphology} (i.e.\ the black data points in Figure~\ref{fig:radprof}). Our observations show good agreement with the analytical predictions from \citet{Jeffreson2018} in some of the radial bins, but diverge from these predictions in others, especially in the outskirts of galaxies (e.g.\ NGC4303 and NGC4535) and in galaxies with low global gas surface densities (e.g.\ NGC3351 and NGC5068), where \tCO\ is in better agreement with the cloud free-fall time or the cloud crossing time. We have looked for environmental factors that may govern this dichotomy between local or global dynamics correlating with the cloud lifetime, and find that the kpc-scale galactic gas surface density might play a key role. By dividing the sample of measurements between `low' and `high' regimes of the area-weighted mean molecular gas surface density in each radial bin (see Figure~\ref{fig:tGMC_tgas_tff_dispersion}), we find that the transition between cloud lifetimes being governed by galactic dynamics versus cloud lifetimes being set by cloud internal dynamics seems to occur at a molecular gas surface density averaged across the galactocentric radial rings (i.e.\ measured on $\ga$~kpc scales) of $\rm \Sigma_{\rm H_2,ring}\approx 8\,M_{\odot}$\,pc$^{-2}$ (see Appendix~\ref{sec:appthres} for details), with galactic dynamics dominating at high surface densities and internal dynamics dominating at low densities.\footnote{One might instead expect a division based on the cloud surface density contrast with respect to the kpc-scale surface density used here, such that galactic dynamics become more important at low density contrasts. This is consistent with our suggestion of a critical large-scale surface density, because the cloud surface density contrast decreases with large-scale gas surface density \citep[see eq.~9 of][]{Kruijssen2015b}.}

\begin{figure*}
\includegraphics[trim=10mm 5mm 5mm 0mm, clip=true, width=\hsize]{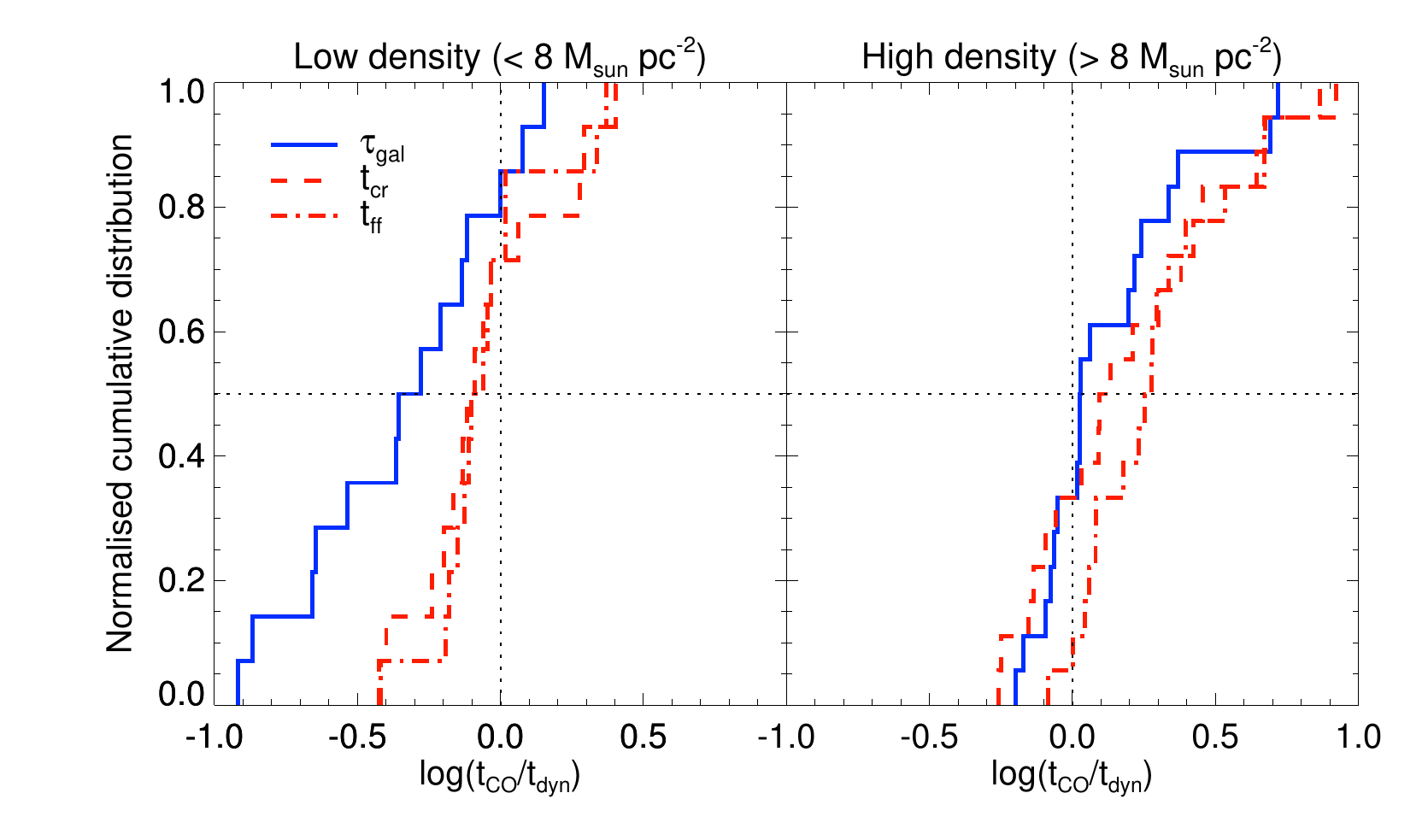}
\caption{Comparison of the measured molecular cloud lifetimes (\tCO) with the predicted cloud lifetimes from galactic dynamics (\tgal, from \citealt{Jeffreson2018}) and internal dynamics (the cloud crossing time, \tcr, and the cloud free-fall time, \tff) for all radial bins with low kpc-scale gas surface densities ($\Sigma_{\rm H_2,ring} < 8\,\Msun$\,pc$^{-2}$, left) and those with high kpc-scale gas surface densities ($\Sigma_{\rm H_2,ring} > 8\,\Msun$\,pc$^{-2}$, right). Shown are the normalised cumulative distributions of the difference in logarithmic space between \tCO\ and each of the predicted dynamical time-scales (see the legend), defined as $\log(t_{\rm CO}/t_{\rm dyn})$. These distributions do not include the radial bins affected by galactic morphology (light blue data points in Figure~\ref{fig:radprof}). In each panel, the horizontal dotted line indicates the median of the distribution and the vertical dotted line indicates perfect agreement between \tCO\ and \tcloud. Better agreement between \tCO\ and any of the three time-scales considered here manifests itself as steeper lines crossing more closely to the intersection of both dotted lines. These panels show that cloud lifetimes in regions with low surface densities correlate best with the time-scale for internal dynamical processes (median offset of $0.06{-}0.09$~dex, as opposed to $0.28$~dex for galactic dynamics), whereas those in regions with high surface densities correlate best with the time-scale for galactic dynamical processes (median offset of $0.03$~dex, as opposed to $0.13{-}0.28$~dex for internal dynamics). See the text and Appendix~\ref{sec:appthres} for details.}
\label{fig:timescales}
\end{figure*}

The different dynamical regimes are illustrated more quantitatively in Figure~\ref{fig:timescales}, where we compare \tCO\ with both the analytical prediction based on galactic dynamical processes from \citet{Jeffreson2018} and the internal dynamical time-scales of the clouds, i.e.\ the cloud free-fall time and the cloud crossing time. In this comparison, we distinguish between the bins where the kpc-scale molecular gas surface density is low ($\Sigma _{\rm H_2,ring}< 8\,\Msun$\,pc$^{-2}$, left panel) or high ($\Sigma _{\rm H_2,ring}> 8\,\Msun$\,pc$^{-2}$, right panel). At low kpc-scale molecular gas surface densities, there is a better agreement between \tCO\ and the internal dynamical time-scales (\tcr\ and \tff, showing median offsets of $0.09$ and $0.06$~dex, with standard deviations of $0.25$ and $0.20$~dex, respectively) than between \tCO\ and \tgal\ (median offset of $0.28$~dex, with a standard deviation of $0.34$~dex). Conversely, at high gas surface densities, there is good agreement between \tCO\ and \tgal\ (median offset of $0.03$~dex, with a standard deviation of $0.27$~dex), whereas the comparison of \tCO\ and the internal dynamical time-scales shows a systematic offset, albeit with a similar spread (for \tcr\ and \tff, the median offsets are $0.13$ and $0.28$~dex, with standard deviations of $0.34$ and $0.26$~dex, respectively). These results do not change significantly for small changes of the critical gas surface density at which the sample is divided into bins of low or high surface density. We test the robustness of these results with respect to the choice of the gas surface density threshold in Figure~\ref{fig:tGMC_tgas_tff_dispersion}, and confirm that the transition between these two regimes occurs between $7{-}9\,\Msun$\,pc$^{-2}$. In addition, we have verified that the result is robust against the removal of outliers in the data. This is not surprising, given that the central parts of the cumulative distributions are relatively steep.

The existence of two regimes\footnote{Of course, these two regimes of internal and galactic dynamics may be subdivided further. This likely requires adding physical dependences beyond the correlation with gas surface density highlighted here. For instance, the internal dynamical processes can be separated into the gravitational free-fall or the crossing time, depending on whether or not a cloud is gravitationally bound. Likewise, the galactic dynamical processes can be separated into the several terms of equation~(\ref{eq:galdyn}), with e.g.\ shear outperforming other processes towards high Toomre $Q$ and shallow rotation curves \citep{Jeffreson2018}.} regulating the molecular cloud lifetime at low and high density can be understood by considering the fraction of the gas reservoir probed by CO. We assume that star formation takes place in compact overdensities, which can be more or less uniquely traced by CO in environments of different gas surface density. At high molecular gas surface densities, CO is visible almost everywhere in the galaxy, including the space in between the compact overdensities in which star formation takes place, and may extend beyond the cloud tidal radii. As a result, even after filtering the diffuse emission on large scales, the remaining reservoir of CO-emitting molecular gas is spatially extended and is more likely to be affected by large-scale galactic dynamical processes. At low molecular gas surface densities, CO traces only the densest parts of the clouds, i.e.\ the overdensities that participate in star formation, and most of the gas reservoir in between is likely to be atomic \citep[e.g.][]{Schruba2011}. The collapse of the CO-emitting part of a cloud is therefore likely decoupled from the galactic dynamics and represents a local process. As a result, the CO-bright cloud lifetimes in low-surface density environments are not expected to be set by the external galactic dynamics, but rather by internal dynamics such as the (CO-)cloud free-fall or crossing time. This regime of `island GMCs' evolving on an internal dynamical time manifests itself e.g.\ in the outskirts of NGC4303 and NGC4535, as well as overall in the low molecular gas surface density galaxies NGC3351 and NGC5068, where the measured \tCO\ is consistent with the cloud free-fall time or the cloud crossing time.

We note that the environments with high gas surface densities exhibit a similar spread of $\log(t_{\rm CO}/t_{\rm dyn})$ for galactic dynamics (0.27~dex) and internal dynamics ($0.26{-}0.34$~dex). However, the medians differ, such that $t_{\rm CO}\approx\tau_{\rm gal}\approx1.3t_{\rm cr}\approx1.9t_{\rm ff}$. Therefore, the observed cloud lifetimes are equally well described as matching the galactic dynamical time-scale, as being equal to 1.9 times the internal free-fall time or 1.3 times the cloud crossing time. This is not unexpected, because the GMC internal dynamical time becomes proportional to the galactic dynamical time in the regime where galactic dynamics set the time-scale for cloud evolution \citep[this is referred to as the `Toomre regime' by][]{Krumholz2012}. Even though the time-scales are proportional to each other in this regime, the fundamental dependence is on galactic dynamics, to which the internal dynamics of the clouds equilibriate.

The analytical predictions from \citet{Jeffreson2018} are based on a simple analytical model, which limits the direct comparison with our measurement. In particular, the model does not distinguish between the different gas phases, but describes the lifetime of the entire gas concentration, irrespective of its phase. As a result, it is not surprising that the analytical predictions over-estimate the lifetime of the CO clouds in regions of low H$_{2}$ density, where a large part of a gas cloud is atomic and not CO-emitting. Several studies have shown that the atomic-to-molecular transition occurs at atomic gas surface densities $\rm \Sigma_{{\rm gas}}\approx 10\,M_{\odot}$\,pc$^{-2}$ at near-solar metallicity \citep[e.g.][]{Wong2002, Leroy2008, Schruba2011, Krumholz2014, Schruba2018}, which is close to the critical molecular gas surface density of $\Sigma_{\rm H_2,ring}=8\,\Msun$\,pc$^{-2}$ below which our cloud lifetimes correlate more strongly with the cloud's internal dynamical time-scales than with the galactic dynamical time-scale from \citet{Jeffreson2018}. This also explains why our results show that \tCO\ often decreases (or stays constant) with increasing galactocentric radius (see Figure~\ref{fig:radprof}), whereas the analytically predicted $\tau_{\rm gal}$ typically gradually increases towards the outskirts of the galaxies (scaling with $\Omega^{-1}$). In the outskirts of galaxies, the gas reservoir often becomes atomic gas-dominated \citep[e.g.][]{Schruba2011}, causing the CO lifetime measurements to only trace the final phase of cloud collapse. As a result, the analytical theory provides an upper limit to the true cloud lifetimes in this atomic-dominated regime.

In the regions where our measurements are well-reproduced by the analytical theory of \citet{Jeffreson2018}, which happens at small-to-intermediate galactocentric radii for most of the galaxies in our sample (NGC628, NGC3627, NGC4254, NGC4303, NGC4321, NGC4535, and NGC5194), the mid-plane free-fall time $\tau_{\rm{ff,g}}$ is generally the shortest of the time-scales included in $\tau_{\rm gal}$. When comparing our measured cloud lifetimes with the other individual characteristic time-scales in equation~(\ref{eq:galdyn}), we can rule out cloud-cloud collisions and spiral arm passages as important mechanisms limiting cloud lifetimes, since they typically act on much longer characteristic time-scales \citep[$\sim100~\myr$; see also][]{Jeffreson2018}. This suggests that the gravitational collapse of the ISM mainly regulates the cloud lifetime in the molecular-dominated discs of star-forming galaxies, and that clouds are not long-lived: they collapse, form stars, and get disrupted by feedback.

\section{Discussion}
\label{sec:discussion}

We now briefly validate our measurements by verifying that all requirements listed in \citet{Kruijssen2018} have been met. In addition, we carry out a comparison to other cloud lifetime measurements from the literature. The section is concluded with a discussion of the physical implications of our results for the GMC lifecycle in galaxies.

\subsection{Accuracy of the results}
\label{sec:requirements}

In Sections~\ref{sec:tgas} and~\ref{sec:others}, we have presented the results from applying our statistical analysis method (using the \textsc{Heisenberg} code) to our sample of disc galaxies. To validate the accuracy of these values, we verify here that we fulfill the requirements listed in sect.~4.4 of \citet{Kruijssen2018}. The following criteria guarantee that the three derived parameters \tgas, \tover, and $\lambda$ are measured with an accuracy of at least 30~per~cent (but often better):
\begin{enumerate}
    \item The durations of the gas and young stellar phases differ by less than an order of magnitude, with $|\log_{10}(t_{\rm star}/t_{\rm gas})| \leq 0.53$ for all galaxies (Tables~\ref{tab:input} and~\ref{tab:results}).
    \item In all cases, we verify that $\lambda \geq l_{\rm ap,min}$, which implies that the region separation length is sufficiently resolved by our observations to obtain a reliable measurement of \tover. Quantitatively, we have $\lambda \geq 1.7 l_{\rm ap,min}$ for all galaxies (Tables~\ref{tab:input} and~\ref{tab:results})
    \item We choose the galactocentric bins to have a minimum of 50 identified emission peaks of each type in each bin and a minimum width of 1~kpc (which is needed to fulfill the assumption of randomly distributed regions on a scale of a few times $\lambda$). A fortiori, we respect the condition $N_{\rm min} \equiv \min(N_{\rm peak,star},N_{\rm peak,gas}) \geq 35$ for each galactocentric bin (and each galaxy) necessary to ensure relative uncertainties of less than 50~per~cent on the derived quantities. For $\geq50$ peaks per tracer, we obtain relative uncertainties of $\leq30$~per~cent (see fig.~25 of \citealt{Kruijssen2018}).
    \item Focusing on an SFR or a gas peak should never lead to a deficit of this tracer relative to a galactic average. This condition is not fulfilled before we filter out diffuse emission. This applies in particular when focusing on \Ha peaks for most galaxies, because the \Ha maps have larger diffuse emission reservoirs than the CO maps, which is also visible directly in Figure~\ref{fig:COmaps}. After the filtering of the diffuse emission (Section~\ref{sec:diffuse}), this criterion is satisfied.
    \item The global star formation histories (SFHs) of the galaxies should be relatively constant over a time interval $\tau$ (i.e.\ $15{-}35~\myr$ for our measurements), so that the evolutionary timelines (including the CO-bright phase) are homogeneously sampled. Unfortunately, the spatially-resolved SFHs of the galaxies in our sample are not known. NGC3627 is interacting with the neighbouring galaxy NGC3628 \citep[e.g.][]{Rots1978, Haynes1979}, but all other galaxies in our sample are not expected to have significant variations of their SFR in the disc during the last $\sim50~\myr$. We note in particular that this is the case for NGC5194 \citep{Eufrasio2017,Tress2019}, despite its relatively recent interaction with NGC5195,$\sim$\,350-500\,Myr ago \citep{Salo2000, MentuchCooper2012,Eufrasio2017}. The condition that the SFR averaged over age intervals with a width of \tstar\ or \tgas\ should not vary by more than 0.2\,dex as function of age for $t \leq \tau$ is therefore highly likely to be satisfied on galactic scales. This ensures that any bias of the measured \tgas\ due to possible SFR variations is less than 50~per~cent. Note that this condition is likely not fulfilled for at least some of the bins covering bars or the tips of bars. The stochasticity and synchronised, bursty nature of the star formation events in these regions therefore leads to large uncertainties or biases on the cloud lifetime. In these regions, the measured \tgas\ becomes dependent on the precise moment of observation (see Section~\ref{sec:morphology}).
\end{enumerate}

The fulfillment of the above criteria guarantees the accuracy of the constraints obtained for \tgas\ and $\lambda$. Additional requirements apply to ensure the accuracy of \tover. While we have verified that these are satisfied, we defer a detailed discussion to the companion paper focusing on the feedback time-scale (M.~Chevance et al.\ in prep.).

\subsection{Comparison with previous work}
\label{sec:literature}

Previous studies of individual galaxies have led to a variety of measured molecular cloud lifetimes, using different techniques to infer these. Cloud lifetimes with similarly short values as in the present study ($10{-}30~\myr$) have been measured by counting clouds or classifying clouds based on their stellar content  \citep[e.g.][]{Elmegreen2000, Hartmann2001, Engargiola2003, Kawamura2009, Meidt2015, Corbelli2017}. Similar values have been obtained by using the spiral arm pattern speed and local circular velocity to convert the offsets between \HII\ regions and molecular clouds into evolutionary time-scales \citep{Egusa2009}. By contrast, much longer values of over 100\,Myr have been suggested by the presence of molecular clouds in the inter-arm regions of nearby spiral galaxies \citep{Scoville1979, Scoville2004, Koda2009}, while a much shorter value of $\sim$ 1\,Myr has been measured in the Central Molecular Zone (i.e.\ the central $\sim$ 500\,pc of the Milky Way), by following clouds along a known gas orbit \citep{Kruijssen2015, Henshaw2016, Barnes2017, Jeffreson2018b}.

These previous studies have made major progress in tackling the fundamental problem of measuring the evolutionary timeline of cloud-scale star formation and feedback. At the same time, they have faced several immediate challenges. First, by their pioneering nature, they targeted single galaxies. Taken together, these studies therefore lack the homogeneity of definitions needed to make direct comparisons between galaxies and determine whether the differences between galaxies are physical in nature or result from differences in experiment design. Related to this, previous works used different weighing schemes for quantifying the average cloud lifecycle, e.g.\ using a number-weighted average or a flux-weighted average (as is done in this paper). Secondly, several of these studies rely on defining and classifying GMCs and \HII regions, which is necessarily subjective. Reliable GMC classifications have often required resolving individual clouds or star-forming regions (of a few tens of parsec), which has so far obstructed systematic measurements of the molecular cloud lifetime outside of the Local Group. While the methodology used in this paper does not differ fundamentally in terms of its broader philosophy, a key change is that it has tried to eliminate the subjectivity of GMC classification and therefore does not require to resolve individual regions, increasing its reproducibility and applicability. Finally, some of these studies primarily investigated the effects of a single dynamical mechanism (e.g.\ spiral arm perturbations) without considering the variety of possible processes affecting clouds.

By using the unified framework provided by \citet{Kruijssen2018}, it is now possible to build on the broad foundation laid by previous studies and probe the variation of the molecular cloud lifetime as a function of environment, both between and within galaxies. In this paper, we present homogeneous measurements of the molecular cloud lifetime for nine different galaxies, finding that they are short, with values between $10{-}30~\myr$ for galaxy-averaged cloud lifetimes and typically a factor of $\gtrsim 2$ variation within galaxies (as presented in Figure~\ref{fig:radprof}). This is consistent with the lifetimes of molecular clouds measured in NGC300 \citep[$10.8^{+2.1}_{-1.7}~\myr$,][]{Kruijssen2019} and M33 \citep[$16.7\pm2.1~\myr$,][]{Hygate2019b} using the same statistical formalism. We do not find any dependence of the measured evolutionary timelines on the strength or the number of spiral arms in the galaxies of our sample. This suggests that, while spiral arms may instigate molecular cloud formation, the subsequent evolution of the clouds is likely governed by the processes identified in this work (i.e.\ dynamics and stellar feedback) irrespectively of the presence of spiral arms. As a result, we suggest that the offsets between molecular clouds and \HII\ regions perpendicular to spiral arms that have been used to infer evolutionary time-scales \citep[e.g.][]{Egusa2009,Meidt2015} are driven primarily by cloud evolution and feedback rather than by dynamical drift alone.

\begin{figure}
\includegraphics[trim=5mm 2mm 5mm 5mm, clip=true, width=\columnwidth]{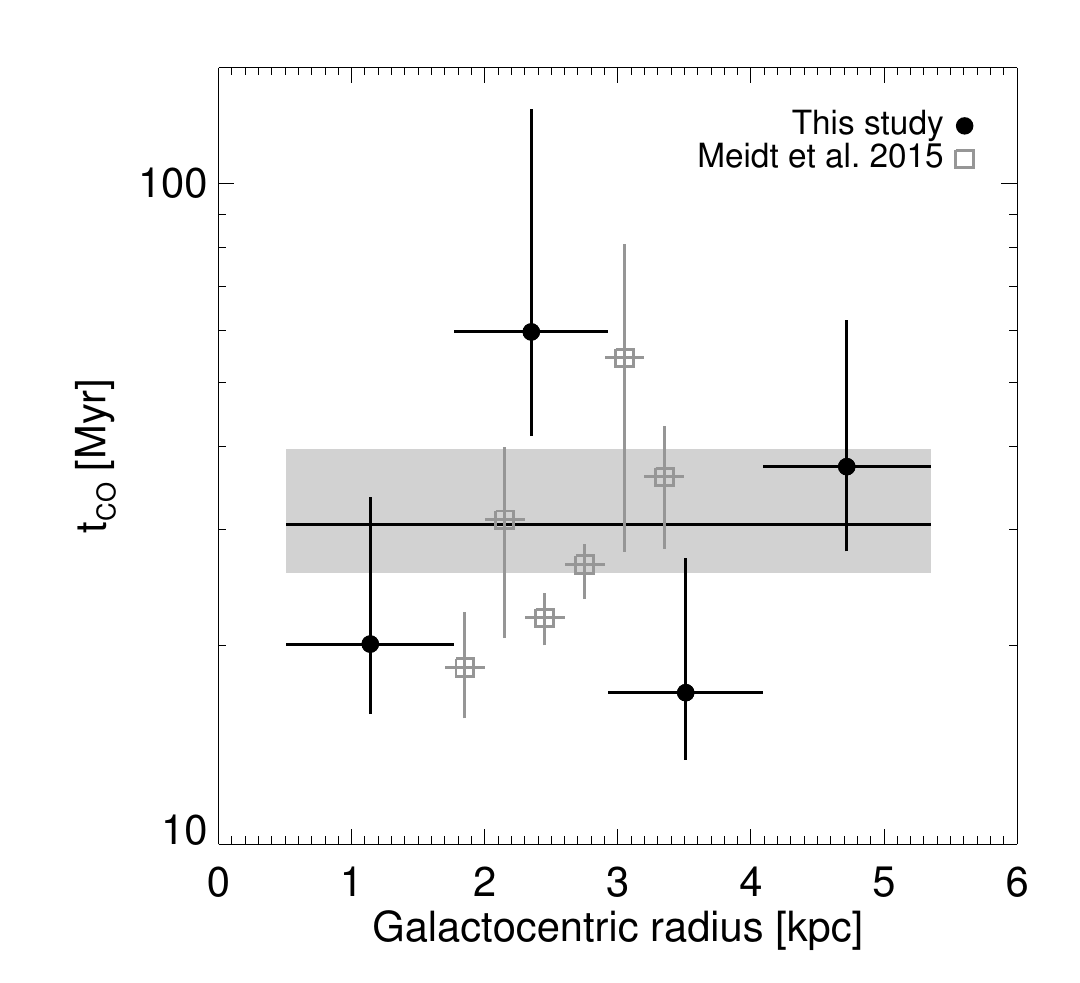}
\caption{Comparison of our GMC lifetime measurements in NGC5194 (black) to those from \citet{Meidt2015} (grey) as a function of galactocentric radius. The horizontal black line represents the average GMC lifetime across the entire galaxy, with the uncertainties indicated by the grey-shaded area.}
\label{fig:Meidt15}
\end{figure}

In Figure~\ref{fig:Meidt15}, we compare the GMC lifetimes measured in this study with the cloud lifetimes inferred by \citet{Meidt2015} in NGC5194 (M51). \citet{Meidt2015} use the variation of the GMC number density as a function of the azimuthal coordinates to estimate the cloud lifetimes as a fraction of the inter-arm travel time. We observe a broad agreement between the range of lifetimes obtained by both methods, although the exact values differ from bin to bin. The discrepancy is the largest in the second radial bin ($1.8{-}2.9~\kpc$) and may exist for two different reasons. First, the galactocentric bins considered by \citet{Meidt2015} are relatively small (width of 0.3\,kpc) and therefore include a small number of clouds (between 4 and 23). This affects how well the different phases of the evolutionary cycle from clouds to young stellar regions are sampled in a given bin, making the results sensitive to stochasticity. Secondly, related to the previous point, the method assumes a constant rate of change in the cloud population with time for each individual bin. While this type of statistical equilibrium may apply across the full cloud population, it is less likely to apply to smaller sub-populations, either due to stochasticity as in the previous point, or due to systematic changes in the local conditions. When using statistical inference to measure the cloud lifetime, these two reasons imply that the dispersion within a bin (and therefore the uncertainties on the measurement) scale directly with the size of the (sub-)population under consideration. This plausibly explains why the cloud lifetimes and their uncertainties differ between both methods.

\begin{figure}
\includegraphics[trim=5mm 2mm 5mm 5mm, clip=true, width=\columnwidth]{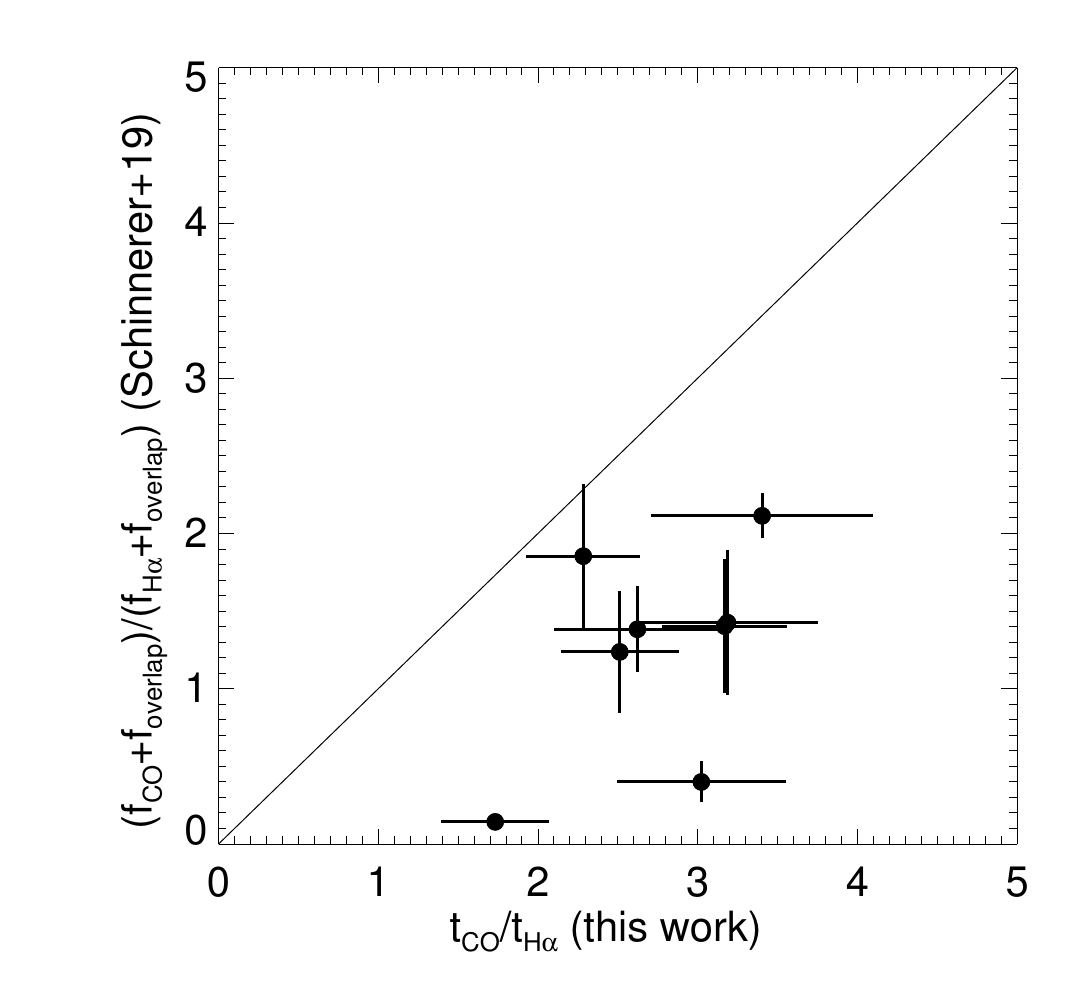}
\caption{Comparison of our measured GMC lifetimes in units of the \Ha\ emission time-scale ($x$-axis) to the estimates based on pixel statistics from table~5 of \citet{Schinnerer2019} ($y$-axis) for the eight full galaxies in common between both studies. The diagonal line shows the 1:1 relation.}
\label{fig:tCOcomp}
\end{figure}

Finally, \citet{Schinnerer2019} present estimates for the duration of the CO-bright phase for eight of the galaxies considered here (all except NGC4303) using pixel statistics. Specifically, that work presents the ratio between the number of CO-bright pixels and the number of \Ha-bright pixels (above a chosen flux density threshold) and discusses how this could be interpreted as the ratio between the visibility time-scales of both tracers. The relative simplicity of pixel statistics has the great advantage that it is highly reproducible, but it also means that it may not be straightforward to translate them directly into time-scales. For this reason, \citet{Schinnerer2019} highlight several of the caveats associated with this temporal interpretation. Figure~\ref{fig:tCOcomp} quantitatively tests this hypothesis by comparing their measurements to the time-scale ratios measured here. The figure shows that the pixel-based approach is in order-of-magnitude agreement with our measurements, but systematically underestimates the GMC lifetime by a factor of $\sim2$ on average. The order-of-magnitude agreement is encouraging, even if the systematic bias and the presence of two strong outliers (NGC3351 and NGC5068) at the bottom of the diagram caution against using pixel statistics as a quantitative tracer of the cloud lifetime.

The difference between our GMC lifetimes and the fractions of CO-bright pixels can be understood as the result of differences in methodology. First, there is a difference in how either approach deals with blending between regions. The analysis of this paper self-consistently accounts for blending effects towards coarser resolutions, and uses these effects to measure the separation scales of the units undergoing evolutionary cycling. Conversely, the number ratio between CO-bright and \Ha-bright pixels by definition approaches unity towards coarser resolutions, implying $t_{\rm CO}/t_{\rm H\alpha}\sim1$ without a physical similarity between the underlying time-scales. Because $t_{\rm H\alpha}<t_{\rm CO}$ for all galaxies, this leads to a systematic bias. Secondly, using the number of pixels bright in either tracer as a proxy for the lifetimes of regions assumes that all regions have the same size (or area). This particularly affects NGC3351 and NGC5068, where the visible extent of the CO clouds is typically smaller than that of the \HII regions. This leads to an underestimation of the time-scales inferred from pixel statistics. The analysis presented here avoids this by accounting for differences in region size between both tracer maps (see Section~\ref{sec:method} and \citealt{Kruijssen2018}). In summary, the quantitative differences between the results of both approaches are a natural result of differences in methodology. This comparison demonstrates that pixel statistics are a good qualitative probe of the GMC lifecycle, but do not perform as well when used as a quantitative metric of the GMC lifetime. Together, the pixel statistics presented in \citet{Schinnerer2019} and the results presented here constitute critical and complementary empirical observables that simulations of galactic-scale star formation will need to reproduce \citep[see e.g.][]{Fujimoto2019}.

\subsection{Implications for the GMC lifecycle in galaxies}

The results presented in Section~\ref{sec:results} reveal that the de-correlation between molecular gas and young stellar regions at $\sim$ 100\,pc scales is ubiquitous across our galaxy sample. The fact that GMCs and \HII\ regions rarely coexist on these small scales indicates the rapid evolutionary cycling between GMCs, star formation and feedback. This has previously been shown in two very nearby ($D<2~\mpc$) galaxies using the same method (e.g.\ NGC300, \citealt{Kruijssen2019}; M33, \citealt{Hygate2019b}) and we can now generalise this result to a much larger sample of galaxies.

The significant variation of the molecular cloud lifetime measured homogeneously across a sample of nine galaxies demonstrates that the cycling between gas and stars is not quantitatively universal, but exhibits a clear environmental dependence. In environments with high kpc-scale molecular gas surface densities ($\Sigma_{\rm H_2,ring}>8$\,\Msun~pc$^{-2}$), our measurements correlate most strongly with the predicted time-scales based on galactic dynamical processes from \citet{Jeffreson2018}. This shows the importance of galactic dynamics in setting the cloud lifetime, and hence its role in regulating the star formation process. In most cases, this predicted dynamical timescale $\tau_{\rm gal}$ is dominated by the time-scale for the gravitational free-fall of the mid-plane ISM ($t_{\rm ff,g}$) and for dispersal by shear ($t_{\beta}$). In environments with low kpc-scale molecular gas surface densities ($\Sigma_{\rm H_2,ring}<8$\,\Msun~pc$^{-2}$), GMCs become decoupled from the large-scale galactic dynamics and the molecular cloud lifetime is consistent with being set by the cloud's internal dynamical time (\tff\ or \tcr).

The short duration of the feedback phase measured in Section~\ref{sec:others} ($t_{\rm fb}=1{-}5~\myr$) lends further support to a highly dynamic view of the ISM. For several galaxies, the feedback time-scale is shorter than the typical minimum time of 4\,Myr for supernovae to explode \citep[e.g.][]{Leitherer14}, indicating that early (stellar) feedback mechanisms are responsible for dispersing the parent molecular cloud within a short time-scale. This means that photoionisation and stellar winds are likely to play an essential role in the rapid destruction of the molecular cloud after the onset of massive star formation. Without a quantitative comparison to theoretical predictions, it is not possible determine whether the parent GMC is destroyed by a phase transition or by kinetic dispersal, i.e.\ whether the remaining molecular gas is photodissociated or merely separated from the young stellar population, potentially broken up in several smaller diffuse clouds. We are currently undertaking such an analysis for the galaxy sample presented here (M.~Chevance et al.\ in prep.), where this question will be addressed in more detail.

\section{Conclusion}
\label{sec:ccl}

We present a systematic measurement of the characteristic time-scales describing the lifecycle of molecular clouds, star formation, and feedback, for a sample of nine nearby star-forming disc galaxies, using cloud-scale ($\sim100~\pc$) resolution imaging of CO and \Ha, obtained as part of the PHANGS collaboration. We employ the multi-scale, multi-wavelength statistical method presented in \citet{KL14} and \citet{Kruijssen2018} to measure the molecular cloud lifetime and the feedback time-scale, which are critical for constraining the physical processes regulating star formation at the cloud scale. These quantities could previously be obtained only for a handful of single galaxies, mostly restricted to the Local Group, and the heterogeneity of methods used did not enable direct comparisons between different studies. As a result, it was unclear if the variety of cloud lifetimes in the literature (ranging between $1~\myr$ and $>100~\myr$) is caused by differences in experiment design or reflects a variety of physical conditions and processes. By applying a rigorous, statistical analysis method homogeneously to a sample of nine galaxies, we are now able to determine the quantities describing the cloud lifecycle systematically across a wide range of galactic environments.

Across our sample of nine star-forming disc galaxies, our analysis method reveals a universal de-correlation of CO and \Ha emission on the cloud scale ($\sim100~\pc$), indicating a rapid evolutionary lifecycle in which star formation is fast and inefficient: molecular clouds live for a (cloud-scale or galactic, see below) dynamical time, form stars, and get disrupted by feedback. Our results show that star-forming disc galaxies can be described as ensembles of independent building blocks, separated by $\lambda=100{-}300~\pc$, undergoing a rapid evolutionary cycle from molecular clouds to young stellar regions. We measure relatively short molecular cloud lifetimes of $t_{\rm CO}=10{-}30~\myr$, with statistically significant variations, both between and within galaxies. The fact that these cloud lifetimes are much shorter than the molecular gas depletion time ($\sim2~\gyr$) implies that the integrated star formation efficiency per star formation event is low; we obtain values in the range of $\epsilon_{\rm sf}=4{-}10$~per~cent. 

Molecular clouds experience a long `inert' or `isolated' phase, taking $75{-}90$~per~cent of their total lifetime, during which they show no signs of massive star formation. When massive stars do emerge, towards the end of the cloud lifecycle, the parent cloud is dispersed within $t_{\rm fb}=1{-}5~\myr$, strongly suggesting that cloud dispersal is driven by stellar feedback. The short duration of this `feedback time-scale', which represents the time between the emergence of the first ionising photons due to massive star formation and the eventual destruction or dispersal of the parent molecular cloud, indicates that early (stellar) feedback such as photoionisation or stellar winds plays a major role in this process, acting before the first supernovae explode. We will present a detailed investigation of the relative importance of different feedback mechanisms (supernovae, photoionisation, stellar winds, and radiation pressure) in GMC dispersal in a companion paper, by comparing the feedback time-scales measured here to theoretical predictions (M.~Chevance et al.\ in prep.).

The above quantities are consistent with the results obtained by applying this method to NGC300 \citep{Kruijssen2019} and M33 \citep{Hygate2019b}, but we extend these to a more representative sample of star-forming main sequence galaxies. In addition, by using a single analysis method to measure the molecular cloud lifetime across a sample of galaxies, we are now able to demonstrate how it varies with the galactic environment, both between galaxies and within them. We distinguish two regimes, in which the GMC lifetime is set by different physical mechanisms. In environments with high kpc-scale molecular gas surface densities ($\Sigma_{\rm H_2,ring}\geq8$\,\Msun pc$^{-2}$), the cloud lifetime is regulated by galactic dynamics, mostly by a combination of the gravitational free-fall of the mid-plane ISM and shear. Spiral arm crossings and cloud-cloud collisions take place on considerably longer ($\sim100~\myr$) time-scales and are too rare to systematically drive cloud evolution across the cloud population. In environments with low kpc-scale molecular gas surface densities ($\Sigma_{\rm H_2,ring}\leq8$\,\Msun pc$^{-2}$), GMCs decouple from the dynamics of the host galaxy, with CO-devoid regions separating them from other GMCs, and the cloud lifetime correlates with the cloud crossing and free-fall times, showing that cloud evolution is regulated by internal dynamics. The division between these two regimes in galactic molecular gas surface density coincides with the atomic-to-molecular gas transition occurring near the above density limit \citep{Wong2002,Leroy2008,Krumholz2014,Schruba2018}. In addition to these general trends, we find that GMC lifetimes can be elevated near the co-rotation radius.

The quantitative variation of the evolutionary timeline describing the cloud lifecycle reveals that the processes that regulate cloud-scale star formation and feedback in galaxies are environmentally dependent. Therefore, to determine the relevant environmental quantities (e.g.\ galactic dynamics, disc structure, ISM pressure) affecting the cycle of cloud evolution, star formation, and feedback, it is necessary to extend the analysis performed in this work to a larger number of galaxies, covering a broad range of environments and morphology. The systematic application of this method to a large fraction of all massive star-forming disc galaxies within 17\,Mpc will soon be possible with the on-going PHANGS-ALMA Large Programme and will be presented in J.~Kim et al.\ (in prep.). This will allow us to quantitatively assess how the efficiency and lifecycle of star-formation and feedback depends on the galactic environment. We expect that this work will contribute to characterising the multi-scale physics driving these lifecycles and move away from a quasi-static picture of star formation in galaxies, instead describing it in terms of the mass flows generated by cloud-scale accretion and stellar feedback. This will represent key empirical input for a predictive theory of how galaxies grow and form stars, as well as for sub-grid models for star formation and feedback in galaxy simulations.

\section*{Acknowledgements}
We thank an anonymous referee for a helpful report, as well as Bruce Elmegreen, Mark Heyer, Benjamin Keller, Jenny (Jaeyeon) Kim, Jeong-Gyu Kim, Eve Ostriker, Mark Krumholz, Jacob Ward, and Brad Whitmore for helpful discussions and/or comments. MC and JMDK gratefully acknowledge funding from the Deutsche Forschungsgemeinschaft (DFG) through an Emmy Noether Research Group (grant number KR4801/1-1) and the DFG Sachbeihilfe (grant number KR4801/2-1). JMDK, APSH, SMRJ, and DTH gratefully acknowledge funding from the European Research Council (ERC) under the European Union's Horizon 2020 research and innovation programme via the ERC Starting Grant MUSTANG (grant agreement number 714907). MC, JMDK, SMRJ, and DTH acknowledge support from the Australia-Germany Joint Research Cooperation Scheme (UA-DAAD, grant number 57387355). APSH, SMRJ, and DTH are fellows of the International Max Planck Research School for Astronomy and Cosmic Physics at the University of Heidelberg (IMPRS-HD). 
BG gratefully acknowledges the support of the Australian Research Council as the recipient of a Future Fellowship (FT140101202).
CNC, AH, and JP acknowledge funding from the Programme National ``Physique et Chimie du Milieu Interstellaire'' (PCMI) of CNRS/INSU with INC/INP, co-funded by CEA and CNES.
AH acknowledges support by the Programme National Cosmology et Galaxies (PNCG) of CNRS/INSU with INP and IN2P3, co-funded by CEA and CNES.
PL, ES, CF, DL and TS acknowledge funding from the ERC under the European Union's Horizon 2020 research and innovation programme (grant agreement No. 694343).
The work of AKL, JS, and DU is partially supported by the National Science Foundation (NSF) under Grants No.~1615105, 1615109,and 1653300. AKL also acknowledges partial support from NASA ADAP grants NNX16AF48G and NNX17AF39G.
ER acknowledges the support of the Natural Sciences and Engineering Research Council of Canada (NSERC), funding reference number RGPIN-2017-03987.
FB acknowledges funding from the ERC under the European Union's Horizon 2020 research and innovation programme (grant agreement No. 726384).
GB is supported by CONICYT/FONDECYT, Programa de Iniciaci\'{o}n, Folio 11150220.
SCOG acknowledges support from the DFG via SFB 881 ``The Milky Way System'' (subprojects B1, B2 and B8) and also via Germany's Excellence Strategy EXC-2181/1--390900948 (the Heidelberg STRUCTURES Excellence Cluster).
K.K.\ gratefully acknowledges funding from the German Research Foundation (DFG) in the form of an Emmy Noether Research Group (grant number KR4598/2-1, PI Kreckel).

This work was carried out as part of the PHANGS collaboration. This paper makes use of the following ALMA data: ADS/JAO.ALMA \#2012.1.00650.S, ADS/JAO.ALMA \#2015.1.00925.S, ADS/JAO.ALMA \#2015.1.00956.S. ALMA is a partnership of ESO (representing its member states), NSF (USA) and NINS (Japan), together with NRC (Canada), NSC and ASIAA (Taiwan), and KASI (Republic of Korea), in cooperation with the Republic of Chile. The Joint ALMA Observatory is operated by ESO, AUI/NRAO and NAOJ. The National Radio Astronomy Observatory is a facility of the National Science Foundation operated under cooperative agreement by Associated Universities, Inc. 
This paper makes use of the PdBI Arcsecond Whirlpool Survey \citep{Schinnerer2013,Pety2013}. The IRAM 30-m telescope and PdBI are run by IRAM, which is supported by INSU/CNRS (France), MPG (Germany) and IGN (Spain).
The results presented in this paper made use of THINGS, `The HI Nearby Galaxy Survey' \citep{Walter2008}.
This work has made use of data from the European Space Agency (ESA) mission
{\it Gaia} (\url{https://www.cosmos.esa.int/gaia}), processed by the {\it Gaia}
Data Processing and Analysis Consortium (DPAC,
\url{https://www.cosmos.esa.int/web/gaia/dpac/consortium}). Funding for the DPAC
has been provided by national institutions, in particular the institutions
participating in the {\it Gaia} Multilateral Agreement.

%%%%%%%%%%%%%%%%%%%% REFERENCES %%%%%%%%%%%%%%%%%%

\bibliographystyle{mnras}
\bibliography{timescales}

%%%%%%%%%%%%%%%%% APPENDICES %%%%%%%%%%%%%%%%%%%%%

\appendix

\section{Fractions of compact and diffuse emission}
\label{sec:app_diffuse}

In order to obtain robust results from our statistical analysis with the \textsc{Heisenberg} code, we need to remove the biasing impact of diffuse emission. We do this by separating the compact \HII\ regions and GMCs from the large-scale diffuse emission in both tracers. The method for doing this has been presented, tested, and validated by \cite{Hygate2019} and a filtering procedure based on filtering in Fourier space has been implemented in \textsc{Heisenberg}. As recommended in \cite{Hygate2019}, we use a Gaussian filter of FWHM $\sim 10 \times \lambda$ to mask the low spatial frequencies in Fourier space (i.e.\ large-scale emission) and filter out the diffuse emission from the compact regions of interest. The Gaussian shape of the filter used limits artefacts compared to a more selective step function. However, contrary to a step function, a Gaussian function extends to infinitely high spatial frequencies, implying that some compact emission is spuriously filtered out. To compensate for this effect, we apply two correction factors defined by \cite{Hygate2019} to the measured fraction of compact emission. The first correction factor, $q_{\rm con}$, compensates for flux loss from the individual compact regions. The second correction factor, $q_{\rm overlap}$, compensates for the flux loss due to overlap between regions. Prescriptions for $q_{\rm con}$ and $q_{\rm overlap}$ are calibrated in \citet{Hygate2019} and \citet{Hygate2019b}, respectively. We then determine the fraction of emission that belongs to compact structures in the \Ha\ map ($f_{\rm H\alpha}$) and the CO map ($f_{\rm CO}$) as: 
\begin{equation}
f_{\rm H\alpha} = \frac{1}{q_{\rm con,H\alpha} q_{\rm overlap,H\alpha}} \frac{F^{\prime}_{\rm H\alpha}}{F_{\rm H\alpha}}
\end{equation}
and 
\begin{equation}
f_{\rm CO} = \frac{1}{q_{\rm con,CO} q_{\rm overlap,CO}} \frac{F^{\prime}_{\rm CO}}{F_{\rm CO}},
\end{equation}
where $F_{\rm H\alpha}$ (respectively $F_{\rm CO}$) is the total flux in the original \Ha\ (respectively CO) map and $F^{\prime}_{\rm H\alpha}$ (respectively $F^{\prime}_{\rm CO}$) is the total flux in the filtered \Ha\ (respectively CO) map. After applying these corrections, we obtain the fractions of compact flux as listed in Table~\ref{tab:diffuse}. The diffuse emission fractions follow as the complement of the compact emission fractions, i.e.\ as $1-f_{\rm H\alpha}$ and $1-f_{\rm CO}$.

As recommended by \citet{Hygate2019}, we ensure that $q_{\rm con}\geq0.9$, so that the correction to be applied to the compact fraction is relatively small. For each galaxy, this is done by setting the FWHM of the Gaussian filter to the smallest multiple of $\lambda$ at which this condition is satisfied. This results in cut-off wavelengths for the Gaussian filters in the range $10{-}12\times\lambda$, as listed in Table~\ref{tab:input}. These values ensure an optimum between maximising the filtering of the diffuse emission, and minimising the spurious filtering of the compact structures. 

For comparison, we note that \cite{Pety2013} find that $50\pm10$~per~cent of the CO(1-0) emission in NGC5194 is distributed on scales larger than 1.3\,kpc, which is close to the size of the Gaussian filter used in our analysis for this galaxy ($\sim$ 1.7\,kpc). This estimate is obtained by comparing the amount of flux recovered by the Plateau de Bure interferometer to the total flux measured by the IRAM-30~m single-dish telescope. This is roughly equivalent to filtering the emission on large scales in Fourier space, without applying the correction factors mentioned above. Before taking into account the correction factors $q_{\rm con} = 0.90_{-0.01}^{+0.01}$ and $q_{\rm overlap} = 0.45_{-0.04}^{+0.06}$, we measure a fraction of diffuse emission of $56\pm2$~per~cent in NGC5194, in agreement with the above estimate by \cite{Pety2013}. \citet{CalduPrimo2015} found (lower) diffuse fractions in the range $8{-}48$~per~cent for two other galaxies, which is consistent with our results listed in Table~\ref{tab:diffuse}. After including the correction factors, we obtain a true diffuse fraction for NGC5194 in a range (representing the $1\sigma$ uncertainty interval) of $0{-}9$~per~cent.

\begin{table*}
\begin{center}
{\def\arraystretch{1.5}
\begin{tabular}{lcccccc}
\hline
                    Galaxy  &                   $f_{\rm H\alpha}$  &                        $f_{\rm CO}$  &               $q_{\rm con,H\alpha}$  &                    $q_{\rm con,CO}$  &              $q_{\rm overlap,H\alpha}$  &                   $q_{\rm overlap,CO}$  \\
                                &                                      &                                      &                                      &                                      &                                      &                                      \\
\hline
                            NGC0628   & $     0.69_{-   0.01}^{+   0.01} $   & $     0.82_{-   0.06}^{+   0.06} $   & $     0.89_{-   0.01}^{+   0.02} $   & $     0.89_{-   0.01}^{+   0.02} $  & $     0.74_{-   0.01}^{+   0.02} $   & $     0.52_{-   0.05}^{+   0.06} $  \\
                            NGC3351   & $     0.31_{-   0.01}^{+   0.01} $   & $     0.83_{-   0.06}^{+   0.06} $   & $     0.89_{-   0.01}^{+   0.01} $   & $     0.90_{-   0.01}^{+   0.01} $   & $     0.73_{-   0.02}^{+   0.02} $   & $     0.56_{-   0.04}^{+   0.05} $  \\
                            NGC3627   & $     0.42_{-   0.01}^{+   0.01} $   & $     0.81_{-   0.06}^{+   0.04} $   & $     0.91_{-   0.01}^{+   0.02} $   & $     0.91_{-   0.01}^{+   0.01} $  & $     0.70_{-   0.02}^{+   0.03} $  & $     0.49_{-   0.04}^{+   0.06} $  \\
                            NGC4254   & $     0.69_{-   0.02}^{+   0.01} $   & $     0.74_{-   0.06}^{+   0.06} $   & $     0.91_{-   0.01}^{+   0.01} $   & $     0.91_{-   0.01}^{+   0.02} $  & $     0.67_{-   0.02}^{+   0.03} $   & $     0.48_{-   0.05}^{+   0.06} $  \\
                            NGC4303   & $     0.68_{-   0.03}^{+   0.01} $   & $     0.82_{-   0.10}^{+   0.08} $   & $     0.89_{-   0.01}^{+   0.02} $   & $     0.90_{-   0.02}^{+   0.02} $    & $     0.62_{-   0.02}^{+   0.05} $   & $     0.46_{-   0.06}^{+   0.09} $  \\
                            NGC4321   & $     0.44_{-   0.01}^{+   0.01} $   & $     0.79_{-   0.05}^{+   0.05} $   & $     0.91_{-   0.01}^{+   0.01} $   & $     0.91_{-   0.01}^{+   0.01} $    & $     0.69_{-   0.02}^{+   0.02} $   & $     0.49_{-   0.04}^{+   0.04} $  \\
                            NGC4535   & $     0.93_{-   0.03}^{+   0.02} $   & $     0.98_{-   0.11}^{+   0.09} $   & $     0.89_{-   0.01}^{+   0.02} $   & $     0.90_{-   0.02}^{+   0.02} $   & $     0.68_{-   0.02}^{+   0.04} $   & $     0.44_{-   0.06}^{+   0.08} $  \\
                            NGC5068   & $     0.64_{-   0.01}^{+   0.01} $   & $     1.35_{-   0.04}^{+   0.04} $   & $     0.92_{-   0.01}^{+   0.01} $   & $     0.95_{-   0.01}^{+   0.01} $    & $     0.74_{-   0.03}^{+   0.04} $    & $     0.76_{-   0.02}^{+   0.03} $  \\
                            NGC5194   & $     0.37_{-   0.01}^{+   0.01} $   & $     1.05_{-   0.09}^{+   0.06} $   & $     0.90_{-   0.01}^{+   0.01} $   & $     0.91_{-   0.01}^{+   0.01} $   & $     0.71_{-   0.01}^{+   0.02} $    & $     0.45_{-   0.04}^{+   0.06} $  \\
\hline
\end{tabular}
}
\caption{Fractions of emission in the \Ha\ and the CO maps that belong to compact structures ($f_{\rm H\alpha}$ and $f_{\rm CO}$, respectively) for each of the nine galaxies in our sample. The diffuse emission fractions follow as $1-f_{\rm H\alpha}$ and $1-f_{\rm CO}$. We also list the associated correction factors. These are $q_{\rm con,H\alpha}$ and $q_{\rm con,CO}$, applied to correct for any over-subtraction of diffuse emission caused by using a Gaussian filter in Fourier space, and $q_{\rm overlap,H\alpha}$ and $q_{\rm overlap,CO}$, applied to correct for any over-subtraction of diffuse emission caused by overlap between regions. See Section~\ref{sec:diffuse} and \citet{Hygate2019b, Hygate2019} for more details.}
\label{tab:diffuse}
\end{center}
\end{table*}

\section{Radial profiles} \label{sec:appradial}

\begin{figure*}
  \begin{minipage}{0.3\linewidth}
    \includegraphics[trim=12mm 2mm 18mm 90mm, clip=true, width=5.5cm]{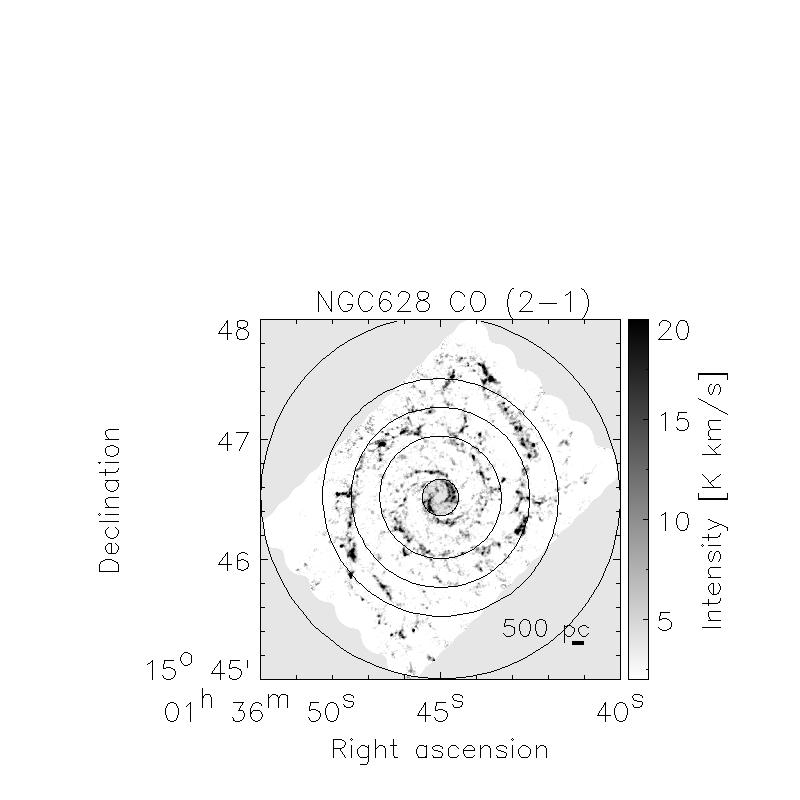}
    \includegraphics[trim=12mm 2mm 18mm 80mm, clip=true, width=5.5cm]{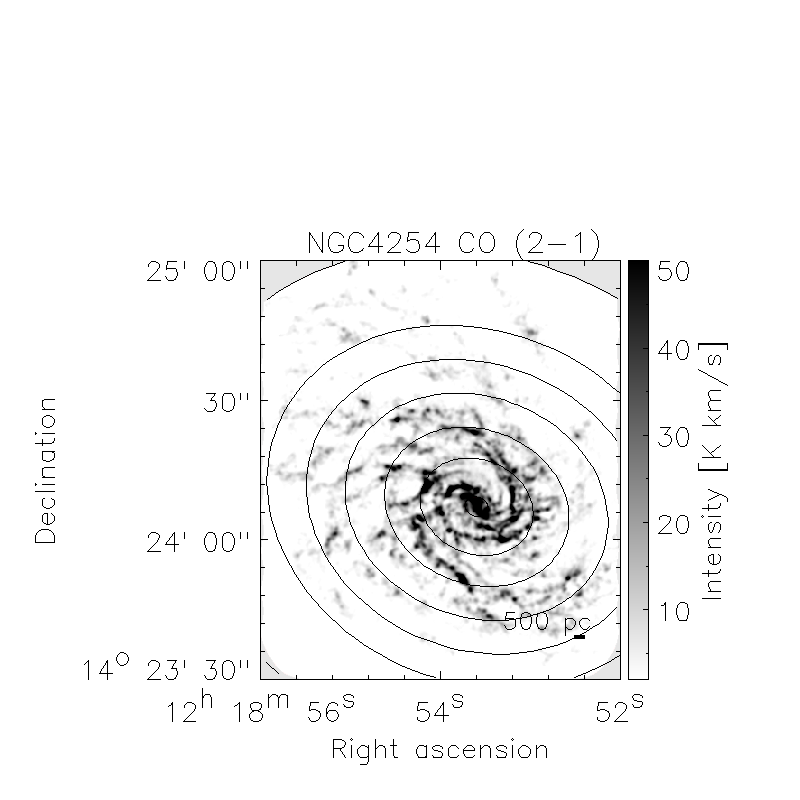}
    \includegraphics[trim=12mm 2mm 18mm 55mm, clip=true, width=5.5cm]{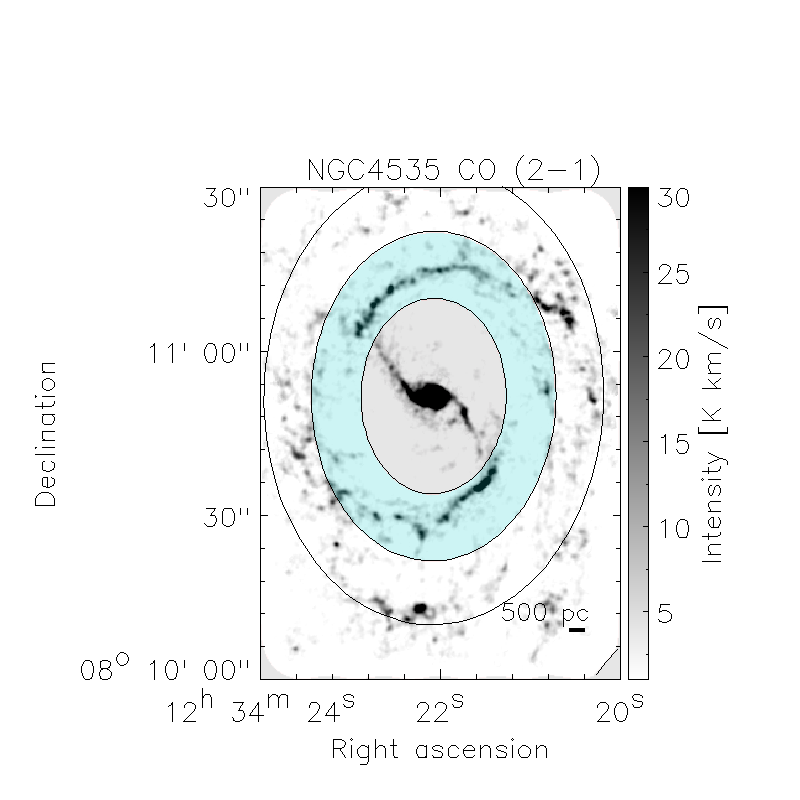}
  \end{minipage}
  \begin{minipage}{0.3\linewidth}
  \vspace{-10mm}
\includegraphics[trim=12mm 2mm 18mm 97mm, clip=true, width=5.5cm]{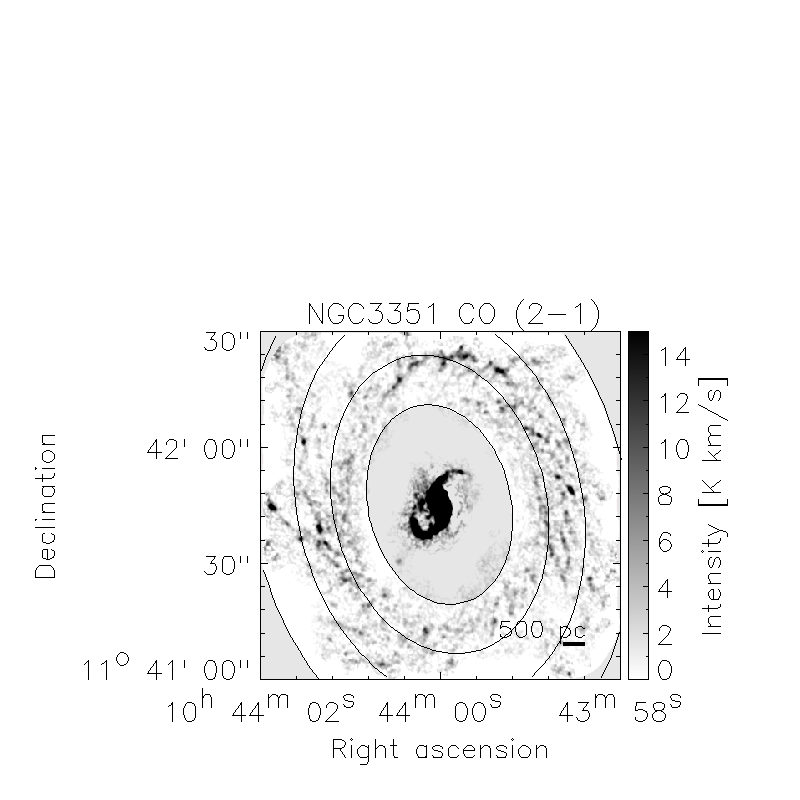}
\includegraphics[trim=12mm 2mm 18mm 100mm, clip=true, width=5.5cm]{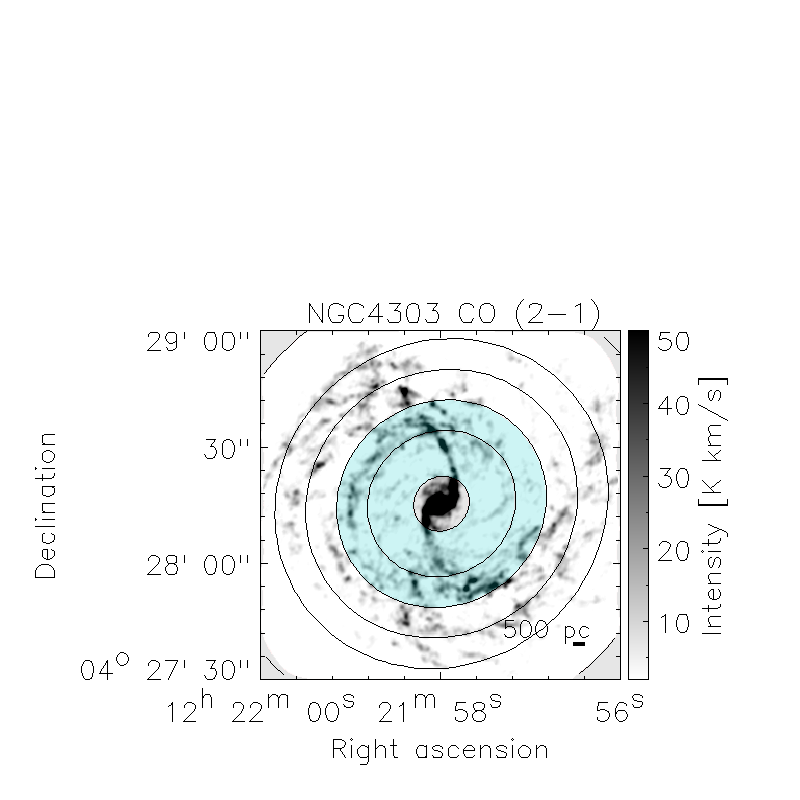}
\includegraphics[trim=12mm 2mm 18mm 70mm, clip=true, width=5.5cm]{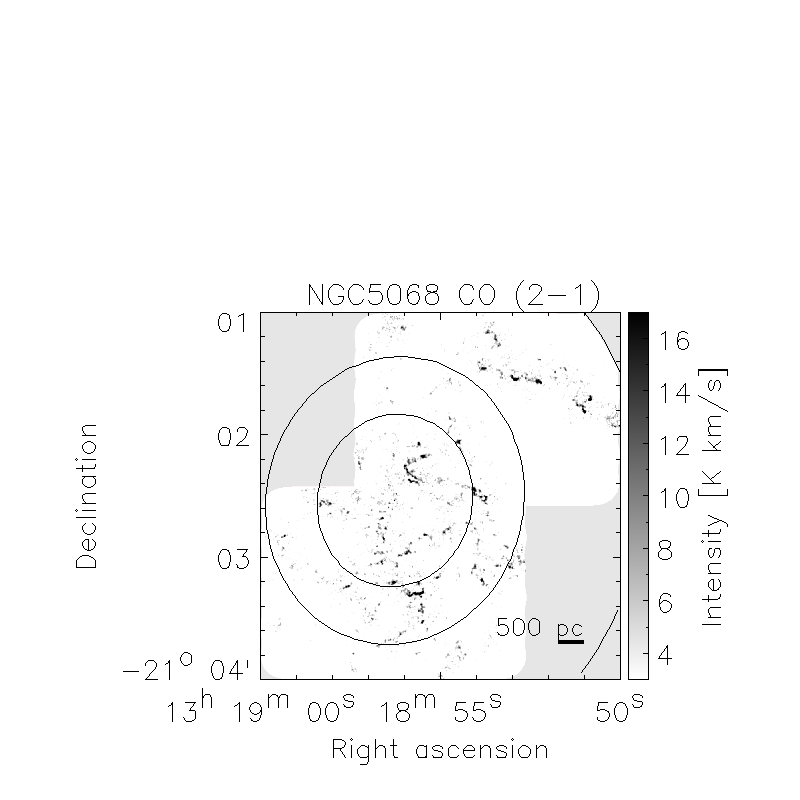}
  \end{minipage}
  \begin{minipage}{0.3\linewidth}
  \vspace{-4mm}
\includegraphics[trim=12mm 2mm 18mm 10mm, clip=true, width=5.5cm]{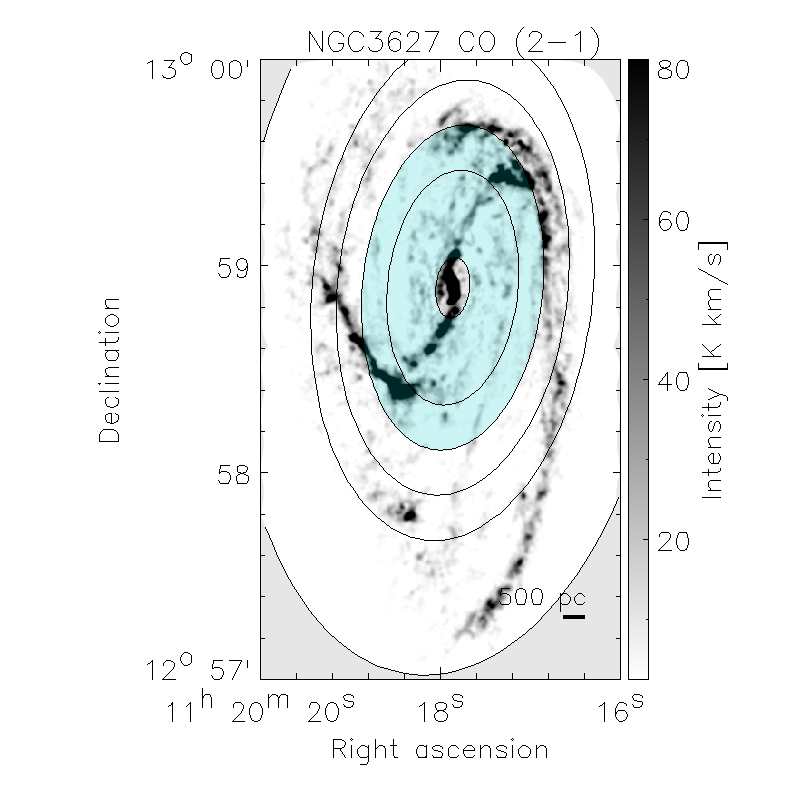}
\includegraphics[trim=12mm 2mm 18mm 115mm, clip=true, width=5.5cm]{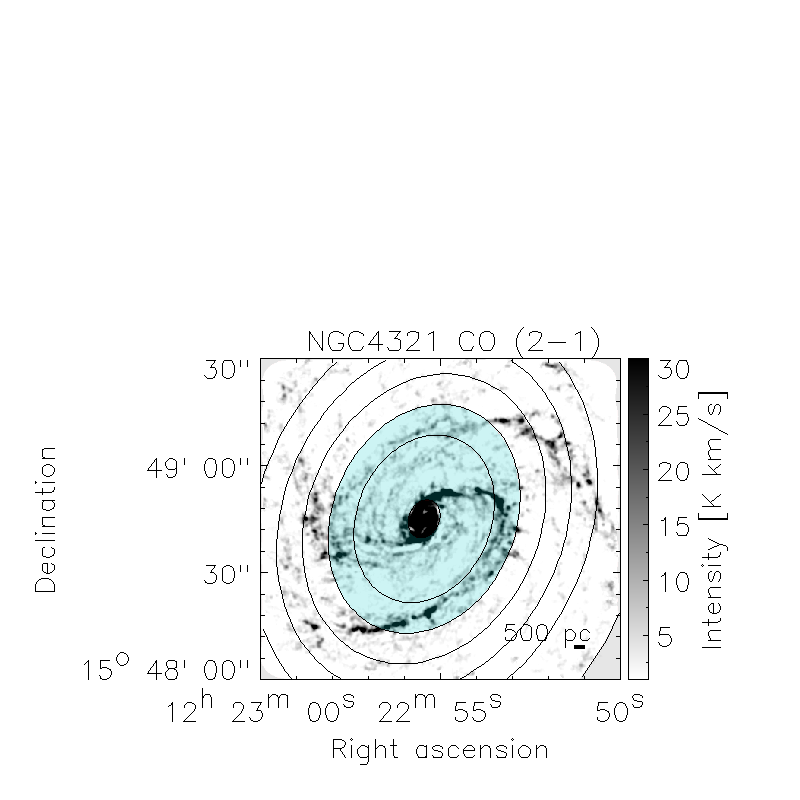}
\includegraphics[trim=12mm 2mm 18mm 140mm, clip=true, width=5.5cm]{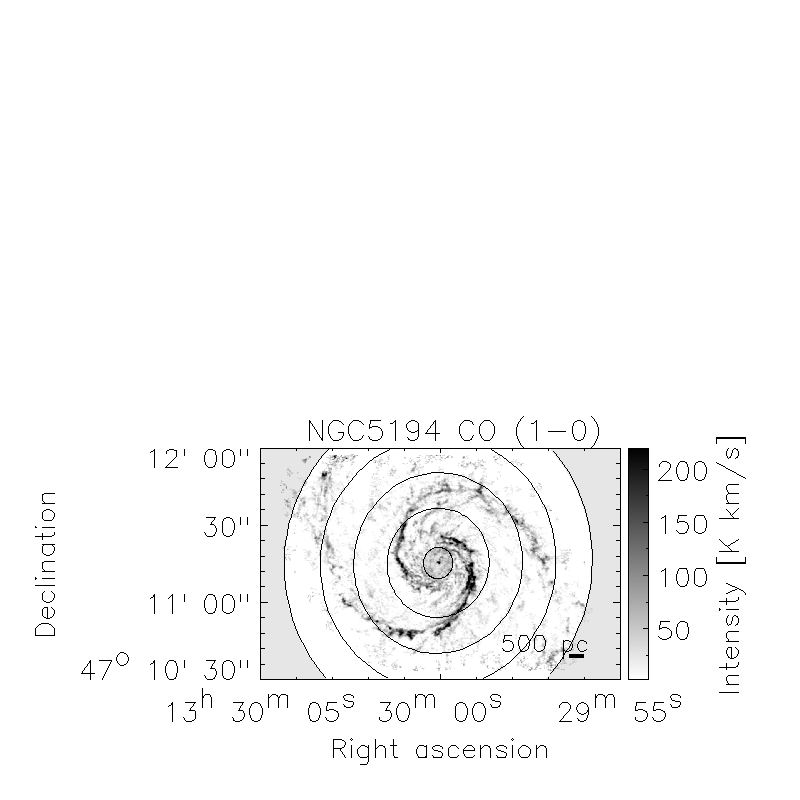}
  \end{minipage}
\caption{Definition of the bins in galactocentric radius for each galaxy, outlined by black ellipses. The grey-shaded centres were identified by eye and have been excluded from the analysis. The same applies to the regions outside the outer ellipse, defined as the outer radius at which an emission peak is identified in either of the two maps. The bins containing a bar or the end of a bar are coloured in light blue. The background images show the CO(2-1) intensity maps [CO(1-0) for NGC5194].}
\label{fig:Bin_images}
\end{figure*}

Here we present the (galactocentric) radial profiles of all quantities necessary to reproduce our analysis of the data and the comparison with analytical models in Figure~\ref{fig:radprof}. The position and width of the radial bins as defined in Section~\ref{sec:binning} are outlined in Figure~\ref{fig:Bin_images}, also highlighting the regions affected by bars (light blue) and the masked central and outer regions (grey). For each of these bins, the reference timescale (\tref), the SFR, and the gas conversion factor (\aco) used as input in the \textsc{Heisenberg} code are presented in Figure~\ref{fig:Bin_inputs}, in addition to the values used when analysing each galaxy in its entirety. For each bin, \tref\ and \aco\ are calculated using the metallicity dependence from \cite{Haydon2019} and \cite{Bolatto2013}, respectively. The galaxy-scale values are based on the CO flux-weighted average metallicity. Note that the galaxy-wide SFR by definition corresponds to the sum of the individual bins. Table~\ref{tab:all_tgas} summarises the measured cloud lifetimes (\tCO), for each galaxy and each individual radial bin.

Figure~\ref{fig:gal_prop} shows the radial profiles of properties describing the galaxies, i.e.\ the molecular gas surface density $\Sigma_{\text{\HH}}$, the stellar surface density $\Sigma_{\text{stars}}$, the SFR surface density $\Sigma_{\text{SFR}}$, the circular velocity, the Toomre $Q$ stability parameter, and the gas phase metallicity [expressed as $12+\log(\rm O/\rm H)$]. We calculate $\Sigma_{\text{\HH}}$ using the CO flux-weighted average \aco\ from Figure~\ref{fig:Bin_inputs}. The stellar mass surface density profiles are derived from S4G/3.6 \mic\ imaging (M.~Querejeta et al.\ in prep.), in a similar way as presented in \cite{Meidt2012, Meidt2014, Querejeta2014}. The SFR surface density profiles are obtained as described in Section~\ref{sec:sfr}. The rotation curves are derived by fitting a model of projected circulation motion to the observed CO velocity fields, as described in detail by \citet{Lang2019}. Where the data quality is not good enough to perform this measurement (due to noise or missing data, which applies to the outskirts of NGC3351, NGC4303, NGC4321, and NGC4535, as well as for NGC5068), we used the fitted rotation curve as a function of the galactocentric radius $R$, in the form $V_{\rm rot} = V_{0} (2/\pi ) \arctan(R/r_{\rm t})$, where $V_{0}$ and $r_{\rm t}$ have been fitted by \cite{Lang2019}. Toomre $Q$ follows from the above input variables in combination with the second moment of the CO maps (A.~K.\ Leroy et al.\ in prep.) to describe the gas velocity dispersion. We note that, strictly speaking, the second moment is an overestimate of $\sigma_{\rm gas}$ due to the fact that beam smearing may blend lines of sight with different first moments (i.e.\ absolute velocities). However, as we are mostly focusing on the flat part of the galaxy rotation curves (Figure~\ref{fig:gal_prop}), the potential effects of beam smearing are limited. Finally, the metallicities are obtained as discussed in Section~\ref{sec:metallicity}. Beyond the quantities shown here, the comparison to theoretical models in Section~\ref{sec:analytical} uses the stellar velocity dispersion, which is determined according to eq.~22 in the Methods section of \citet{Kruijssen2019}, and the spiral arm pattern speeds from the references listed in Table~\ref{tab:co-rotation}.

Figure~\ref{fig:GMC_prop} shows the radial profiles of the CO flux-weighted average properties of the cloud population, i.e.\ the radius, velocity dispersion $\sigma_{\rm GMC}$, mass, surface density $\Sigma_{\rm GMC}$, volume density $\rho_{\text{\HH}}$, and virial parameter $\alpha_{\rm vir}$. The GMC radii and masses are derived using the output from \textsc{Heisenberg} as described and motivated in Section~\ref{sec:analytical}. The GMC velocity dispersions are taken from the \textsc{CPROPS} GMC catalogues of these galaxies. More details about the application of CPROPS and the properties of the cloud population in this sample of galaxies can be found in E.~Rosolowsky et al.\ (in prep.). The surface densities, volume densities, and virial parameters are derived from the first three quantities.

\begin{figure*}
\includegraphics[width=0.3\linewidth]{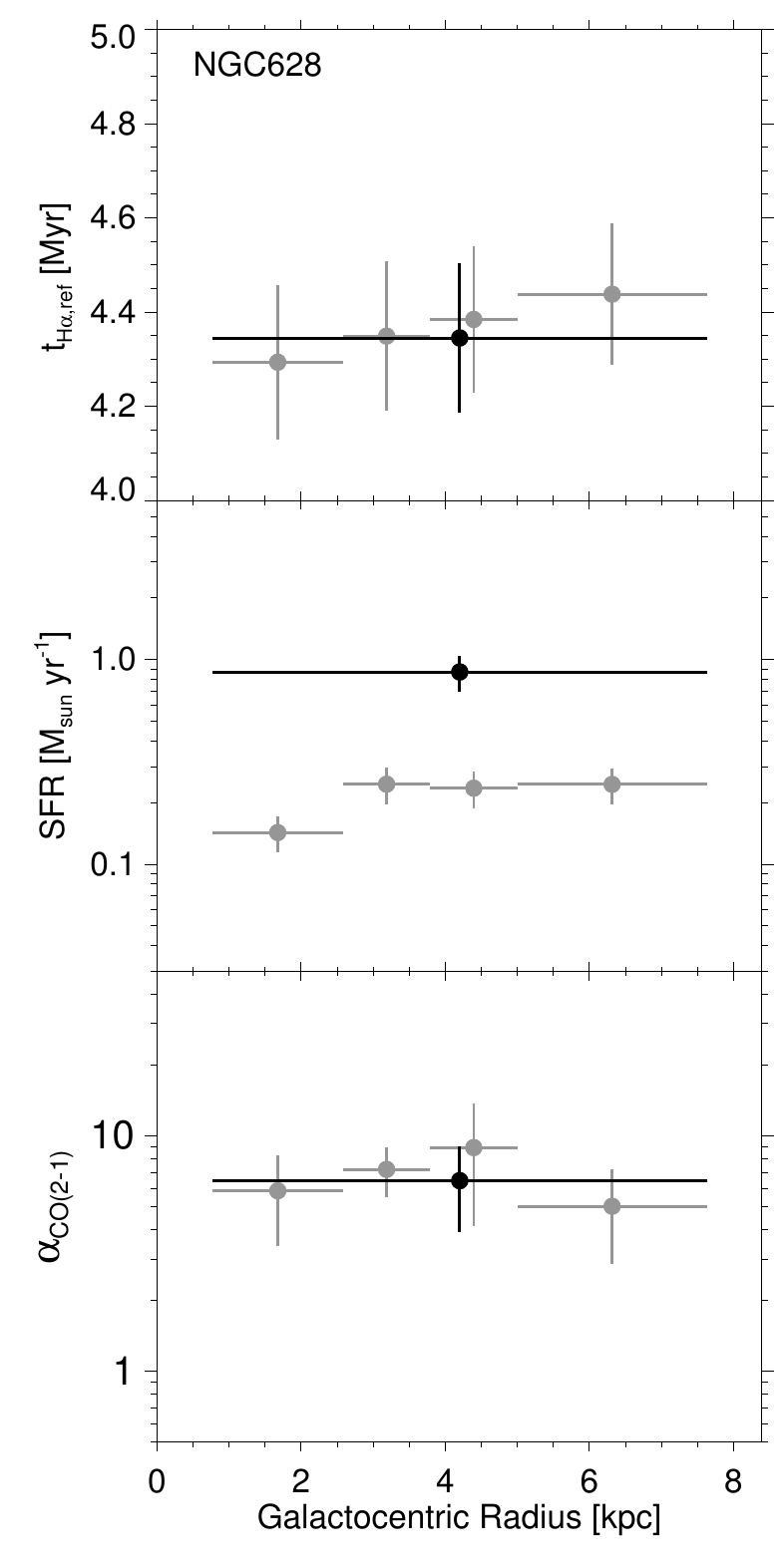}
\includegraphics[width=0.3\linewidth]{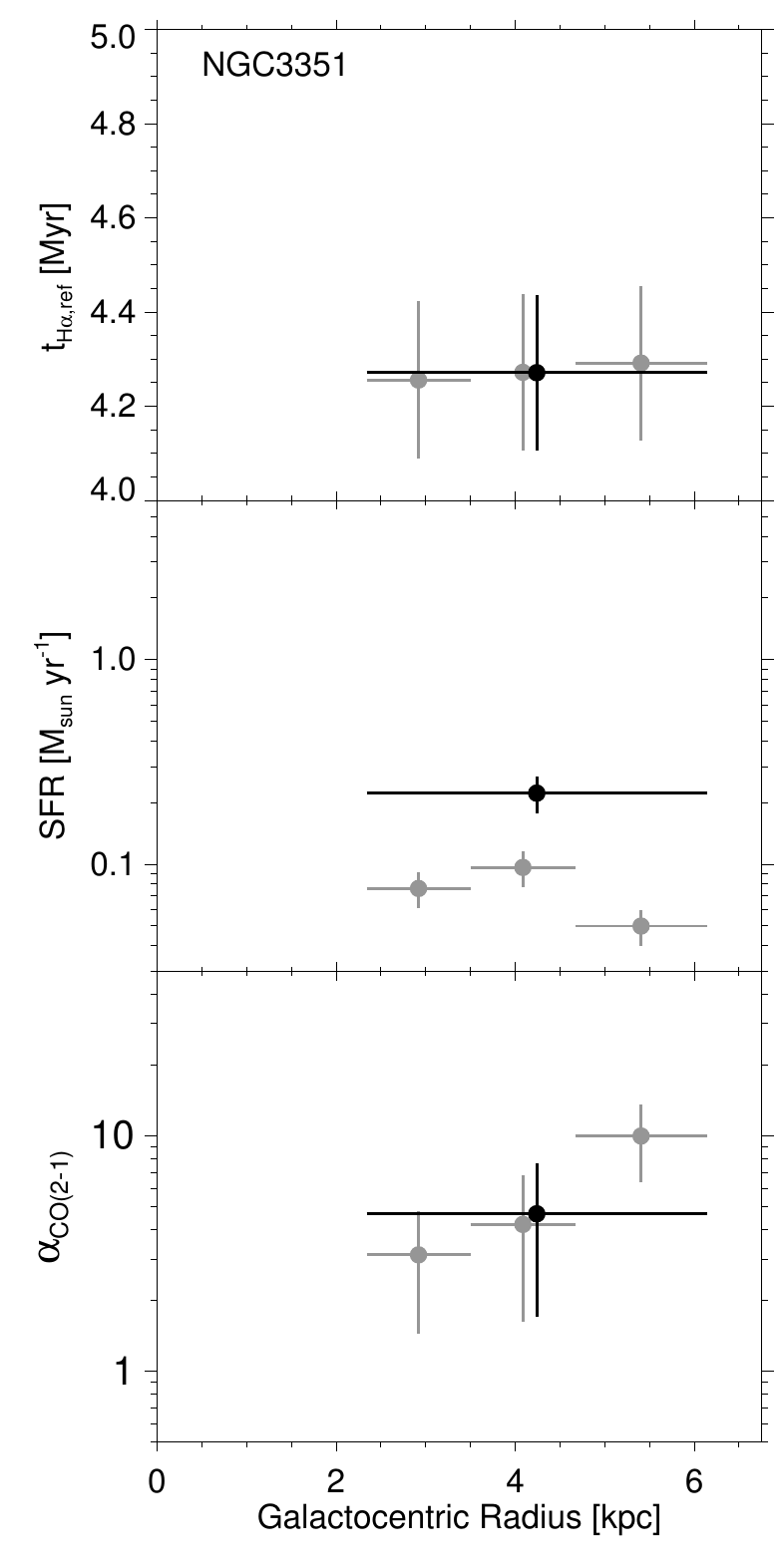}
\includegraphics[width=0.3\linewidth]{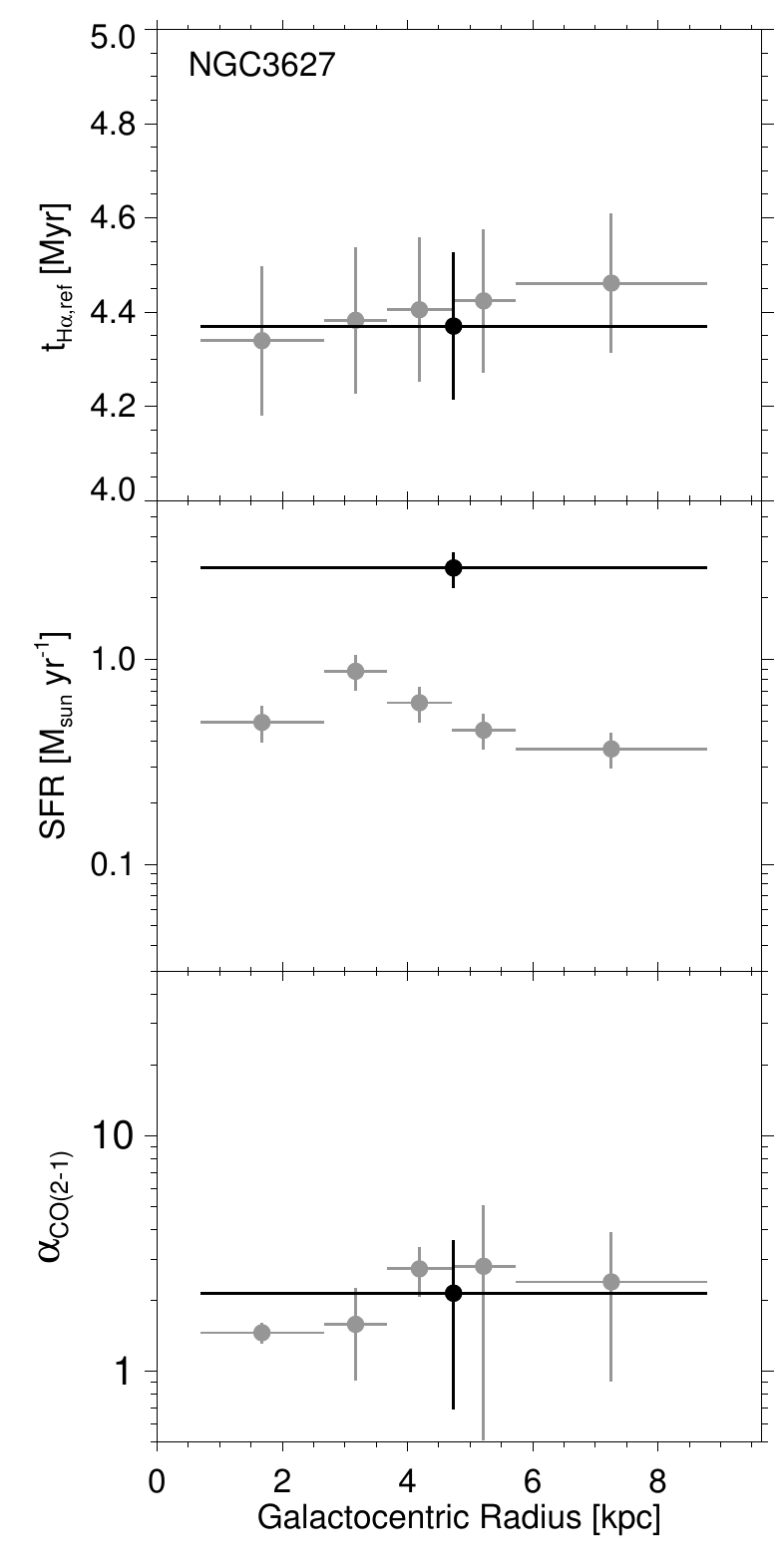}
\includegraphics[width=0.3\linewidth]{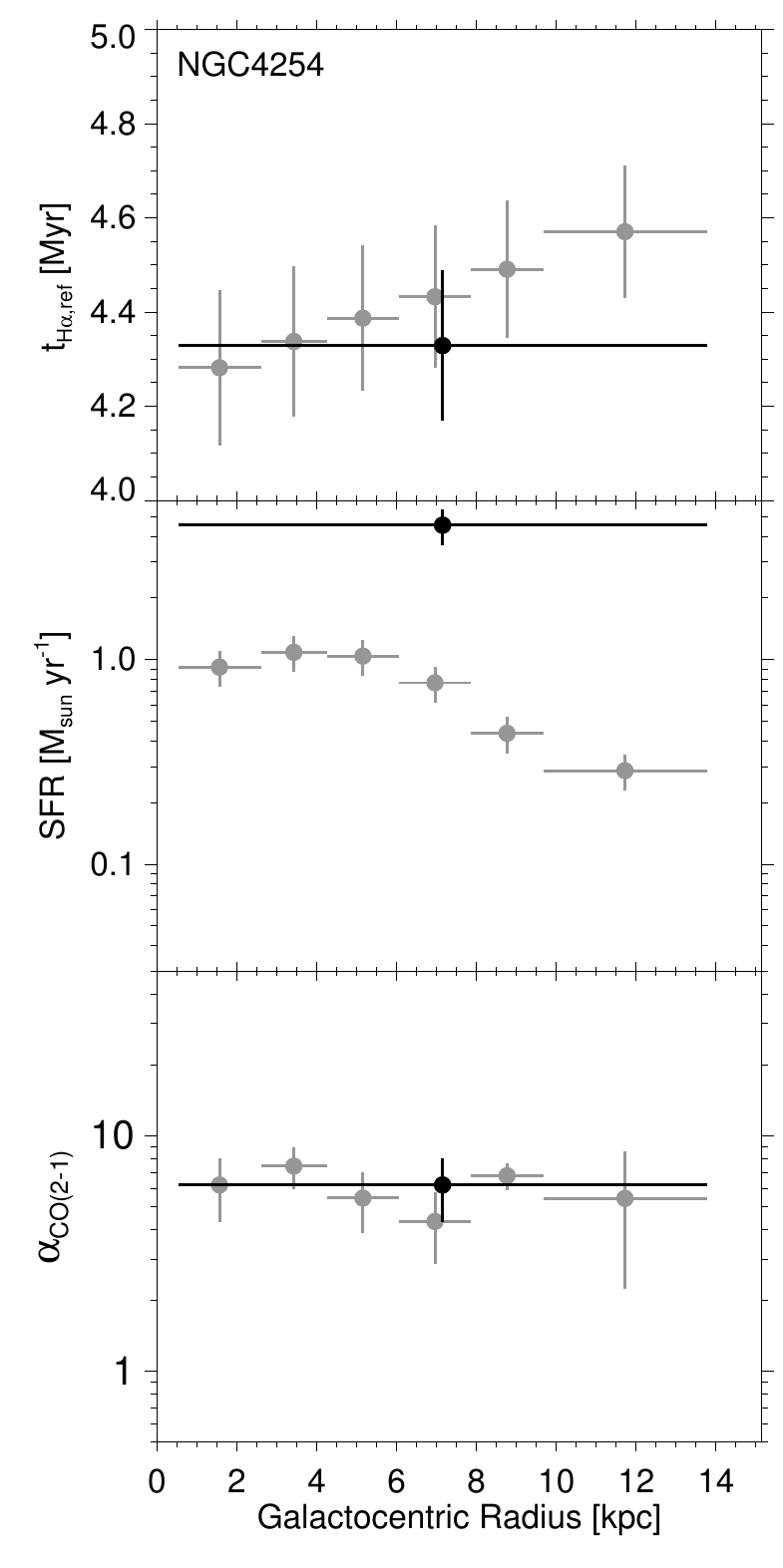}
\includegraphics[width=0.3\linewidth]{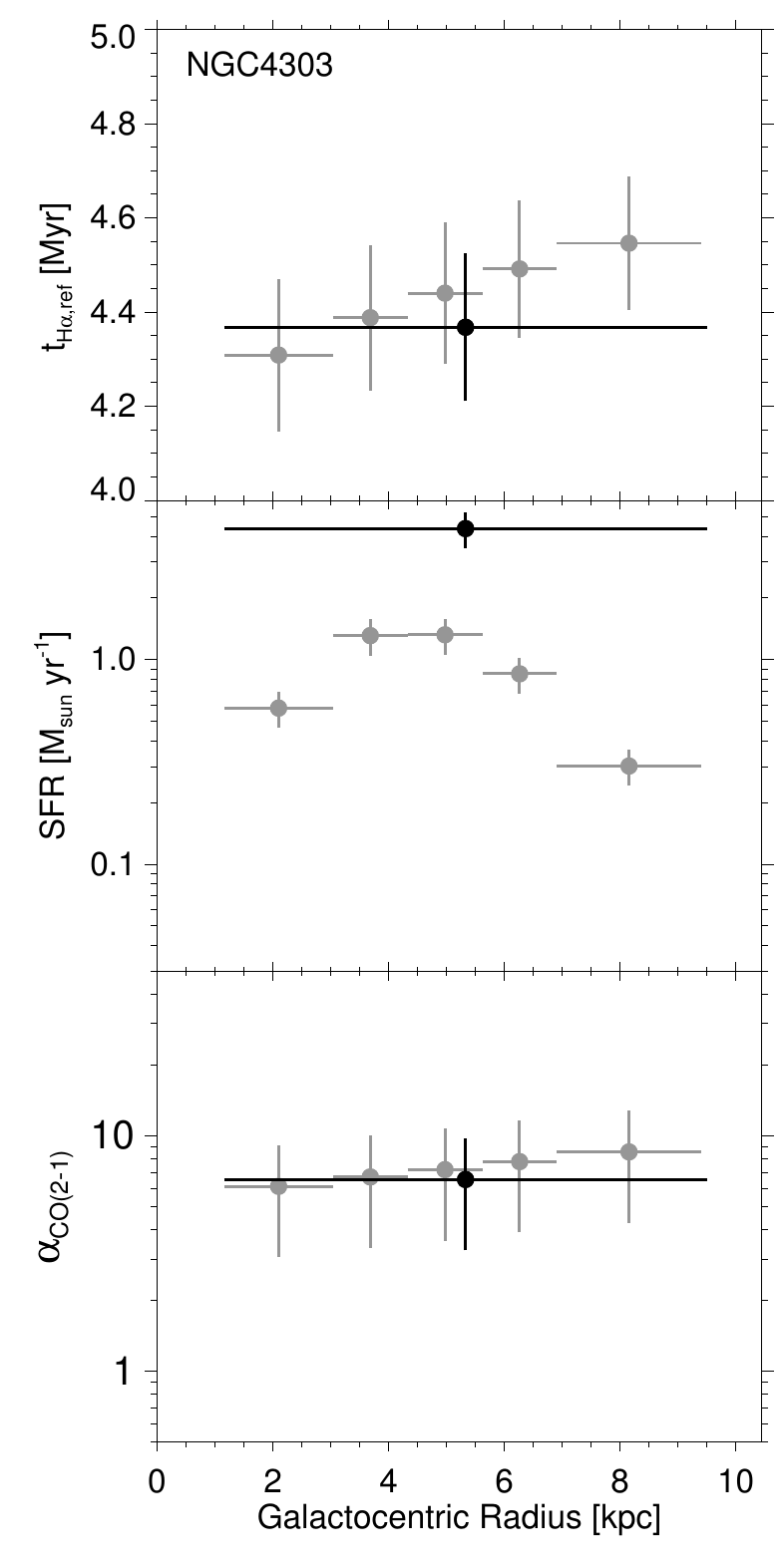}
\includegraphics[width=0.3\linewidth]{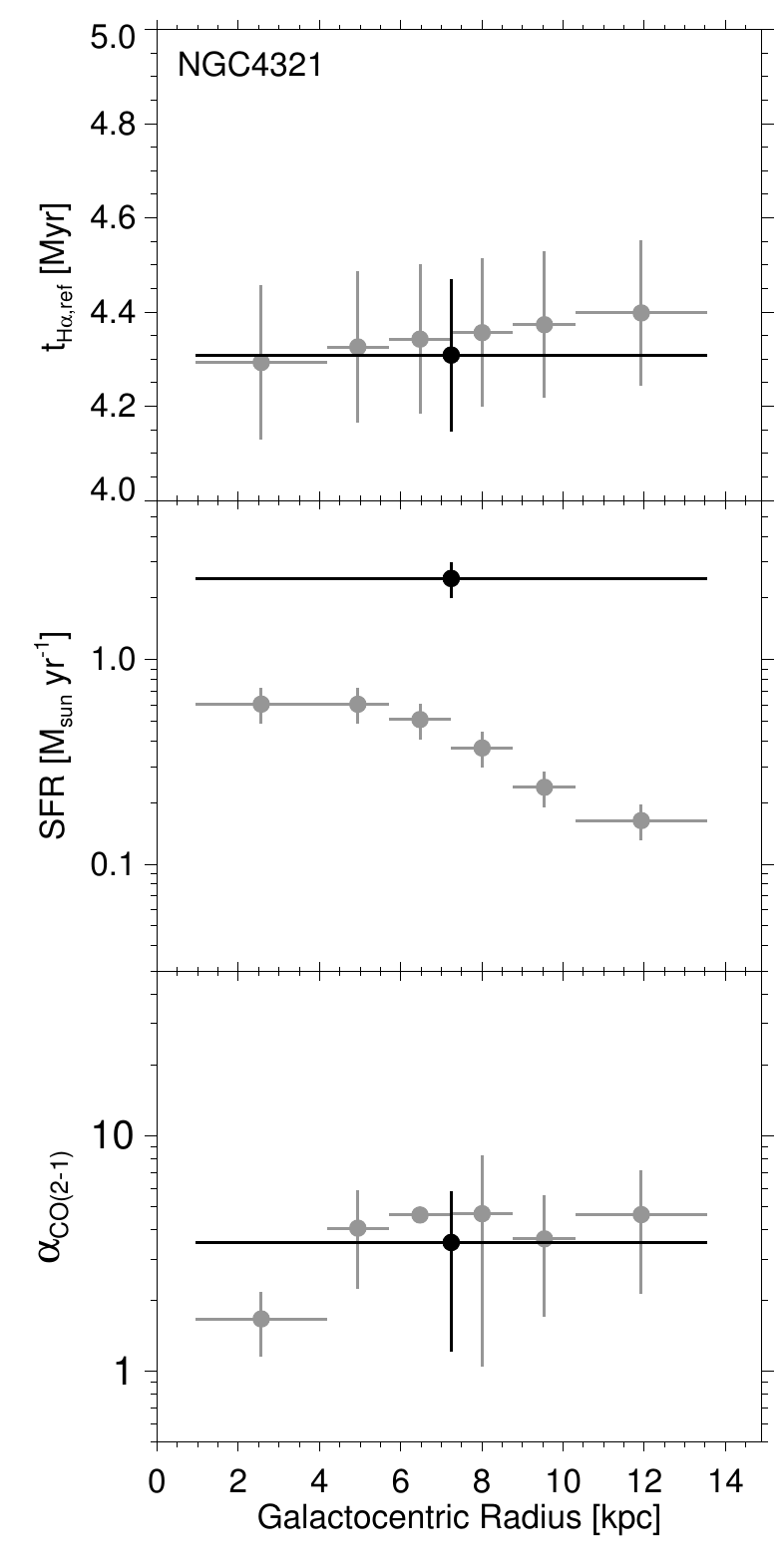}
\caption{Radial profiles of the input quantities of our analysis, i.e.\ \tHaref, absolute SFR and \aco. The black data points show the values for the entire galaxy (which are averages for \tHaref\ and \aco, and the total for the SFR), whereas the grey data points show the same for each individual bin of galactocentric radius. For each data point, the horizontal bar represents the range of radii within which these quantities are measured and the vertical bar represents the $1\sigma$ uncertainties.}
\label{fig:Bin_inputs}
\end{figure*}
\begin{figure*}

\includegraphics[width=0.3\linewidth]{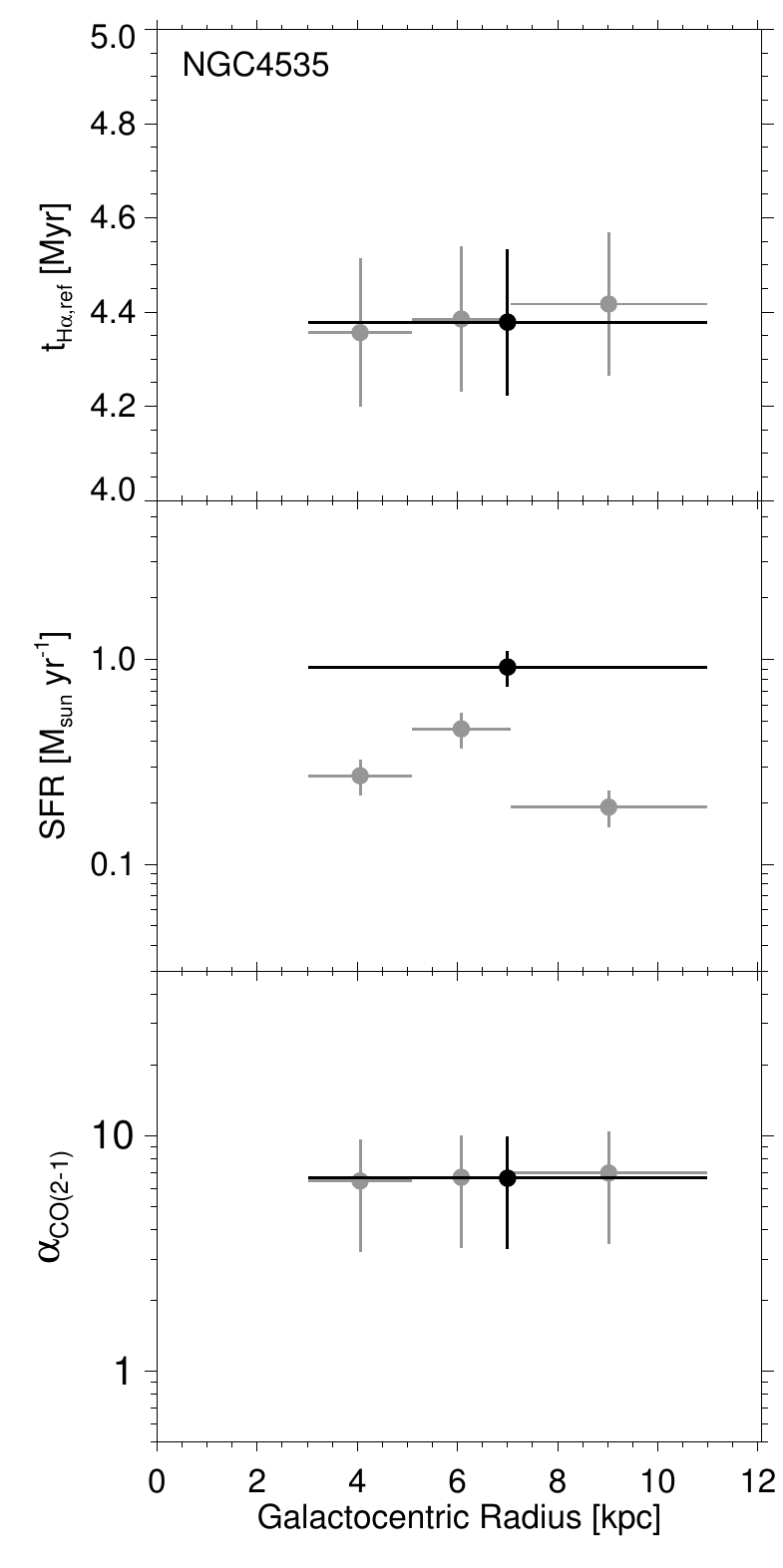}
\includegraphics[width=0.3\linewidth]{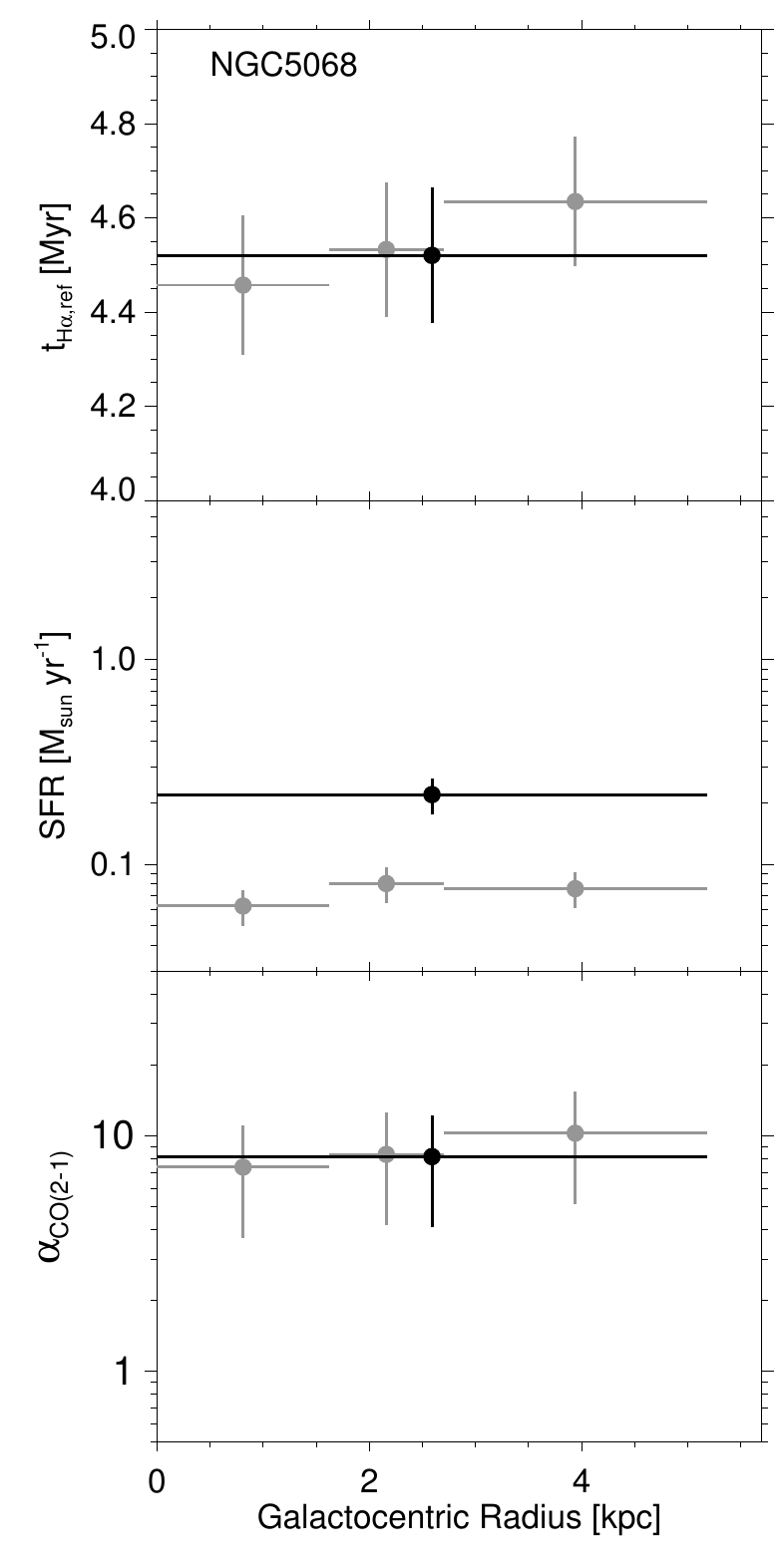}
\includegraphics[width=0.3\linewidth]{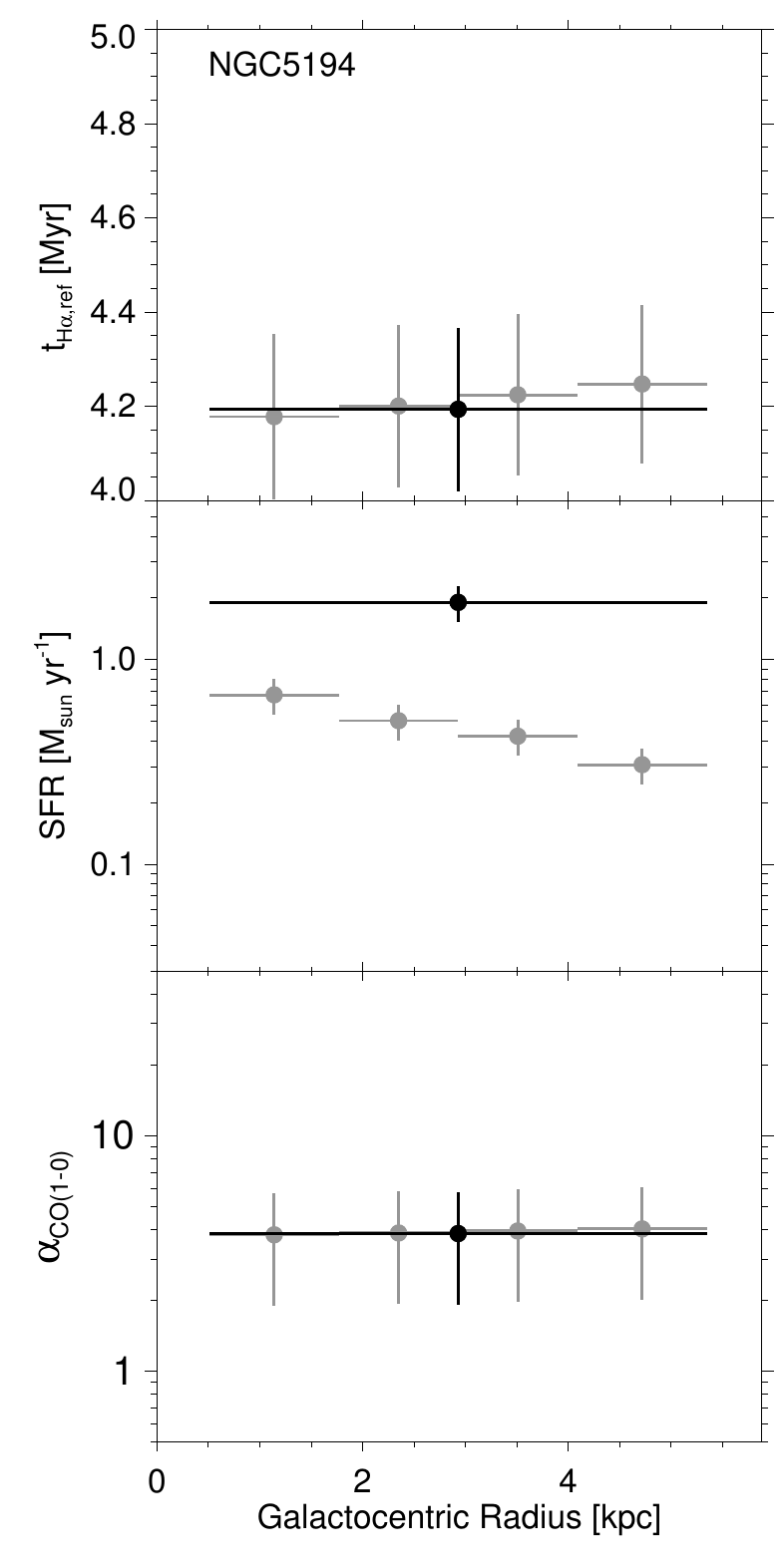}
\contcaption{}
\end{figure*}

\begin{table*}
\begin{center}
\begin{tabular}{lllllllll}
\hline
                    NGC0628  &      Radial interval [kpc]  &                     entire  &                  0.77-2.58  &                  2.58-3.79  &                  3.79-5.00  &                  5.00-7.63  &                             &                             \\
                             &        t$_{\rm GMC}$ [Myr]  &       $24.0^{+3.6}_{-2.5}$  &       $20.0^{+5.1}_{-3.6}$  &       $16.4^{+4.3}_{-2.8}$  &       $14.7^{+3.3}_{-2.9}$  &      $24.9^{+10.3}_{-4.9}$  &                             &                             \\[2.5ex]
                    NGC3351  &      Radial interval [kpc]  &                     entire  &                  2.34-3.50  &                  3.50-4.67  &                  4.67-6.14  &                             &                             &                             \\
                             &        t$_{\rm GMC}$ [Myr]  &       $20.6^{+3.4}_{-3.0}$  &       $26.3^{+8.5}_{-5.7}$  &       $16.4^{+3.9}_{-2.8}$  &       $15.3^{+8.7}_{-2.9}$  &                             &                             &                             \\[2.5ex]
                    NGC3627  &      Radial interval [kpc]  &                     entire  &                  0.69-2.66  &                  2.66-3.68  &                  3.68-4.70  &                  4.70-5.73  &                  5.73-8.78  &                             \\
                             &        t$_{\rm GMC}$ [Myr]  &       $18.9^{+3.4}_{-3.2}$  &       $20.0^{+7.9}_{-4.3}$  &        $3.8^{+2.6}_{-0.8}$  &     $43.2^{+87.7}_{-11.5}$  &     $45.7^{+14.8}_{-11.6}$  &      $28.7^{+12.0}_{-6.3}$  &                             \\[2.5ex]
                    NGC4254  &      Radial interval [kpc]  &                     entire  &                  0.53-2.60  &                  2.60-4.25  &                  4.25-6.06  &                  6.06-7.86  &                  7.86-9.67  &                 9.67-13.77  \\
                             &        t$_{\rm GMC}$ [Myr]  &       $20.9^{+3.9}_{-2.3}$  &      $17.6^{+36.9}_{-4.1}$  &       $14.8^{+6.9}_{-2.3}$  &       $19.1^{+5.0}_{-2.7}$  &    $105.4^{+55.1}_{-23.4}$  &       $17.3^{+5.5}_{-3.7}$  &       $21.5^{+7.4}_{-4.3}$  \\[2.5ex]
                    NGC4303  &      Radial interval [kpc]  &                     entire  &                  1.16-3.10  &                  3.10-4.39  &                  4.39-5.68  &                  5.68-6.97  &                  6.97-9.50  &                             \\
                             &        t$_{\rm GMC}$ [Myr]  &       $16.9^{+4.6}_{-2.2}$  &        $7.9^{+4.8}_{-1.9}$  &      $17.8^{+13.7}_{-4.0}$  &      $15.6^{+15.8}_{-2.8}$  &      $21.6^{+20.5}_{-4.7}$  &       $11.3^{+3.1}_{-2.1}$  &                             \\[2.5ex]
                    NGC4321  &      Radial interval [kpc]  &                     entire  &                  0.95-4.18  &                  4.18-5.71  &                  5.71-7.24  &                  7.24-8.77  &                 8.77-10.31  &                10.31-13.54  \\
                             &        t$_{\rm GMC}$ [Myr]  &       $19.1^{+2.3}_{-2.2}$  &      $31.2^{+19.3}_{-6.1}$  &       $16.0^{+3.6}_{-2.7}$  &       $16.2^{+5.5}_{-2.2}$  &       $20.2^{+4.2}_{-3.9}$  &       $11.7^{+4.2}_{-2.0}$  &      $29.3^{+20.1}_{-7.3}$  \\[2.5ex]
                    NGC4535  &      Radial interval [kpc]  &                     entire  &                  3.02-5.09  &                  5.09-7.06  &                 7.06-10.98  &                             &                             &                             \\
                             &        t$_{\rm GMC}$ [Myr]  &       $26.4^{+4.7}_{-3.6}$  &     $61.3^{+92.4}_{-19.2}$  &       $25.9^{+5.0}_{-3.2}$  &       $16.7^{+8.1}_{-3.7}$  &                             &                             &                             \\[2.5ex]
                    NGC5068  &      Radial interval [kpc]  &                     entire  &                  0.00-1.62  &                  1.62-2.70  &                  2.70-5.18  &                             &                             &                             \\
                             &        t$_{\rm GMC}$ [Myr]  &        $9.6^{+2.9}_{-1.8}$  &       $17.2^{+3.7}_{-2.6}$  &        $7.7^{+9.0}_{-2.6}$  &        $5.7^{+3.0}_{-1.1}$  &                             &                             &                             \\[2.5ex]
                    NGC5194  &      Radial interval [kpc]  &                     entire  &                  0.51-1.77  &                  1.77-2.93  &                  2.93-4.09  &                  4.09-5.35  &                             &                             \\
                             &        t$_{\rm GMC}$ [Myr]  &       $30.5^{+9.2}_{-4.8}$  &      $20.1^{+13.5}_{-4.4}$  &     $59.7^{+70.3}_{-18.2}$  &      $17.0^{+10.2}_{-3.5}$  &      $37.3^{+24.9}_{-9.5}$  &                             &                             \\[2.5ex]
\hline
\end{tabular}
\caption{Measured molecular cloud lifetimes for each galaxy in its entirety, as well as in each individual radial bin.}
\label{tab:all_tgas}
\end{center}
\end{table*}

\begin{figure*}
\includegraphics[trim=0mm 0mm 0mm 0mm, clip=true]{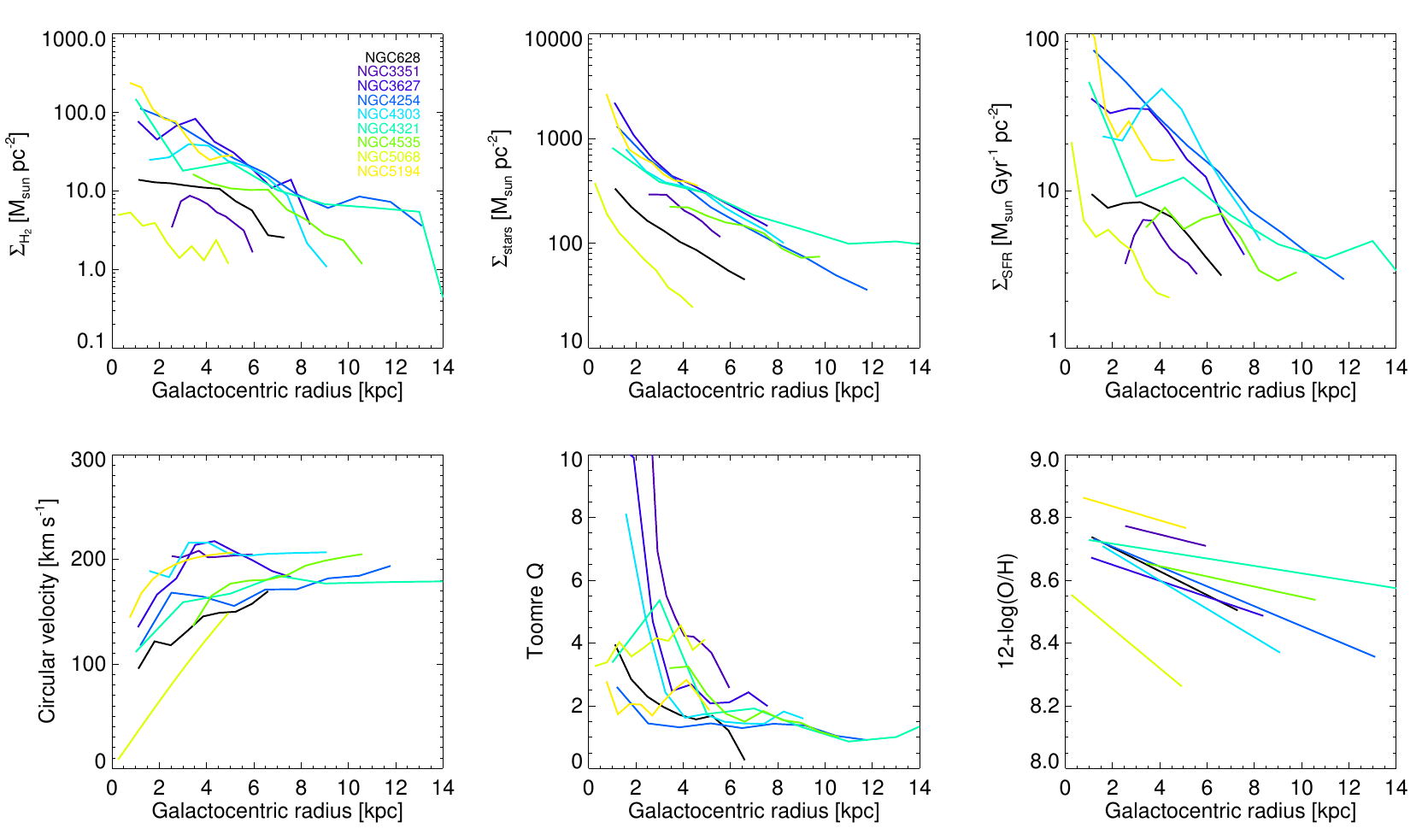}
\caption{Radial profiles of properties describing the galaxies in our sample. From left to right, the top row shows the molecular gas surface density (using the conversion factor from Table~\ref{tab:input}), the stellar mass surface density and the SFR surface density (calculated as described in Section~\ref{sec:sfr}). The bottom row shows the rotation curve \citep{Lang2019}, the Toomre Q parameter and the gas-phase metallicity gradient \citep[see Section~\ref{sec:metallicity}]{Pilyugin2014}.}
\label{fig:gal_prop}
\end{figure*}

\begin{figure*}
\includegraphics[trim=0mm 0mm 0mm 0mm, clip=true]{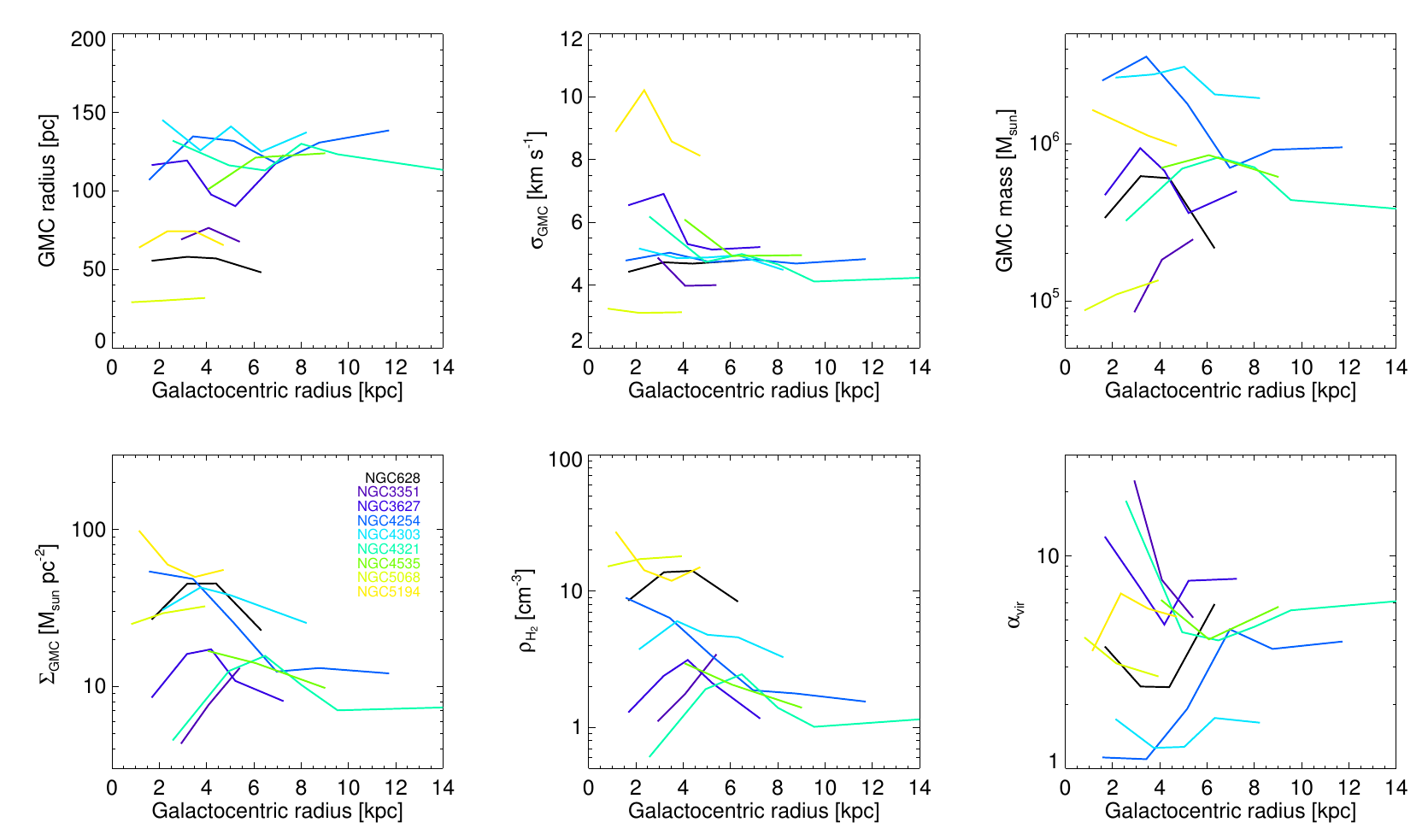}
\caption{Radial profiles of the CO flux-weighted average properties of the molecular cloud population of each galaxy. From left to right, the top row shows the CO flux-weighted average cloud radius, velocity dispersion, and luminous mass as a function of galactocentric radius. The bottom row shows the CO flux-weighted average molecular gas surface density of clouds, the \HH\ number density and the virial parameter. The quantities in the top row are derived from the output of \textsc{Heisenberg} (for the GMC radius and mass) and the \textsc{CPROPS} GMC catalogue (E.~Rosolowsky et al.\ in prep.; for the velocity dispersion). The quantities in the bottom row are derived from those in the top row.}
\label{fig:GMC_prop}
\end{figure*}

\section{Gas surface density threshold separating galactic and cloud-scale dynamics} \label{sec:appthres}
In Section~\ref{sec:galactic dynamics}, we investigate whether cloud lifetimes are set by internal dynamics (i.e.\ the cloud crossing time or free-fall time) or galactic dynamics (i.e.\ the combination of mechanisms considered by \citealt{Jeffreson2018}, see Section~\ref{sec:analytical}). By characterising the properties of the radial bins in Figure~\ref{fig:radprof} where the measured cloud lifetimes better agree with the red (internal) and blue (galactic) lines, we find that both situations can occur and seem to occupy different ranges of the large-scale molecular gas surface density (i.e.\ averaged on kpc scales, across the entire radial bin). At low surface densities, cloud lifetimes follow the cloud crossing time or free-fall time, whereas at high surface densities, they match the galactic dynamical time-scale. Figure~\ref{fig:timescales} quantifies this statement by dividing the sample of radial bins in which the cloud lifetime is measured at a surface density of $\Sigma_{\rm H_2,ring}=8\,M_{\odot}$\,pc$^{-2}$ and considering the difference between the measured cloud lifetime and the galactic and internal dynamical time-scales. To arrive at this critical surface density threshold separating both regimes, we have systematically varied the density threshold at which the sample of radial bins is divided into low and high densities.

\begin{figure}
\includegraphics[trim=0mm 0mm 0mm 0mm, clip=true, width=1\linewidth]{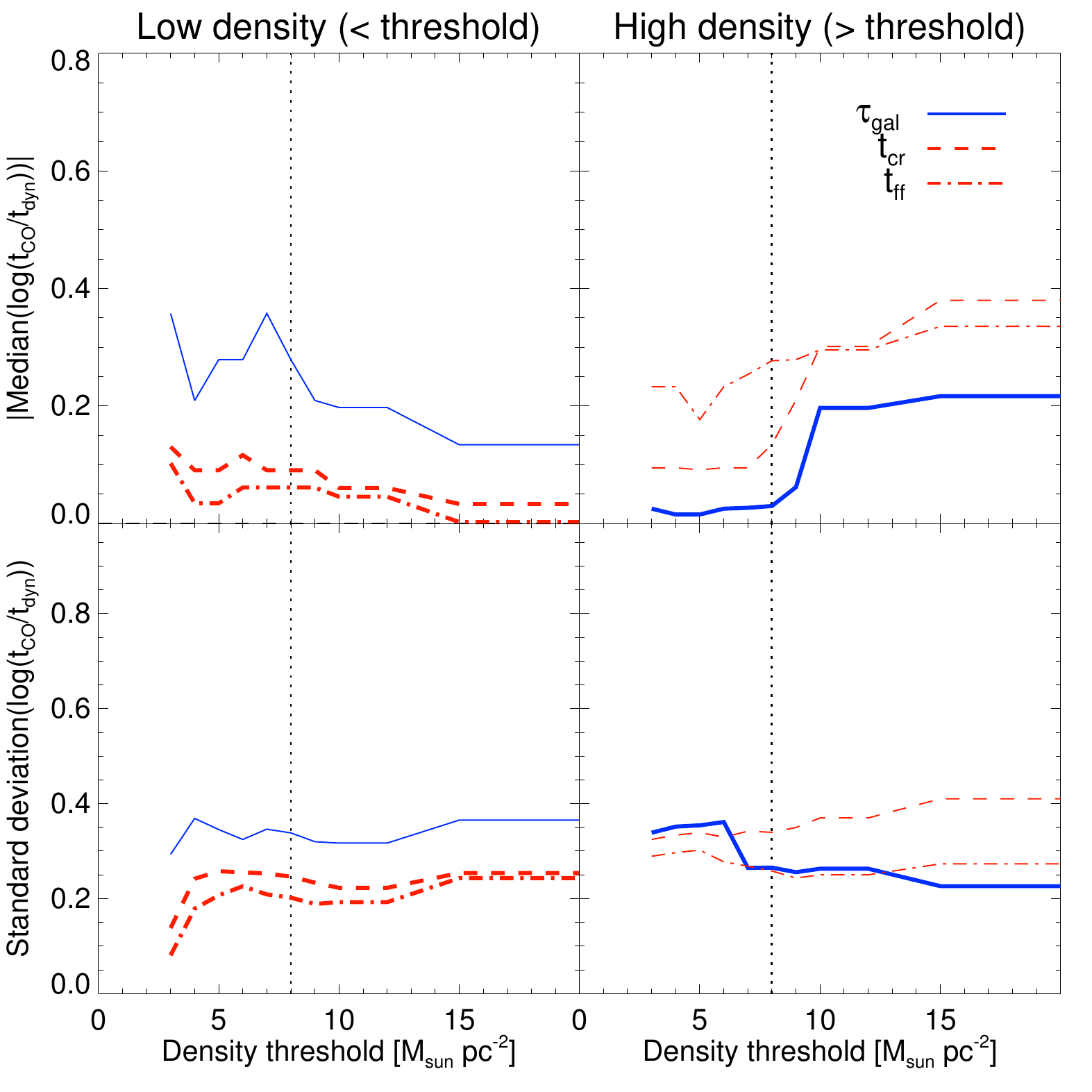}
\caption{Median absolute logarithmic offset between the measured cloud lifetime (\tCO) and the predicted internal (\tcr\ and \tff) and galactic (\tgal) dynamical times (top panels), as well as the standard deviation of this offset (bottom panels), for clouds in environments of low (left-hand panels) and high (right-hand panels) kpc-scale molecular gas surface densities, as a function of the density threshold used to divide the sample into `low' and `high' gas surface densities. See Figure~\ref{fig:timescales} for reference. A perfect correlation corresponds to a median and scatter of zero. We find that the cloud lifetime in high density regions (respectively low density regions) correlates best with \tgal\ (respectively \tcr\ and \tff) when dividing the regions at a density threshold of $\Sigma_{\rm H_2,ring}=8\,M_{\odot}$\,pc$^{-2}$, indicated by the vertical dotted line. Figure~\ref{fig:timescales} compares \tCO, \tgal, \tcr, and \tff\ for the two `low' and `high' gas surface density regimes separated by this density threshold.}
\label{fig:tGMC_tgas_tff_dispersion}
\end{figure}

Figure~\ref{fig:tGMC_tgas_tff_dispersion} shows how the choice of the surface density threshold affects the median absolute logarithmic offset between the measured cloud lifetime (\tCO) and the predicted internal (\tcr\ and \tff) and galactic (\tgal) dynamical times, as well as the standard deviation of this offset. While the cloud lifetimes in low-density regions are relatively insensitive to the choice of threshold and generally correlate well with \tcr\ and \tff\ (left-hand panels), the high-density regions only correlate well with the galactic dynamical time-scale if a sufficiently high threshold density is used to define `high density' environments (i.e.\ $\Sigma_{{\rm H}_2}>8\,M_{\odot}$\,pc$^{-2}$). The corresponding scatter exhibits a steep decrease for $\Sigma_{{\rm H}_2} \geq 7\,M_{\odot}$\,pc$^{-2}$ (bottom-right panel), which is also where the median offset between $t_{\rm CO}$ and $t_{\rm ff}$ starts to drop (top-left panel). For threshold values higher than $9\,M_{\odot}$\,pc$^{-2}$, the prediction due to galactic dynamics develops an offset from the observed cloud lifetimes (top-right panel). Finally, the cloud lifetimes in low-density environments start matching the prediction for galactic dynamics to within 0.2~dex when adopting threshold densities of $\Sigma_{{\rm H}_2}>12\,M_{\odot}$\,pc$^{-2}$ (top-left panel). In the light of these observations, we adopt a threshold value of $\Sigma_{{\rm H}_2}=8\,M_{\odot}$\,pc$^{-2}$.

\begin{figure}
\includegraphics[trim=0mm 0mm 0mm 0mm, clip=true, width=1\linewidth]{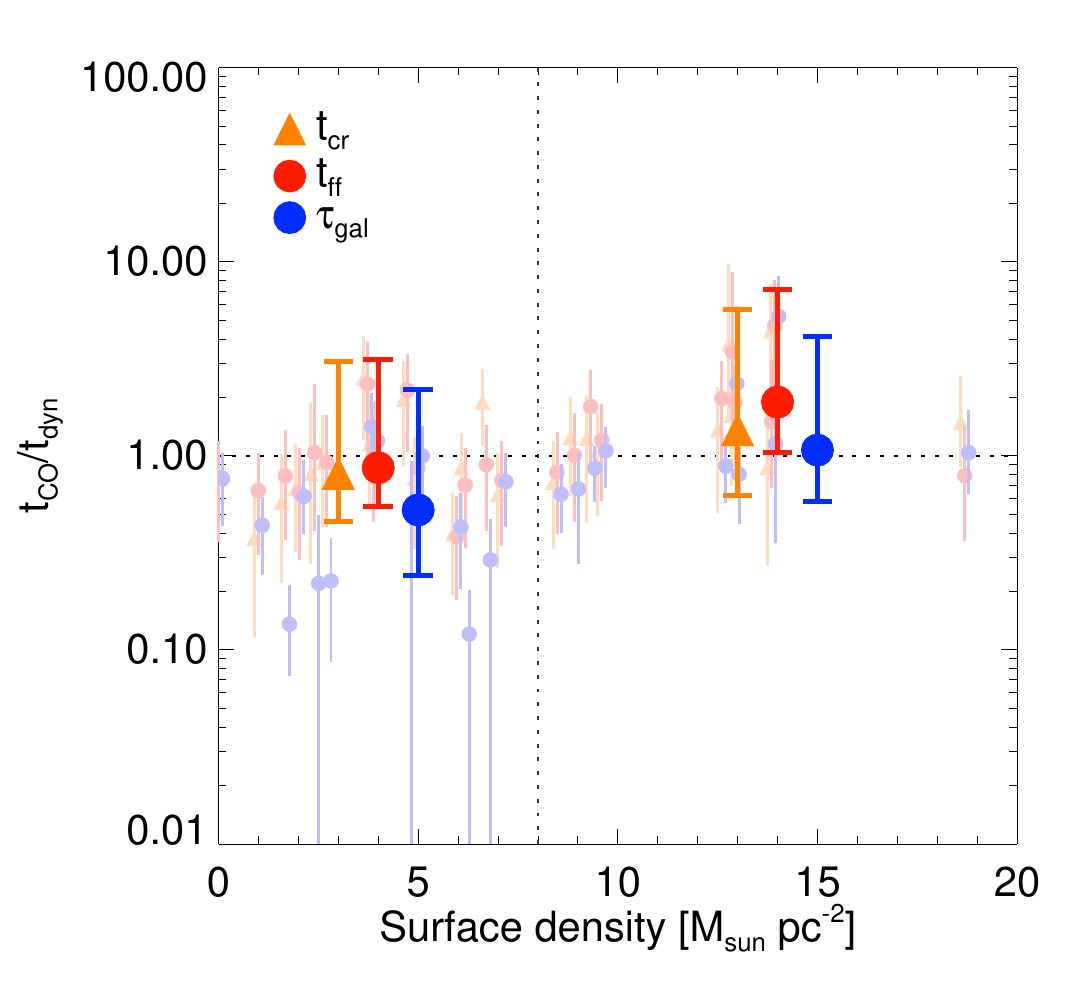}
\caption{Ratio of the measured cloud lifetime \tCO\ over the dynamical time (\tgal\ in pale blue dots, \tcr\ in pale orange triangles, \tff\ in pale red dots) as a function of the galactic gas surface density $\Sigma_{\rm H_2,ring}$. On either side of $\Sigma_{\rm H_2,ring} = 8$\,\Msun\ pc$^{-2}$ (vertical dotted line), the median and standard deviation of the ratios are indicated by the large, bright symbols, respectively. At surface densities lower than this surface density threshold, we note a larger dispersion of \tCO/\tgal\ compared to \tCO/\tcloud, as well as a median of $t_{\rm CO}/\tau_{\rm gal}\approx0.5$, whereas $t_{\rm CO}/t_{\rm cr}\approx t_{\rm CO}/t_{\rm ff}\approx1$. By contrast, at surface densities higher than this threshold, the dispersions of the ratios are all similar, with absolute values of $t_{\rm CO}/\tau_{\rm gal}\approx1$, $t_{\rm CO}/t_{\rm cr}\approx 1.9$ and $t_{\rm CO}/t_{\rm cr}\approx1.3$. See Section~\ref{sec:galactic dynamics} for a detailed discussion and physical interpretation of these results.}
\label{fig:tGMC_tgas_tcloud_vs_surfdens}
\end{figure}

An alternative to Figure~\ref{fig:timescales} for visualising the two regimes is shown in Figure~\ref{fig:tGMC_tgas_tcloud_vs_surfdens}, where we demonstrate how the ratios \tCO/\tgal\, \tCO/\tcr\, and \tCO/\tff\ depend on the kpc-scale gas surface density for all of our measurements. At surface densities $\Sigma_{\rm H_2,ring} < 8$\,\Msun pc$^{-2}$, the figure shows that \tCO\ is systematically offset from \tgal\, by a factor of 0.5, and that the ratio between the two quantities shows considerable scatter. By contrast, the median \tCO/\tcr\ and \tCO/\tff\ are close to unity in this regime, with modest scatter. This implies that internal dynamics set the GMC lifetime at low gas surface densities. At surface densities $\Sigma_{\rm H_2,ring} > 8$\,\Msun pc$^{-2}$, the ratio \tCO/\tgal\ is unity, whereas \tCO/\tcr\ and \tCO/\tff\ are now systematically offset from unity, by a factor of 1.3 and 1.9, respectively. The similar scatter of all ratios at high kpc-scale gas surface densities means that the cloud lifetime matches the time-scale for galactic dynamics, as well as a fixed multiple of ($1.3{-}1.9$ times) the internal dynamical time-scale. As discussed in Section~\ref{sec:galactic dynamics}, this close agreement with both time-scales is expected when galactic dynamics set the time-scale for cloud evolution \citep[this is referred to as the `Toomre regime' by][]{Krumholz2012}, because the cloud dynamics adjust to the galactic dynamics in this regime. In summary, Figure~\ref{fig:tGMC_tgas_tff_dispersion} and~\ref{fig:tGMC_tgas_tcloud_vs_surfdens} substantiate the rough division made of our sample into two regimes of kpc-scale molecular gas surface density. Future work with a larger sample of galaxies will need to refine this division.

%%%%%%%%%%%%%%%%%%%%%%%%%%%%%%%%%%%%%%%%%%%%%%%%%%

\vspace{4mm}

\noindent {\it
$^{1}$Astronomisches Rechen-Institut, Zentrum f\"ur Astronomie der Universit\"at Heidelberg, M\"onchhofstra\ss e 12-14, 69120 Heidelberg, Germany\\
$^{2}$Max-Planck Institut f\"ur Astronomie, K\"onigstuhl 17, 69117 Heidelberg, Germany\\
$^{3}$Max-Planck Institut f\"ur Extraterrestrische Physik, Giessenbachstra\ss e 1, 85748 Garching, Germany\\
$^{4}$Astrophysics Research Institute, Liverpool John Moores University, IC2, Liverpool Science Park, 146 Brownlow Hill, Liverpool L3 5RF, UK\\
$^{5}$Research School of Astronomy and Astrophysics, Australian National University, Canberra, ACT 2611, Australia\\
$^{6}$IRAM, 300 rue de la Piscine, 38406 Saint Martin d'H\`eres, France\\
$^{7}$CNRS, IRAP, 9 Av. du Colonel Roche, BP 44346, F-31028 Toulouse cedex 4, France\\
$^{8}$Universit\'{e} de Toulouse, UPS-OMP, IRAP, F-31028 Toulouse cedex 4, France\\
$^{9}$Department of Astronomy, The Ohio State University, 140 West 18th Ave, Columbus, OH 43210, USA\\
$^{10}$Sterrenkundig Observatorium, Universiteit Gent, Krijgslaan 281 S9, B-9000 Gent, Belgium\\
$^{11}$Sorbonne Universit\'e, Observatoire de Paris, Universit\'e PSL, CNRS, LERMA, F-75005, Paris, France\\
$^{12}$Departamento de Astronom\'{i}a, Universidad de Chile, Casilla 36-D, Santiago, Chile\\
$^{13}$4-183 CCIS, University of Alberta, Edmonton, Alberta, Canada\\
$^{14}$Argelander-Institut f\"{u}r Astronomie, Universit\"{a}t Bonn, Auf dem H\"{u}gel 71, 53121 Bonn, Germany\\
$^{15}$The Observatories of the Carnegie Institution for Science, 813 Santa Barbara Street, Pasadena, CA 91101, USA\\
$^{16}$European Southern Observatory, Karl-Schwarzschild-Stra{\ss}e 2, D-85748 Garching bei M\"{u}nchen, Germany\\
$^{17}$Univ.\ Lyon, Univ.\ Lyon1, ENS de Lyon, CNRS, Centre de Recherche Astrophysique de Lyon UMR5574, F-69230 Saint-Genis-Laval France\\
$^{18}$Instit\"ut  f\"{u}r Theoretische Astrophysik, Zentrum f\"{u}r Astronomie der Universit\"{a}t Heidelberg, Albert-Ueberle-Strasse 2, 69120 Heidelberg, Germany\\
$^{19}$Caltech/IPAC, California Institute of Technology, Pasadena, CA, USA\\
$^{20}$Observatorio Astron{\'o}mico Nacional (IGN), C/Alfonso XII 3, Madrid E-28014, Spain
}

\bsp
\label{lastpage}
\end{document}